%% file: root.tex
\documentclass[b5paper,12pt,fleqn]{book}
\usepackage{fancyhdr}
\renewcommand{\headrulewidth}{0.pt}
\renewcommand{\footrulewidth}{0.pt}
\usepackage{rotating}
\usepackage[english]{babel}
\usepackage[T1]{fontenc}
\usepackage[left=2.3cm,right=2.3cm,bottom=3cm,top=2cm]{geometry}
\usepackage{multirow}
\usepackage{amsmath, amssymb}
\usepackage{empheq}
\usepackage{graphicx}
\usepackage{listings}
\usepackage{wrapfig}
\usepackage[toc,page]{appendix}
\usepackage{picture}
\usepackage{emptypage}
\usepackage{siunitx}
\usepackage[version=4]{mhchem}
\usepackage{placeins}
\usepackage{dsfont}
\usepackage{bm}
\usepackage{xfrac}
\usepackage{setspace}
\usepackage{slashed}
\usepackage{stmaryrd}
\usepackage[super,nomove]{cite}
\usepackage[pagebackref=false]{hyperref}
\usepackage{mathrsfs}
\usepackage{fancyref}
\usepackage{varioref}
\usepackage{booktabs}
\usepackage{color}
\usepackage{wrapfig}
\usepackage{caption}
\usepackage{pdfpages}
\usepackage{feynmp}
\usepackage{float}
\usepackage[vcentermath]{youngtab}
\usepackage{tocvsec2}
\usepackage{resizegather}
\usepackage{adjustbox}
\usepackage{ragged2e}
\usepackage{subcaption}

\usepackage{palatino}

\maxtocdepth{subsection}
\graphicspath{{figures/}}
\usepackage{bm}
\DeclareSIUnit\molar{M}
\makeatletter

\newcommand{\tr}[1]{\textrm{#1}}
\newcommand{\ket}[1]{|{#1}\rangle}
\newcommand{\bra}[1]{\langle{#1}|}
\newcommand{\epv}[1]{\langle{#1}\rangle}
\renewcommand{\@makechapterhead}[1]{%
\vspace*{50 pt}%
{\setlength{\parindent}{0pt} \raggedright \normalfont
\bfseries\Huge
\ifnum \value{secnumdepth}>1 
   \if@mainmatter\chaptername~\thechapter.\ \fi%
\fi
#1\par\nobreak\vspace{40 pt}}}
\makeatother
\DeclareMathAlphabet{\mathpzc}{OT1}{pzc}{m}{it}
\definecolor{au}{cmyk}{1,0.69,0,0.11}
\title{Pulse Sequence Design in solid-state NMR}
\author{Ravi Shankar Palani}

\begin{document}

\begin{titlepage}
	\begin{center}

	{\huge \textbf{Pulse Design in Solid-State}}\\[4pt]
	{\huge \textbf{Nuclear Magnetic Resonance}}\\[4pt]
	{\textsc{Study and Design of Dipolar Recoupling Experiments}}\\
	{\textsc{in Spin-$1/2$ Nuclei}}\\[4cm]

	\includegraphics[width=0.3\textwidth]{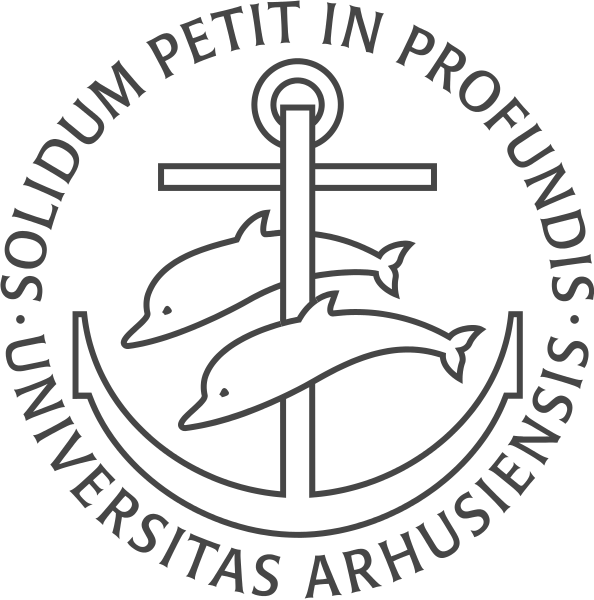}\\[1.0cm]
	
	\textsc{Ravi Shankar Palani}\\[1cm]

	{Ph.D. Dissertation,} \\
	{under the guidance of Prof. Niels Christian Nielsen,}\\
	{Interdisciplinary Nanoscience Center (iNANO),}\\
	{Faculty of Science, Aarhus University.}\\[1cm]
	
	{\textsc{October 2016}}\\
	\begin{center}
		\includegraphics[width=0.3\linewidth]{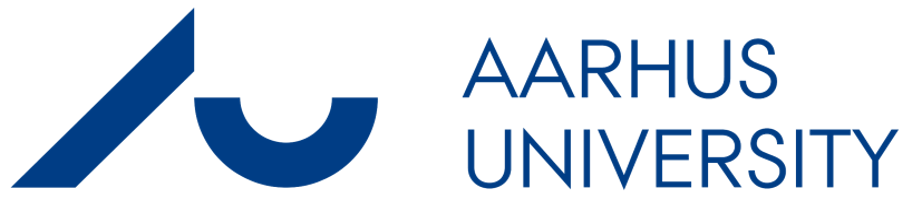}
	\end{center}
	
	\end{center}
\end{titlepage}

\pagestyle{fancy}
\frontmatter
{\fancyhead[LO]{}
\fancyhead[RE]{}
\fancyhead[LE,RO]{}
\fancyfoot[C]{\footnotesize \thepage}
\include{Chap0_Framematter/Abstract}
\include{Chap0_Framematter/Resume}
\include{Chap0_Framematter/Acknowledgements}
\include{Chap0_Framematter/Preface}
}

\tableofcontents

\mainmatter
\renewcommand{\headrulewidth}{0.4pt}
\renewcommand{\footrulewidth}{0.pt}
\fancyhead[LE,RO]{\footnotesize \thepage}
\fancyhead[LO]{\slshape\footnotesize\nouppercase \rightmark}
\fancyhead[RE]{\slshape\footnotesize\nouppercase \leftmark}
\fancyfoot[C]{}
\include{Chapters/chap1_Introduction}
\include{Chapters/chap2_NMRtheory}
\include{Chapters/chap3_DesignPrinciples}
\include{Chapters/chap4_AmpModSeq}
\include{Chapters/chap5_GenSeq}
\include{Chapters/chap6_Concl}
{\footnotesize
\bibliographystyle{unsrt}
\addcontentsline{toc}{chapter}{Bibliography}
\bibliography{bibl}
}

\newpage


\begin{appendices}
	{
		\makeatletter\@openrightfalse\makeatother
		\include{Chap99_Appendix/Append_adRFDR_sourcecode}
	}
\end{appendices}

\end{document}

%% file: Chap0_Framematter/Abstract.tex
\chapter*{Abstract}
\addcontentsline{toc}{chapter}{\numberline{}Abstract}
	The work presented in this dissertation is centred on the theory of experimental methods in solid-state Nuclear Magnetic Resonance (NMR) spectroscopy, which deals with interaction of electromagnetic radiation with nuclei in a magnetic field and possessing a fundamental quantum mechanical property called spin. Nuclei with a spin number $1/2$ is the focus here. Unlike liquid-state, the spin interactions that are dependent on the orientation of the sample with respect to the external magnetic field, are not averaged out in solid-state and therefore lead to broad indistinct signals. To avoid this, the solid sample is spun rapidly about a particular axis thereby averaging out the orientation-dependent interactions. However, these interactions are also typically a major source of structural information and a means to improve spectral resolution and sensitivity. Therefore it is crucial to be able to reintroduce them as and when needed, in the course of an experiment. This is achieved using radio frequency pulses, a control oscillating magnetic field, applied transverse to the external magnetic field. The class of pulse sequences that specifically reintroduces space-mediated spin-spin (termed dipolar coupling) interactions are referred to as dipolar recoupling pulse sequences and forms the subject of this dissertation.
	
	NMR experiments, that involve repeating (periodic) pulse sequences, are generally understood by finding an average or effective Hamiltonian (interaction, in the language of physics), which approximates the spin dynamics to a great extent and often offers insights into the workings of the experiment that a full numerical simulation of the spin dynamics do not. The average Hamiltonian is found using either the intuitively named average Hamiltonian theory (AHT) or the Floquet theory, where the former is the most widely used approach in NMR. AHT and Floquet theory, which use mathematical tools namely, Magnus expansion and Van Vleck transformations respectively, yield the same effective Hamiltonian that is valid for stroboscopic observation. The equivalence between the theories has been discussed in the past in literature. Generalised expressions for the effective Hamiltonian using AHT are derived in the frequency domain in this dissertation, to allow for appreciation of the equivalence with Floquet theory mediated effective Hamiltonian. The derivation relies on the ability to express the time dependency of the interactions with a finite set of fundamental frequencies, which has long been sought and the lack of which has, at times, been misunderstood as limitations of AHT.
	
	A formalism to represent any periodic time-dependent interaction in the Fourier space with no more than two fundamental frequencies for every involved spin, is developed and presented here, which allows for the computation of effective Hamiltonian for any pulsed experiment. The formalism has been applied to understand established dipolar recoupling pulse sequences, namely Radio-Frequency-Driven Recoupling (RFDR), Rotor Echo Short Pulse IRrAdiaTION mediated Cross Polarisation ($^{\tr{RESPIRATION}}$CP) and sevenfold symmetric C7 pulse schemes. Limitations of the pulse sequences, in particular their sensitivity to isotropic chemical shift which is a measure of the electron cloud surrounding the nuclei of interest, are addressed by designing novel variants of the pulse sequences, aided by insights gained through the effective Hamiltonian description.

%% file: Chap0_Framematter/Resume.tex
\chapter*{Resum\'e}
\addcontentsline{toc}{chapter}{\numberline{}Resum\'e}
	Arbejdet, der pr{\ae}senteres i denne afhandling, omhandler den teori, der ligger til grund for eksperimentelle metoder anvendt indenfor faststof kernemagnetisk resonans (NMR) spektroskopi. Denne gren af spektroskopi besk{\ae}ftiger sig med interaktioner imellem elektromagnetisk str{\aa}ling og atomare kerne, der besidder den kvantemekaniske egenskab spin og er placeret i et magnetfelt. Det prim{\ae}re fokus er p{\aa} atomare kerne med spintal 1/2. I mods{\ae}tning til i v{\ae}skefasen findes der i fast stof spininteraktioner, hvis st{\o}rrelse er afh{\ae}ngig af pr{\o}vens orientering i forhold til det eksterne magnetfelt, og disse giver anledning til brede og ofte overlappende signaler. For at undg{\aa} dette, roteres den faste pr{\o}ve med h{\o}j hastighed omkring en specifik akse s{\aa}ledes, at de orienteringsafh{\ae}ngige interaktioner midles. Interaktionerne er dog vigtige kilder til strukturel information, og det er derfor af afg{\o}rende vigtighed at kunne aktivere dem igen efter behov i l{\o}bet af et eksperiment. Dette opn{\aa}s ved brug af radiofrekvens pulse -- et kontrolleret, oscillerende magnetfelt -- som p{\aa}f{\o}res vinkelret p{\aa} det eksterne magnetfelt. Typen af pulssekvenser, der specifikt genintroducerer rumligt medierede spin-spin interaktioner (dipol koblinger), kaldes dipol{\ae}re rekoblings pulssekvenser og er emnet for denne afhandling.
	
	NMR eksperimenter, der benytter periodiske pulssekvenser, forst{\aa}s generelt ved at bestemme en gennemsnitlig eller effektiv Hamilton operator (eller interaktions operator), som i vid udstr{\ae}kning approksimerer spin dynamikken og ofte giver indsigt i, hvordan eksperimentet virker, hvilket end ikke en komplet numerisk simulering af dynamikken kan levere. Den gennemsnitlige Hamilton operator findes ved hj{\ae}lp af enten den intuitivt navngivne Average Hamiltonian Theory (AHT) eller Floquet teori, som benytter sig af henholdsvis Magnus ekspansionen og Van Vleck transformationer. Begge producerer den samme effektive Hamilton operator, som beskriver systemet n{\o}jagtigt under foruds{\ae}tning af stroboskopisk observation. Ækvivalensen imellem de to teorier er tidligere blevet diskuteret i litteraturen, men den bliver typisk glemt indenfor NMR feltet. I denne afhandling bliver generaliserede udtryk for effektive Hamilton operatorer udledt ved hj{\ae}lp af AHT i frekvensdom{\ae}net s{\aa}ledes, at {\ae}kvivalensen med de Floquet udledte effektive Hamilton operatorer bliver tydeliggjort. Udledningen afh{\ae}nger af en evne til at udtrykke tidsafh{\ae}ngigheden af interaktionerne med et endeligt s{\ae}t af grundl{\ae}ggende frekvenser, hvilket l{\ae}nge har v{\ae}ret eftertragtet, og manglen p{\aa} denne evne er til tider blevet fejltolket som en begr{\ae}nsning i AHT. En formalisme, der kan repr{\ae}sentere en vilk{\aa}rlig, periodisk, tidsafh{\ae}ngig interaktion i Fourierrummet med h{\o}jest to grundl{\ae}ggende frekvenser per involveret spin, udvikles og pr{\ae}senteres her.
	
	Formalismen er blevet anvendt til at forst{\aa} eksisterende dipol{\ae}re rekoblings pulssekvenser. Specifikt Radio-Frequency-Driven Recoupling (RFDR), Rotor Echo Short Pulse IRrAdiaTION mediated Cross polarisation ($^{\tr{RESPIRATION}}$CP) og syvfoldigt symmetriske C7 pulssekvenser. Pulssekvensernes begr{\ae}nsninger -- specielt deres f{\o}lsomhed overfor isotropt kemisk skift, som er et m{\aa}l for elektront{\ae}thedsskyen, der omgiver den unders{\o}gte kerne -- bliver adresseret ved at designe nye varianter med forbedret tolerance ved hj{\ae}lp af indsigt leveret af effektiv Hamilton operator beskrivelsen.

%% file: Chap0_Framematter/Acknowledgements.tex
\chapter*{Acknowledgements}
\addcontentsline{toc}{chapter}{\numberline{}Acknowledgements}
	First off, I must thank my supervisor, Professor Niels Christian Nielsen for his cheerful positivity and the workaholic that he is, which has had a profound impact on me. Being an night owl that I am, he was quite liberal with when and how I did things. Just like how an interaction frame transformation hides away the uninteresting rf Hamiltonian, being a student of Niels Chr., put away my non-scientific worries and hugely helped me focus on the actual science (yes, I had to crack this!).
	
	I acknowledge my co-supervisor Professor P.K. Madhu for introducing me to Niels Chr., and for the understanding mentor that he has been all through. His wonderful rapport with leading scientists in the field from all over the world has been of immense help and inspiring. I must also thank my mentors at IISER Pune, Professors T. S. Mahesh and M. S. Santhanam, for my continuation in academic research is a testament to my first full-time research experience with them.
	
	I am grateful to Sheetal Kumar Jain, for patiently teaching me the basics of solid-state NMR, answering all my silly questions and for helping define my project goals during the early stage of my Ph.D.
	
	Anders Bodholt Nielsen has guided me through, when I failed to see thesis-worthy projects shaping up, like beacons enabling aircrafts to land when pilots fail to see the runway. I am immensely thankful to him for that and teaching me advanced concepts in NMR theory. With regard to projects, his focus on the big picture is a huge learning for a guy like me, who focuses on the details and is often lost with it.
	
	Zden{\v{e}}k To{\v{s}}ner has been fantastic in hosting me at Charles University, Prague, for an intellectually stimulating few weeks where we tested some great ideas on effective Hamiltonian optimisations of pulse sequences. His thorough understanding and drive to learn new concepts have been inspiring.
	
	I thank Professor Thomas Vosegaard, for all the exciting and insightful discussions on the projects and for his numerous administrative help over the years. Lasse has taught me practical NMR and I consider his no-compromise-on-quality work an important lesson. I am thankful to Asif, for the collaborative work on decoupling and proof-reading my thesis. Kristoffer has been resourceful with his expertise on experiments and translating the abstract here to Danish. It was fortunate to be part of the interdisciplinary group and I thank all the members, including but not limited to, MoKS, Marcus Ullisch, Frans Mulder, Yuichi Yoshimura, Morten Bjerring, Mads Vinding, Dennis Juhl, Natalia Kulmiskaya, Camilla Andersen and Nicholas Balsgart.
	
	Finding Aarhus Cricket Club was a pleasant surprise and along with Poker evenings with wonderful people, my weekends served as amazing stress busters. I am thankful to my new-found Tamil and Indian friends in Aarhus for making me feel less homesick during my stay here.
	
	Finally I would like to thank my supportive family, in particular my Dad, for instilling the scientific way of thought in me, at a young and impressionable age.

%% file: Chap0_Framematter/Preface.tex
\chapter*{Preface}
\addcontentsline{toc}{chapter}{\numberline{}Preface}
This Ph.D. dissertation is founded on research performed at Aarhus University, over a period of three years. The work was supervised by professor Niels Christian Nielsen and postdoc Anders Bodholt Nielsen. The work was performed in collaboration with members of BioNMR group at Aarhus University, ETH Zurich and TIFR Mumbai. The focus has been on developing a theoretical description to find an effective Hamiltonian for any general dipolar recoupling pulse sequence, in solid-state NMR. The developed description and insights gained through it, along with novel pulse sequences designed owing to the insights, are explained. For some of the findings presented here, manuscripts have been submitted to peer-reviewed scientific journals and are appended with the dissertation.

\section*{Publications}
\begin{enumerate}
	\item Anders B. Nielsen, Kong Ooi Tan, \underline{Ravi Shankar}, Susanne Penzel, Riccardo Cadalbert, Ago Samoson, Beat H. Meier, and Matthias Ernst: "Theoretical description of $^{\tr{RESPIRATION}}$CP", \textit{Chemical Physics Letters, 645:150-156, 2016}
	\item Lasse A Straas\o, \underline{Ravi Shankar}, Kong Ooi Tan, Johannes Hellwagner, Beat H. Meier, Michael Ryan Hansen, Niels Chr. Nielsen, Thomas Vosegaard, Matthias Ernst, and Anders B. Nielsen: "Improved transfer efficiencies in radio-frequency driven recoupling solid-state NMR by adiabatic sweep through the dipolar recoupling condition", \textit{The Journal of Chemical Physics, 145(3):034201, 2016}
	\item  Kristoffer Basse$^*$, \underline{Ravi Shankar}$^*$, Morten Bjerring, Thomas Vosegaard, Niels Chr.	Nielsen, and Anders B. Nielsen: "Handling the influence of chemical shift in amplitude-modulated heteronuclear dipolar recoupling solid-state NMR", \textit{The Journal of Chemical Physics, 145(9):094202, 2016.}\\
	$^{*}$ contributed equally to the work.
	\item Asif Equbal, \underline{Ravi Shankar}, Michal Leskes, Shimon Vega, Niels Chr. Nielsen, P. K. Madhu: "Significance of symmetry in the nuclear spin Hamiltonian for efficient heteronuclear dipolar decoupling in solid-state NMR: A Floquet description of supercycled rCW schemes". \textit{Manuscript submitted to The Journal of Chemical Physics.}
	\item \underline{Ravi Shankar}, Matthias Ernst, Beat H. Meier, P. K. Madhu, Thomas Vosegaard, Niels Chr. Nielsen and Anders B. Nielsen: "A General Theoretical Description of the Influence of Isotropic Chemical Shift under Dipolar Recoupling Experiments for Solid-State NMR". \textit{Manuscript submitted to The Journal of Chemical Physics.}
\end{enumerate}

\section*{Overview}
\noindent Chapter 1: A brief introduction to the presented work.\\

\noindent Chapter 2: Basic concepts of quantum mechanics relevant to understand NMR and description of the interactions encountered in NMR are detailed.\\

\noindent Chapter 3: Average Hamiltonian theory employed to obtain effective Hamiltonian for a pulsed experiment, along with discussion on finding propagators for multiples of a time shorter than the periodic time of the interaction frame Hamiltonian, are explained. The author recommends a careful reading of this intense chapter, for easier understanding of the following chapters.\\

\noindent Chapter 4: Theoretical description to find effective Hamiltonian for amplitude modulated pulse sequences, together with applications to explain and design variants of a homonuclear and a heteronuclear dipolar recoupling experiment are elaborated.\\

\noindent Chapter 5: Theoretical description to find effective Hamiltonian for a general (amplitude and phase modulated) pulse sequence is detailed, along with an application to describe C-symmetry homonuclear dipolar recoupling experiments.

%% file: Chapters/chap1_Introduction.tex
\chapter{Introduction}
	Spectroscopy is the study of absorption and emission of electromagnetic radiation by matter. Nuclear magnetic resonance (NMR)\cite{bloch1946nuclear,purcell1946resonance} is a phenomenon, that exploits a fundamental property of certain atomic nuclei, called \textit{spin}, and occurs when the nuclei experience magnetic field. NMR spectroscopy is a tool utilised to study physical, chemical and biological properties of matter, and finds numerous applications in natural sciences and medicine\cite{lauterbur1973image}, including but not limited to study of polymers, inorganic materials and biological macromolecules\cite{duer2008solid} like proteins\cite{castellani2002structure} and Deoxyribonucleic acid (DNA)\cite{wuthrich1986nmr}. It is the preferred choice of study tool for membrane proteins\cite{bertelsen2007membrane}, amyloid fibrils\cite{nielsen2009unique} and the like, that do not form high-quality crystals, which is a necessary requirement for X-ray crystallography\cite{drenth2007principles}. Investigations of samples using NMR are non-destructive and offers atomic-resolution structural information, advantages that are fundamental to its versatility.\\
	
	Sensitivity of NMR experiments is intrinsically low due to the relatively small energy difference between consecutive spin states, which is the fundamental measure and is in the radio-frequency (rf) range. For certain nuclei, this is worsened by low natural abundance of NMR active isotopes. The issue is tackled by transferring polarisation from high to low gyromagnetic nuclei and by enhancing the abundance of NMR active isotopic nuclei through isotopic labelling techniques in sample preparation. The significant advantage and challenge of solid-state NMR, as compared to liquid-state NMR, is the direct spectral presence of orientation-dependent (anisotropic) interactions, resulting in low resolution of spectra. Such interactions are averaged out in samples in liquid-state by fast isotropic molecular motion, and as a result liquid-state NMR spectra are highly resolved. The solid-state NMR spectrum is improved by rapidly spinning the sample about a specific axis, known as magic-angle spinning (MAS)\cite{andrew1958nuclear,lowe1959free}, to average out the anisotropic interactions. RF pulse sequences are an additional oscillating magnetic field applied transverse to the dominant external magnetic field. Sequences developed specifically to suppress the effects of interactions not completely averaged out by MAS, called decoupling pulse sequences, also help in this regard\cite{bloom1955effects}. However, anisotropic interactions contain information about the structure, dynamics, and orientation of the spin system under study\cite{andronesi2005determination}, and it is therefore crucial to reintroduce the interaction or recouple, as and when needed. In the case of dipole-dipole interaction, the recoupled interaction can be used to transfer magnetisation from one to another, which is instrumental in distance measurements between nuclei\cite{griffin1998dipolar,ladizhansky2009homonuclear}, together with sensitivity enhancement through polarisation transfers between nuclei and improved resolution in multi-dimensional experiments. Pulse sequences that are specifically designed to address this are known as recoupling pulse sequences, and forms the subject of this dissertation. Such techniques are used to establish correlations among spins, yielding essential information about the structure of the system, as the rate of magnetisation transfer depends on the spatial proximities and the relative orientations of the spins.\\
	
	NMR spectroscopy is a theoretically complex tool, and the interactions are often time-dependent, brought in by MAS and rf irradiation. The spin dynamics of small spin systems can be numerically simulated, however, it rarely offers any understanding or insights into the workings of an experiment. An NMR experiment is typically understood by finding an effective or average Hamiltonian that approximates the active interactions present in the spin system under study. The effective Hamiltonian offers better comprehension and aids in the development of pulse sequences to enable or suppress desired interactions. The two most approved ways of treating such time-dependent Hamiltonians, in order to obtain a propagator for a spin system, are average Hamiltonian theory (AHT)\cite{waugh1968js,haeberlen1968coherent,burum1981magnus,klarsfeld1989recursive} and Floquet theory\cite{floquet,shirley1965solution,maricq1982application,shimon1996}. AHT is the most widely adopted method in NMR owing to its intuitive approach in arriving at a time-independent Hamiltonian, valid for use at multiples of a certain defined cycle time. Floquet theory transforms the Hamiltonian to a so-called Floquet space, that is infinite in dimension and which is a combined Fourier and Hilbert space, where the time-dependencies are implicitly seen through frequencies. To obtain a time-independent effective Hamiltonian, the infinite-dimensional matrix is diagonalised, often using Van Vleck\cite{van1948dipolar} transformation. The effective Hamiltonian, so obtained, is identical to the effective Hamiltonian obtained using AHT, as shown by Llor\cite{llor1992equivalence}. However, to show the equivalence for any general pulse sequence, the time-dependent Hamiltonian has to be written as a Fourier series\cite{fourier1822theorie} with finite number of fundamental frequencies, before the averaging using AHT can be applied. The lack of which has often led to people limiting AHT to only cases, where the interaction frame time-dependent Hamiltonian is simple and expressed by using a single fundamental frequency defined by the MAS rate. A description to enable expressing time-dependent Hamiltonian modulated by any general pulse sequence is developed in this work and its usefulness in obtaining a time-independent effective Hamiltonian using AHT is detailed. The average Hamiltonian helps describing an experiment with an approximate overall Hamiltonian present in the spin system in relation to the control rf field parameters. This enables the study of different features of the experiment and helps design better pulse sequences.\\
	
	The above description is put to use studying established homonuclear dipolar recoupling experiments, Radio-Frequency-Driven Recoupling (RFDR), sevenfold symmetric C7 pulse schemes and heteronuclear dipolar recoupling experiment, Rotor Echo Short Pulse IRrAdiaTION mediated Cross Polarisation ($^{\tr{RESPIRATION}}$CP). The analysis is used to explain the effect of chemical shift, in particular the isotropic component, on polarisation transfer and design novel variants to minimise the adverse effects of the same. Additionally, a simpler calculation of effective flip angles about an axis imparted by the combined rf field and isotropic chemical shift are shown to qualitatively explain the effects, without the need for the full calculation of the effective Hamiltonian.

%% file: Chapters/chap2_NMRtheory.tex

\newcommand{\epc}{$\frac{1}{2}(1+c_\beta)$}
\newcommand{\emc}{$\frac{1}{2}(1-c_\beta)$}
\newcommand{\sts}{$\frac{1}{\sqrt{2}}s_\beta$}
\newcommand{\cb}{$c_\beta$}
\newcommand{\sbe}{$s_\beta$}
\newcommand{\sbs}{$\sqrt{\frac{3}{8}} s_\beta^2$}
\newcommand{\sbt}{$\sqrt{\frac{3}{8}} s_{2 \beta}$}
\newcommand{\cop}{$\frac{1}{4}(1+c_\beta)^2$}
\newcommand{\com}{$\frac{1}{4}(1-c_\beta)^2$}
\newcommand{\ctb}{$c_\beta^2$}
\newcommand{\slp}{$\frac{1}{2}(3c_\beta^2$-1)}

\chapter{Theory of Nuclear Magnetic Resonance}
	This chapter will cover the fundamentals of NMR by introducing quantum mechanical concepts pertaining to NMR, followed by mathematical representation of interactions present in a spin system. The concepts are readily found in numerous introductory textbooks\cite{sakurai2011modern,levitt2001spin,mehring2012principles} on quantum mechanics and NMR spectroscopy, and this chapter is intended only as a compendium of relevant concepts.
	
	\section{Quantum Mechanical description}
		\subsection{Angular momentum and Spin}
			\label{ssec:spin}
			A rotating object has \textit{angular momentum}. Unlike classical mechanics, the angular momentum in quantum mechanics is \textit{quantized}\cite{pauli1933einige,dirac1930principles,gerlach1922_1,gerlach1922_2,gerlach1922_3}, resulting in a set of allowed discrete stable rotational states (say, for a rotating molecule) specified by a quantum number $\textbf{J}$, which can take any value from the set $\{0,1,2\dots\}$. The total angular momentum $\lvert \vec{L} \rvert$ and the quantum number $\textbf{J}$ are related by $\lvert \vec{L} \rvert^2 = \textbf{J}(\textbf{J}+1)\hbar^2$, where $\hbar = 1.0545718 \times 10^{-34}$J$\cdot$s is a fundamental constant known as the reduced Planck constant. The z-component of the angular momentum is given by, $L_z = m \hbar$, where $m$ takes one of the $2\textbf{J}+1$ values, $m \in \{-\textbf{J}, -\textbf{J}+1 \dots, \textbf{J}-1, \textbf{J}\}$.\\

			\textit{Spin} is similar to angular momentum in the sense that maths of spin is reminiscent to that of angular momentum, however the physics is not.  Spin is an \textit{intrinsic} property of the particle and not a consequence of any rotational motion. It is present even at a temperature of absolute zero, unlike rotational angular momentum which is a function of temperature. Every elementary particle has a particular spin quantum number $\textbf{S}$, which takes an integer or a half integer value. The angular momentum $\lvert \vec{L} \rvert$ of a particle due to its spin, is related to the spin quantum number $\textbf{S}$ by $ \lvert \vec{L} \rvert^2 = \textbf{S}(\textbf{S}+1)\hbar^2$. Spin also possesses a magnetic moment $\vec{\mu}$ that is related to the spin quantum number by $\vec{\mu} = \gamma \vec{L}$, where $\gamma$ is defined as the ratio of magnetic moment to angular momentum of the particle of interest, and is called the gyromagnetic ratio. Again, there are $2\textbf{S}+1$ possible states with the same $\textbf{S}$ but different $m$ value that describes the z-component of spin angular momentum $L_z = m\hbar$. These states are all degenerate unless an external magnetic field is applied. In the presence of an external magnetic field (along the arbitrary z-direction), the associated potential energy of the state is given by 
			\begin{eqnarray}
				\begin{aligned}
					U_{pot} &= \int |\vec{\mu} \times \vec{B}_0| d\theta &= - |\vec{\mu}| |\vec{B}_0| \cos\theta &= -m \hbar\gamma |\vec{B}_0|.
				\end{aligned}
				\label{eq:zeeman_E}
			\end{eqnarray}
			i.e., states with same $\vec{L}$, but different $m$ values are separated by an energy gap $|\Delta E| = \Delta m \hbar\gamma |\vec{B}_0|$. This splitting of degenerate energy levels on application of magnetic field is generally known as Zeeman splitting, and the energy difference falls in the radio frequency range for nuclear spins.\\
			
			For an object with only a magnetic moment, say a compass, the external magnetic field produces a torque that aligns the magnetic moment along the magnetic field. However for spins, which possess both a magnetic moment and a proportional angular momentum, the case is different. The torque $\vec{\tau}$ resulting from the action of the external magnetic field $\vec{B}_0$ on the magnetic moment $\vec{\mu}$ changes the direction of the angular momentum $\vec{L}$ fulfilling $\vec{\tau} = \frac{d\vec{L}}{dt}$. This motion is referred to as \textit{precession} and the rate of precession, also called the \textit{Larmor frequency}\cite{levitt2001spin} is then given by,
			\begin{eqnarray}
				\omega = \frac{d\phi}{dt} = \frac{1}{|\vec{L}|\sin\theta }\frac{d\vec{L}}{dt} = \frac{-\gamma|\vec{\mu}||\vec{B}_0|\sin\theta}{|\vec{\mu}|\sin\theta} = -\gamma|\vec{B}_0|.
				\label{eq:larmor}
			\end{eqnarray}
			
			In the \textit{Standard Model}\cite{oerter2006theory}, all matter in the universe is made of elementary particles, the two basic types of which are quarks and leptons - six of each kind, all of which are spin-$\frac{1}{2}$. The most stable lepton, \textit{electron}, has an electric charge \textit{-e} along with a spin value of $\frac{1}{2}$. Proton and neutron, each of which is composed of three quarks with only two of them aligned in parallel\footnote{At high-energies, it is possible that all three quarks are aligned parallel to result in a spin-$\frac{3}{2}$ state for proton/neutron. But these are beyond the case-space of magnetic resonance.}, have a net spin-$\frac{1}{2}$ and net electric charge of \textit{e} and \textit{0}.\\
	
		\subsection{Spin states}
			In quantum mechanics, the state of a quantum system is described by a time-dependent wave function $\psi(t)$. This can be represented in Dirac notation as a linear combination or quantum superposition of an orthonormal basis set $\{\ket{n}\}$,
			\begin{equation}
				\begin{aligned}
					\ket{\psi(t)} = \sum_n c_n(t) \ket{n},
				\end{aligned}
			\end{equation}
			where $c_n(t)$ are complex amplitude as functions of time. The amplitude squared, $|c_n(t)|^2$ gives the probability of measuring the spin to be in the state $\ket{n}$ at time $t$. This is a fundamental law of quantum mechanics - Born rule\cite{born1926} - that has not yet been derived from the other postulates of quantum mechanics\cite{LaFlamme2012}. A state that can be described by a single ket, like above, is called a \textit{pure state}. For a system that is a statistical ensemble of numerous pure states, it is not possible to represent the system with a single ket. Such a state is termed a \textit{mixed state}. It is worth illustrating the difference between a quantum superposition of pure states and a mixed state: a detector set to measure $\frac{1}{2}(\ket{\alpha} + \ket{\beta})$ in a spin-$\frac{1}{2}$ nucleus that is in a pure state described as a linear combination in equal measures of states $\ket{\alpha}$ and $\ket{\beta}$ (i.e., $\frac{1}{2}(\ket{\alpha} + \ket{\beta})$), returns 1 whereas the same detector when used on a mixture of two spin-$\frac{1}{2}$ nuclei where one spin is in state $\ket{\alpha}$ and the other is in $\ket{\beta}$, returns $\frac{1}{2}$. To accommodate mixed states in the formalism, \textit{density operators} are used.\\
			
			Density operator, that describes a quantum state as a function of time, is defined as
			\begin{equation}
				\hat{\rho}(t) = \sum_{n} p_n \ket{\psi_n(t)}\bra{\psi_n(t)},
			\end{equation}
			where $p_n$ is statistical probability of the state $\psi_n(t)$ in the ensemble/mixture. The density operator for a pure state would have $p_{n'} = 1$ for one particular $n'$ in the set $\{n\}$ and 0 for rest of the set. For an NMR system in thermal equilibrium, the population distribution of spin states is Boltzmann. The density operator for such an N-spin system is,
			\begin{eqnarray}
				\begin{aligned}
					\hat{\rho}_{eq} = \frac{1}{2^N}\Big(\mathds{1} + \frac{\hbar \gamma B_0}{kT} \sum_{j=1}^{N}I_{jz}\Big)
				\end{aligned}
			\end{eqnarray}
			where $k = 1.38064852 \times 10^{−23}$J.K$^{-1}$ is the Boltzmann constant and T is temperature of the system.\\
			
			The expectation value of any operator $\hat{\mathcal{O}}$, which is defined for a state $\ket{\psi}$ as
			\begin{equation}
			\begin{aligned}
				\epv{\hat{\mathcal{O}}} = \bra{\psi}\hat{\mathcal{O}}\ket{\psi},
			\end{aligned}
			\end{equation}
			can be reformulated to suit the density operator formalism by making use of the identity $\sum\limits_i \ket{x_i}\bra{x_i} = 1$.
			\begin{eqnarray}
				\epv{\hat{\mathcal{O}}} & = & \sum_{i}\sum_{j}\epv{\psi|x_i}\epv{x_i|\hat{\mathcal{O}}|x_j}\epv{x_j|\psi} \nonumber \\
				& = & \sum_{i}\sum_{j}\epv{x_j|\psi}\epv{\psi|x_i}\epv{x_i|\hat{\mathcal{O}}|x_j} \nonumber \\
				& = & \sum_{i}\sum_{j}\bra{x_j}\hat{\rho}\ket{x_i}\epv{x_i|\hat{\mathcal{O}}|x_j} \nonumber \\
				& = & \mathrm{Tr}\{\hat{\rho}\hat{\mathcal{O}}\}, \label{eq:exp_trace}
			\end{eqnarray}
			where Tr\{\dots\} refers to tracing over the diagonal elements of the matrix.
		\subsection{Temporal dynamics}
			The time evolution of a state $\ket{\psi(t)}$ of a quantum system is described by Schr{\"o}dinger equation\cite{schrodinger_eqn}, 
			\begin{eqnarray}
			i\hbar \frac{\partial \ket{\psi(t)}}{\partial t} & = & \hat{\mathcal{H}}(t)\ket{\psi(t)}.
			\label{eq:sch}
			\end{eqnarray}
			This can be extrapolated to describe time evolution of density operators, which is of interest, to describe NMR systems.
			\begin{eqnarray}
				i\hbar \frac{\partial \hat{\rho}(t)}{\partial t} & = & i\hbar \frac{\partial \ket{\psi(t)}}{\partial t} \bra{\psi(t)} + \ket{\psi(t)} \frac{\partial \bra{\psi(t)}}{\partial t} i\hbar \nonumber \\
				& = & \hat{\mathcal{H}}(t) \ket{\psi(t)} + \bra{\psi(t)}\hat{\mathcal{H}}(t)  \label{eq:vonNeu_Herm}\\
				\frac{\partial\hat{\rho}}{\partial t}(t) & = & \lbrack \hat{\mathcal{H}}(t),\rho(t) \rbrack, 
				\label{eq:vonNeu}
			\end{eqnarray}
			where $\hat{\mathcal{H}}(t)$ is the time-dependent Hamiltonian operator whose Hermitian property is used in Eq. \ref{eq:vonNeu_Herm}, $\lbrack\dots\rbrack$ refers to the commutation operator. Eq. \ref{eq:vonNeu} is known as the Liouville-von Neumann equation and governs spin evolution\footnote{ignoring the phenomenon of Relaxation.}. The solution to Eq. \ref{eq:vonNeu}, for \textit{time-independent} Hamiltonian, is
			\begin{eqnarray}
				\hat{\rho}(t) = e^{-i\hat{\mathcal{H}}t}\hat{\rho}(0)e^{i\hat{\mathcal{H}}t}.
				\label{eq:rhot_time_indep}
			\end{eqnarray}
			The solution to time-dependent Hamiltonian can be found by assuming piece-wise time-independency for $\hat{\mathcal{H}}(t)$, thus resulting in the solution
			\begin{eqnarray}
				\hat{\rho}(t) & = & \hat{U}(t)\hat{\rho}(0)\hat{U}^{\dagger}(t). \label{eq:neuSol}
			\end{eqnarray}
			The operator $\hat{U}(t)$ in Eq. \ref{eq:neuSol} is referred as the \textit{propagator} and is described by 
			\begin{eqnarray}
				 \hat{U}(t) & = & e^{-i\hat{\mathcal{H}}_1\int_{t_{n-1}}^{t_n}dt'}\dots e^{-i\hat{\mathcal{H}}_1\int_{t_1}^{t_2}dt'} e^{-i\hat{\mathcal{H}}_1\int_{0}^{t_1}dt'} \nonumber \\
				 & = & \hat{\mathcal{T}}e^{-i\int_{0}^{t}\hat{\mathcal{H}}(t')dt'},
			\label{eq:dyson}
			\end{eqnarray}
			where $\hat{\mathcal{T}}$ is the Dyson time-ordering operator\cite{dyson1952divergence}. It is worth noting here that $\ket{\psi(t)} = \hat{U}(t)\ket{\psi(0)}$, which upon derivation and using Eq. \ref{eq:sch} gives,
			\begin{eqnarray}
				\hat{\mathcal{H}}\ket{\psi(t)} & = & i\hbar\frac{\partial \hat{U}}{\partial t}(t)\ket{(\psi(0))} \nonumber \\
				\hat{\mathcal{H}}\hat{U}(t)\ket{\psi(0)} & = & i\hbar\frac{\partial \hat{U}}{\partial t}(t)\ket{(\psi(0))} \nonumber\\ 
				\implies \frac{\partial \hat{U}}{\partial t} & = & \hat{\mathcal{H}}\hat{U} \label{eq:dUdt}
			\end{eqnarray}
		
		\subsection{Change of reference frame}
		\label{sec:int_frame}
			In the analysis of NMR experiments, it often proves useful to choose a rotating frame for description. In general, say the spin system is under the effect of Hamiltonian, $\hat{\mathcal{H}}(t)$ given by
			\begin{eqnarray}
				\hat{\mathcal{H}}(t) & = & \hat{\mathcal{H}}_{\textrm{int}}(t) + \hat{\mathcal{H}}_{\textrm{ext}}(t),
			\end{eqnarray}
			where $\hat{\mathcal{H}}_{\textrm{ext}}(t)$ is the dominant Hamiltonian and $\hat{\mathcal{H}}_{\textrm{int}}(t)$ is the internal Hamiltonian, whose effect under $\hat{\mathcal{H}}_{\textrm{ext}}(t)$ on the spin system is of interest. The propagator, $\hat{U}(t)$ corresponding to $\hat{\mathcal{H}}(t)$ can be decomposed as
			\begin{eqnarray}
				\hat{U}(t) & = & \hat{U}_{\textrm{ext}}(t)\hat{\tilde{U}}(t).
			\end{eqnarray}
			Taking time-derivative on both sides\footnote{Time dependency (t) is dropped for better readability},
			\begin{eqnarray}
				\frac{\partial \hat{U}}{\partial t} & = & \frac{\partial \hat{U}_{\textrm{ext}}}{\partial t}\hat{\tilde{U}} +  \hat{U}_{\textrm{ext}}\frac{\partial \hat{\tilde{U}}}{\partial t} \nonumber\\
				\hat{\mathcal{H}}\hat{U} & = & \hat{\mathcal{H}}_{\textrm{ext}}\hat{U}_{\textrm{ext}}\hat{\tilde{U}} + \hat{U}_{\textrm{ext}}\frac{\partial \hat{\tilde{U}}}{\partial t} \label{eq:intFr1}\\
				\hat{U}_{\textrm{ext}}^{-1}(\hat{\mathcal{H}}_{\textrm{ext}}+\hat{\mathcal{H}}_{\textrm{int}})\hat{U}_{\textrm{big}}\hat{\tilde{U}}  & = & \hat{U}_{\textrm{ext}}^{-1}\hat{\mathcal{H}}_{\textrm{ext}}\hat{U}_{\textrm{ext}}\hat{\tilde{U}} + \frac{\partial \hat{\tilde{U}}}{\partial t} \label{eq:intFr2}\\
				\frac{\partial \hat{\tilde{U}}}{\partial t} & = & \hat{U}_{\textrm{ext}}^{-1}\hat{\mathcal{H}}_{\textrm{int}}\hat{U}_{\textrm{ext}}\hat{\tilde{U}}, \label{eq:intFr3}
			\end{eqnarray}
			where Eq. \ref{eq:intFr1} follows from Eq. \ref{eq:dUdt}. Drawing parallels from Eq. \ref{eq:dUdt}, Eq. \ref{eq:intFr3} can be rewritten as $\frac{\partial \hat{\tilde{U}}}{\partial t} = \hat{\tilde{\mathcal{H}}}_{\textrm{int}}\hat{\tilde{U}}$, where $\hat{\tilde{\mathcal{H}}}_{\textrm{int}} = \hat{U}_{\textrm{ext}}^{-1}\hat{\mathcal{H}}_{\textrm{int}}\hat{U}_{\textrm{ext}}$. Any operator $\hat{\mathcal{O}}$ can be written in the interaction frame of $\hat{\mathcal{H}}_{\textrm{ext}}$ as $\hat{\tilde{\mathcal{O}}} = \hat{U}_{\textrm{ext}}^{-1}\hat{\mathcal{O}}\hat{U}_{\textrm{ext}}$. The Liouville-von Neumann equation (Eq. \ref{eq:vonNeu}), expressed in this frame as $\frac{\partial \hat{\tilde{\rho}}}{\partial t}(t) = \lbrack \hat{\tilde{\mathcal{H}}},\hat{\tilde{\rho}}(t) \rbrack$, is still valid.

	\section{System Hamiltonians}
		The Hamiltonian in NMR takes the form
		\begin{equation}
		\hat{\mathcal{H}}(t) = \hat{\mathcal{H}}_{\textrm{int}}(t) + \hat{\mathcal{H}}_{\textrm{ext}}(t)
		\end{equation}
		where $\hat{\mathcal{H}}_{\textrm{ext}}$ represents the Zeeman interaction $\hat{\mathcal{H}}_z$ - the largest of NMR interactions and is the one between the static external magnetic field and the nuclear spins. $\hat{\mathcal{H}}_{\textrm{ext}}$ also includes the user controlled rf interaction $\hat{\mathcal{H}}_{\textrm{rf}}$, a time-dependent magnetic field applied along the plane transverse to the static magnetic field to manipulate nuclear spin polarisations. $\hat{\mathcal{H}}_{\textrm{int}}$ represents a host of internal interactions. This chapter details the mathematical representation of these interactions.\\
		
		\subsection{Tensor representation}
		\label{ssect:2_tensor_Hamil}
			A spin-spin interaction Hamiltonian in NMR can generally be represented, in the Cartesian basis as
			\begin{equation}		
				\begin{alignedat}{3}
					\hat{\mathcal{H}}_{\lambda} &= \vec{\hat{I}}_k\cdot\lambda\cdot\vec{\hat{I}}_n &= 
					\begin{bmatrix}
					\hat{I}_{kx} & \hat{I}_{ky} & \hat{I}_{kz}
					\end{bmatrix} 
					\begin{bmatrix}
					\lambda_{xx} & \lambda_{xy} & \lambda_{xz} \\
					\lambda_{yx} & \lambda_{yy} & \lambda_{yz} \\
					\lambda_{zx} & \lambda_{zy} & \lambda_{zz}
					\end{bmatrix}
					\begin{bmatrix}
					\hat{I}_{nx} \\
					\hat{I}_{ny} \\
					\hat{I}_{nz}
					\end{bmatrix}
				\end{alignedat}
			\label{eq:genHamil}
			\end{equation}
			where $\vec{\hat{I}}_k$ and $\vec{\hat{I}}_n$ represent spins and the matrix $\lambda$ represents the strength and spatial dependency of the interaction as a reducible rank-2 tensor. A spin-field interaction Hamiltonian can be represented as
			\begin{equation}		
				\begin{alignedat}{3}
					\hat{\mathcal{H}}_{\lambda} &= \vec{\hat{I}}_k\cdot\lambda\cdot\vec{B}_0 &= 
					\begin{bmatrix}
					\hat{I}_{kx} & \hat{I}_{ky} & \hat{I}_{kz}
					\end{bmatrix} 
					\begin{bmatrix}
					\lambda_{xx} & \lambda_{xy} & \lambda_{xz} \\
					\lambda_{yx} & \lambda_{yy} & \lambda_{yz} \\
					\lambda_{zx} & \lambda_{zy} & \lambda_{zz}
					\end{bmatrix}
					\begin{bmatrix}
					B_{0x} \\
					B_{0y} \\
					B_{0z}
					\end{bmatrix}
				\end{alignedat}
			\label{eq:genHamil2}
			\end{equation}
			The reducible rank-2 tensor $\lambda$ can be expressed as a sum of three components - a diagonal $\lambda_0$, an antisymmetric $\lambda_1$ and a traceless symmetric $\lambda_2$ components, i.e., three \textit{irreducible} tensors\cite{fano1959} $\lambda_k$ of rank $k = \{0,1,2\}$ as
			\begin{eqnarray}
				\lambda & = & \lambda_{0} + \lambda_{1} + \lambda_{2} \label{eq:Red_sp_int_tensor}
			\end{eqnarray}
			with
			\begin{equation}
				\begin{alignedat}{2}
					\lambda_{0} &=
					\begin{bmatrix}
					\lambda_{0,0} & 0 & 0 \\
					0 & \lambda_{0,0} & 0 \\
					0 & 0 & \lambda_{0,0}
					\end{bmatrix}\\
					\lambda_{1} &=
					\begin{bmatrix}
					0 & \lambda_{xy}^a & \lambda_{xz}^a \\
					-\lambda_{xy}^a & 0 & \lambda_{yz}^a \\
					-\lambda_{xz}^a & -\lambda_{yz}^a & 0
					\end{bmatrix}\\
					\lambda_{2} &=
					\begin{bmatrix}
					\lambda_{xx}^s-\lambda_0 & \lambda_{xy}^s & \lambda_{xz}^s \\
					\lambda_{xy}^s & \lambda_{yy}^s-\lambda_0 & \lambda_{yz}^s \\
					\lambda_{xz}^s & \lambda_{yz}^s & \lambda_{zz}^s-\lambda_0
					\end{bmatrix}
				\end{alignedat}
			\label{eq:cart_decomp}
			\end{equation}
			where $\lambda_{0,0} = \frac{1}{3}\sum_i \lambda_{ii}^s$, $\lambda_{ij}^a = \frac{1}{2}(\lambda_{ij}-\lambda_{ji})$ and $\lambda_{ij}^s = \frac{1}{2}(\lambda_{ij}+\lambda_{ji})$. It is evident that the number of independent components in an irreducible tensor of rank $k$ is $2k+1$. The three irreducible tensor components in Eq. \ref{eq:Red_sp_int_tensor} transform \textit{differently} and \textit{independently} under rotations.\\
			
			\noindent A general rank-2 tensor, like $\lambda$, transforms under rotation as $\lambda' = R\lambda R^{-1}$, where $R$ is a 3x3 rotation matrix. The spatial tensor $\lambda$ can also be represented as a nine-dimensional vector
			\begin{equation}
				\vec{\lambda} = (\lambda_{xx},\lambda_{xy},\lambda_{xz},\lambda_{yx},\lambda_{yy},\lambda_{yz},\lambda_{zx},\lambda_{zy},\lambda_{zz})
			\end{equation}
			in which case, Eq. \ref{eq:genHamil} and \ref{eq:genHamil2} are represented as a scalar product of two vectors
			\begin{equation}
				\begin{alignedat}{2}
					\hat{\mathcal{H}}_{\lambda} &= \vec{\lambda}\cdot\vec{\hat{I}}
				\end{alignedat}
				\label{eq:H_vector_rep}
			\end{equation}
			with $\vec{\hat{I}} = \vec{\hat{I}}_k \otimes \vec{\hat{I}}_n$ for spin-spin interactions and $\vec{\hat{I}} = \vec{\hat{I}}_k \otimes \vec{B}_0$ for spin-field interactions. The nine-dimensional vector form of the spatial tensor transforms under a rotation as $\vec{\lambda}' = R\vec{\lambda}$, where $R$ is a full 9x9 matrix. However it is known from Eq. \ref{eq:Red_sp_int_tensor} that $\vec{\lambda}$ can be written in a rank-separated basis as
			\begin{equation}
			\begin{alignedat}{2}
				\vec{\lambda} &= (\lambda_{0},\lambda^a_{xy},\lambda^a_{xz},\lambda^a_{yz},\lambda^s_{xx}-\lambda_0,\lambda^s_{yy}-\lambda_0,\lambda^s_{xy},\lambda^s_{xz},\lambda^s_{yz})\\
				&= (\lambda_{0,0},\lambda_{1,-1},\lambda_{1,0},\lambda_{1,1},\lambda_{2,-2},\lambda_{2,-1},\lambda_{2,0},\lambda_{2,1},\lambda_{2,2}).
			\end{alignedat}
				\label{eq:ranksep}
			\end{equation}
			The rotation matrix $\mathfrak{R}$ is a block diagonal matrix in the rank-separated basis, as the components of three different ranked tensors do not mix. The transformation can therefore be written as
			{\setlength{\mathindent}{0cm}
			\begin{equation}
			\begin{tabular}{|c|c|c|c|c|c|c|c|c|c|c|c|c|}  
						\cline{1-1} \cline{3-11} \cline{13-13}
						$\lambda'_{0,0}$ & & $\mathfrak{R}^0_{0,0}$ & & & & & & & & & & $\lambda_{0,0}$\\
						\cline{1-1} \cline{3-11} \cline{13-13}
						$\lambda'_{1,-1}$& & & $\mathfrak{R}^1_{-1,-1}$ & $\mathfrak{R}^1_{0,-1}$ & $\mathfrak{R}^1_{1,-1}$ & & & & & & & $\lambda_{1,-1}$\\
						\cline{1-1} \cline{3-11} \cline{13-13}
						$\lambda'_{1,0}$ & & & $\mathfrak{R}^1_{-1,0}$ & $\mathfrak{R}^1_{0,0}$ & $\mathfrak{R}^1_{1,0}$ & & & & & & & $\lambda_{1,0}$\\						
						\cline{1-1} \cline{3-11} \cline{13-13}
						$\lambda'_{1,1}$ & & & $\mathfrak{R}^1_{-1,1}$ & $\mathfrak{R}^1_{0,1}$ & $\mathfrak{R}^1_{1,1}$ & & & & & & & $\lambda_{1,1}$\\						
						\cline{1-1} \cline{3-11} \cline{13-13}
						$\lambda'_{2,-2}$ & $=$ & & & & & $\mathfrak{R}^2_{-2,-2}$ & $\mathfrak{R}^2_{-2,-1}$ & $\mathfrak{R}^2_{-2,0}$ & $\mathfrak{R}^2_{-2,1}$ & $\mathfrak{R}^2_{-2,2}$ & $\cdot$ & $\lambda_{2,-2}$\\						
						\cline{1-1} \cline{3-11} \cline{13-13}
						$\lambda'_{2,-1}$ & & & & & & $\mathfrak{R}^2_{-1,-2}$ & $\mathfrak{R}^2_{-1,-1}$ & $\mathfrak{R}^2_{-1,0}$ & $\mathfrak{R}^2_{-1,1}$ & $\mathfrak{R}^2_{-1,2}$ & & $\lambda_{2,-1}$\\						
						\cline{1-1} \cline{3-11} \cline{13-13}
						$\lambda'_{2,0}$ & & & & & & $\mathfrak{R}^2_{0,-2}$ & $\mathfrak{R}^2_{0,-1}$ & $\mathfrak{R}^2_{0,0}$ & $\mathfrak{R}^2_{0,1}$ & $\mathfrak{R}^2_{0,2}$ & & $\lambda_{2,0}$\\						
						\cline{1-1} \cline{3-11} \cline{13-13}
						$\lambda'_{2,1}$ & & & & & & $\mathfrak{R}^2_{1,-2}$ & $\mathfrak{R}^2_{1,-1}$ & $\mathfrak{R}^2_{1,0}$ & $\mathfrak{R}^2_{1,1}$ & $\mathfrak{R}^2_{1,2}$ & & $\lambda_{2,1}$\\						
						\cline{1-1} \cline{3-11} \cline{13-13}
						$\lambda'_{2,2}$ & & & & & & $\mathfrak{R}^2_{2,-2}$ & $\mathfrak{R}^2_{2,-1}$ & $\mathfrak{R}^2_{2,0}$ & $\mathfrak{R}^2_{2,1}$ & $\mathfrak{R}^2_{2,2}$ & & $\lambda_{2,2}$\\						
						\cline{1-1} \cline{3-11} \cline{13-13}
				\end{tabular}
				\label{eq:tensorrot}
			\end{equation}}\\
				
			As the predominant transformation in NMR is a rotation, it is convenient to express the interaction tensors in spherical tensor basis, rather than in Cartesian basis. The interaction tensor represented in spherical tensor basis, $\mathcal{R}_l$ of rank $l$, can be expressed as a vector with $2l+1$ elements, $\{\mathcal{R}_{l,-l},\mathcal{R}_{l,-l+1},...,\mathcal{R}_{l,l}\}$. In general, the elements transform under a rotation as
			\begin{equation}
				\mathcal{R}_{lm}^{(new)} = \sum_{m'=-l}^{l}\mathfrak{D}^{l}_{m'm}(\Omega)\mathcal{R}_{lm'}^{(old)}.
			\label{eq:spTensorTransf}
			\end{equation}
			where $\mathfrak{D}_{m'm}^{l}(\Omega) = e^{-i\alpha m'}d_{m'm}^{(l)}(\beta) e^{-i\gamma m}$ represents the Wigner rotation matrix elements for a rotation represented by three Euler angles ($\Omega$ = \{$\alpha$, $\beta$, $\gamma$\}). This is consistent with Eq. \ref{eq:tensorrot} where the rotation matrix $\mathfrak{D}$ is denoted in the Cartesian basis by $\mathfrak{R}$. The reduced Wigner matrix\cite{wigner1939unitary} elements, $d_{m'm}^{(l)}$ are given in Table \ref{tab:wig}.\\
			
\begin{figure}[h]
				\centering
				\fbox{\includegraphics[width=\linewidth]{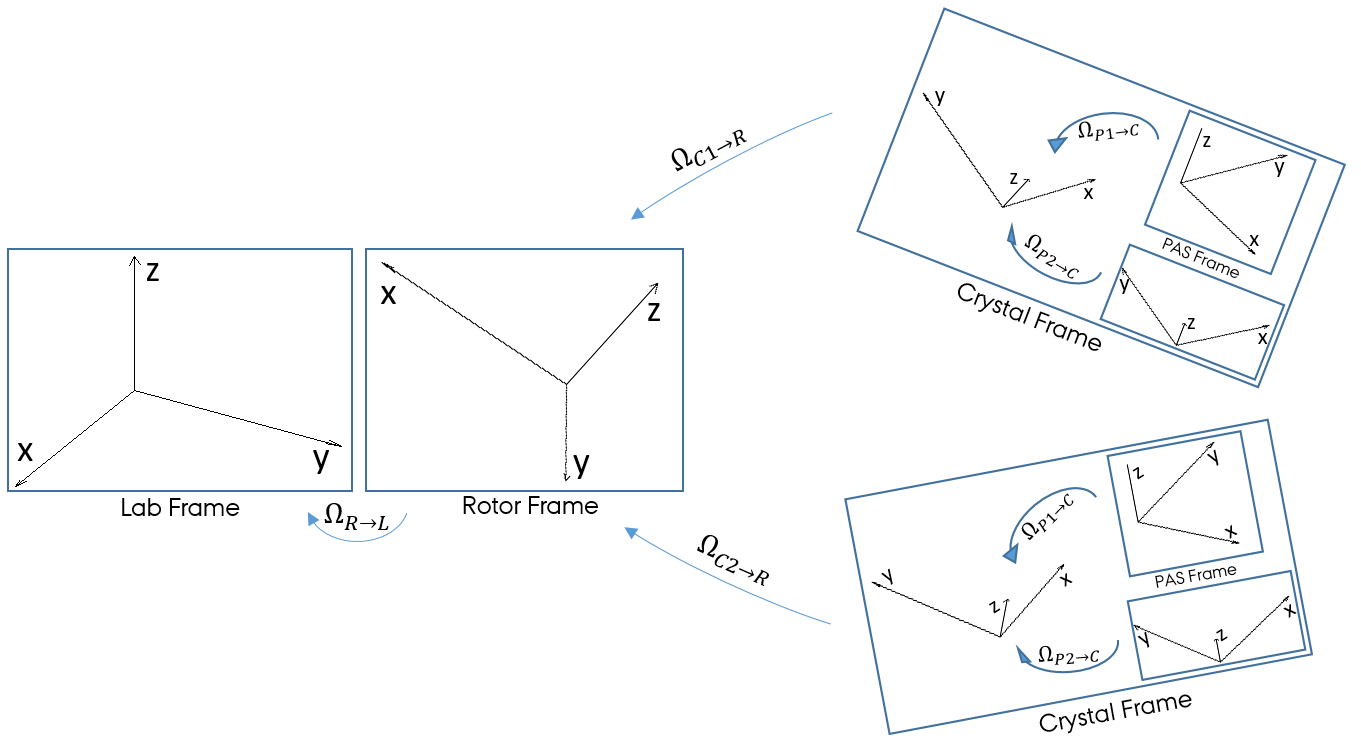}}
				\caption{Frame transformations for spatial interaction tensor. Principal Axis frame of an interaction is transformed to a molecular/crystal frame followed by a transformation to the rotor and finally a transformation to account for the angle the rotor axis makes with the external magnetic field.}
				\label{fig:pcr}
			\end{figure}

			The irreducible spatial spherical tensors $\mathcal{R}_k$ of rank $k$ and their components $\mathcal{R}_{k,l}$ are related to the reducible Cartesian spatial tensor components $\lambda_{ij}$ as
			\begin{equation}
				\begin{alignedat}{2}
					\mathcal{R}_{0,0} &= \frac{-1}{\sqrt{3}} (\lambda_{xx} + \lambda_{yy} + \lambda_{zz})\\
					\mathcal{R}_{1,0} &= \frac{i}{\sqrt{2}}(\lambda_{xy}-\lambda_{yx})\\
					\mathcal{R}_{1,\pm1} &= \frac{1}{2}(\lambda_{zx}-\lambda_{xz} \pm i(\lambda_{zy}-\lambda_{yz}))\\
					\mathcal{R}_{2,0} &= \frac{1}{\sqrt{6}}(2\lambda_{zz}-\lambda_{xx}-\lambda_{yy})\\
					\mathcal{R}_{2,\pm1} &= \mp\frac{1}{2}(\lambda_{xz}+\lambda_{zx} \pm i(\lambda_{yz}+\lambda_{zy}))\\
					\mathcal{R}_{2,\pm2} &= \frac{1}{2}(\lambda_{xx}-\lambda_{yy} \pm i(\lambda_{xy}+\lambda_{yx}))\\
				\end{alignedat}
			\label{eq:cart2sph}
			\end{equation}
			\begin{table}[!h] 
				\footnotesize 
				\caption{\footnotesize{Reduced Wigner matrix 
						elements $d^{(j)}_{m',m}(\beta)$ for $j$ = 1 and 2.$^a$}} 
				\begin{tabular*}{\hsize}{ccccccc}  
					\hline 
					\hline
					$j$ & $m' \setminus m$  &  -2        &   -1        &   0   &   1         &   2   \\         
					\hline   
					1   &         -1        &            &   \epc      & \sts  & \emc        &       \\  
					&          0        &            &  -\sts      &  \cb  & \sts        &       \\         
					&          1        &            &   \emc      & -\sts & \epc        &       \\         
					\hline       
					2   &         -2        & \cop       & \epc \sbe   & \sbs  & \emc \sbe   &  \com \\         
					&         -1        & -\epc \sbe & \ctb - \emc & \sbt  & \epc - \ctb & \emc \sbe\\         
					&          0        &    \sbs    &   -\sbt     & \slp  &  \sbt       & \sbs \\         
					&          1        & -\emc \sbe & \epc - \ctb & -\sbt & \ctb - \emc & \epc \sbe \\         
					&          2        & \com       & -\emc \sbe  & \sbs  & -\epc \sbe  &  \cop \\         
					
					\hline   
					\hline
				\end{tabular*}   
				\footnotesize{$^a$ Abbreviations: \cb = $\cos \beta$, \sbe = $\sin \beta$  .}  
				\label{tab:wig}
				\normalsize
			\end{table}			
			For every interaction in a spin system, it is possible to find a reference frame, known as the principal axis frame (PAS), denoted with a superscript $P$, in which the symmetric component ($\lambda_2$) of the interaction tensor is a diagonal matrix and is fully described by $\lambda^P_{xx}, \lambda^P_{yy}$ and $\lambda^P_{zz}$. The anti-symmetric component is however not diagonal and is described by $\lambda^P_{xy}$, $\lambda^P_{xz}$ and $\lambda^P_{yz}$\cite{saito2010chemical}. It is common in NMR to represent the components of the Cartesian tensor with the isotropic average $\delta_{iso}$, the anisotropy $\delta_{aniso}$ and the asymmetry parameter $\eta$ of the tensor as,
			\begin{equation}
				\begin{alignedat}{2}
					\delta_{iso} &= \frac{1}{3}(\lambda^P_{xx} + \lambda^P_{yy} + \lambda^P_{zz})\\
					\delta_{aniso} &= \lambda^P_{zz}-\delta_{iso}\\
					\eta &= \frac{\lambda^P_{yy}-\lambda^P_{xx}}{\delta_{aniso}}.
				\end{alignedat}
			\label{eq:2-35}
			\end{equation}
			such that $\delta_{iso}$ describe the zeroth-rank tensor component, while $\delta_{aniso}$ and $\eta$ describe the second-rank tensor component. Eq. \ref{eq:cart2sph} in PAS frame, then is simply
			\begin{equation}
				\begin{alignedat}{2}
					\mathcal{R}_{0,0}^P &= -\sqrt{3} \delta_{iso},\\
					\mathcal{R}_{1,0}^P &= i\sqrt{2}\lambda_{xy},\\
					\mathcal{R}_{1,\pm1}^P &= -\lambda_{xz} \mp i\lambda_{yz},\\
					\mathcal{R}_{2,0}^P &= \sqrt{\frac{3}{2}}\delta_{aniso},\\
					\mathcal{R}_{2,\pm1}^P &= 0,\\
					\mathcal{R}_{2,\pm2}^P &= \frac{-1}{2}\delta_{aniso}\eta.
				\end{alignedat}
			\label{eq:cart2sph_PAS}
			\end{equation}
			A set of transformations is needed to relate the spatial dependency tensor in PAS frame to lab frame. The transformations are governed by Eq. \ref{eq:spTensorTransf}. The PAS frame interaction tensor is first transformed to the molecular frame (also called a crystal frame). As the molecules in an NMR sample are in a multitude of possible orientations, crystal frames are transformed to the frame of rotor that contains the sample. The rotor is typically aligned at a specific angle with respect to the external magnetic field (lab frame) and the last transformation concerns this. The entire sequence of transformations is shown in Fig. \ref{fig:pcr}.
			The spin part in Eq. \ref{eq:H_vector_rep} can also be represented as spherical spin tensor operators, in a similar fashion. The one-spin spherical spin-tensor operators for a spin-$\frac{1}{2}$ nucleus are
			\begin{equation}
				\begin{alignedat}{3}
					\mathcal{T}_{0,0} &= \mathds{1} &\\
					\mathcal{T}_{1,0} &= \hat{I}_z\\
					\mathcal{T}_{1,1} &= \frac{-1}{\sqrt{2}}\hat{I}^+ &= \frac{-1}{\sqrt{2}}(\hat{I}_x + i\hat{I}_y)\\
					\mathcal{T}_{1,-1} &= \frac{1}{\sqrt{2}}\hat{I}^- &= \frac{1}{\sqrt{2}}(\hat{I}_x - i\hat{I}_y).
				\end{alignedat}
			\end{equation}
			This allows Eq. \ref{eq:H_vector_rep} to be rewritten in lab frame as
			\begin{eqnarray}
			\mathcal{H}_{\lambda} & = & C^{\lambda}\sum_{j=0}^{2} \sum_{m=-j}^{j}(-1)^m\mathcal{R}_{j,-m}^{\lambda}\mathcal{T}_{j,m}^{\lambda}
			\end{eqnarray}
			in spherical tensor basis.\\
			
			The components $\mathcal{R}^{P}_{j,m}$ and $\mathcal{T}_{j,m}$ for NMR interactions are summarized in Table \ref{tab:sphTensorComp}, where the spatial component are in PAS frame and the spin components are in lab frame. The set of transformations to obtain the spatial components in lab frame is performed as described above, which could also incorporate sample spinning.
			\begin{table}[h] 
				\footnotesize 
				\caption{\footnotesize{Spherical tensor elements of spatial and spin parts of the interaction Hamiltonian. The spatial components are given in the PAS frame while the spin part components are given in the lab frame.}} 
				\begin{tabular*}{\hsize}{c||ccc}  
					\hline 
					\hline
					$\lambda$ & CS & J & D\\
					\hline
					$C^{\lambda}$ & $\gamma_i$ & 1 & $-2\hbar\gamma_i\gamma_j$\\
					
					$\mathcal{R}_{0,0}^{\lambda,P}$ & $\delta_{iso}$ & $J_{iso}$ & 0\\
					$\mathcal{R}_{2,0}^{\lambda,P}$ & $\sqrt{\frac{3}{2}}\delta_{aniso}$ & $\sqrt{\frac{3}{2}}J_{aniso}$ & $\sqrt{\frac{3}{2}}\frac{\mu_0}{4\pi} |\vec{r}_{ij}|^{-3}$\\
					$\mathcal{R}_{2,\pm1}^{\lambda,P}$ & 0 & 0 & 0 \\
					$\mathcal{R}_{2,\pm2}^{\lambda,P}$ & $-\delta_{aniso}\frac{\eta_{\textrm{CS}}}{2}$ & $-\textrm{J}_{aniso}\frac{\eta_{\textrm{J}}}{2}$ & 0\\
					
					$\mathcal{T}_{0,0}^{\lambda}$ & $\hat{I}_{iz} B_0$ & $\vec{\hat{I}}_i\cdot\vec{\hat{I}}_j$ & 0 \\
					$\mathcal{T}_{2,0}^{\lambda}$ & $\sqrt{\frac{2}{3}}\hat{I}_{iz}B_0$ & $\frac{1}{\sqrt{6}}(3\hat{I}_{iz}\hat{I}_{jz}-\vec{\hat{I}}_i\cdot\vec{\hat{I}}_j)$ & $\frac{1}{\sqrt{6}}(3\hat{I}_{iz}\hat{I}_{jz}-\vec{\hat{I}}_i\cdot\vec{\hat{I}}_j)$\\
					$\mathcal{T}_{2,\pm1}^{\lambda}$ & $\mp\hat{I}_i^{\pm}B_0$ & $\mp\frac{1}{2}(\hat{I}_i^{\pm}\hat{I}_{jz} + \hat{I}_{iz}\hat{I}_j^{\pm})$ & $\mp\frac{1}{2}(\hat{I}_i^{\pm}\hat{I}_{jz} + \hat{I}_{iz}\hat{I}_j^{\pm})$\\
					$\mathcal{T}_{2,\pm2}^{\lambda}$ & 0 & $\frac{1}{2}\hat{I}_i^{\pm}\hat{I}_j^{\pm}$ & $\frac{1}{2}\hat{I}_i^{\pm}\hat{I}_j^{\pm}$\\
					\hline   
					\hline
				\end{tabular*}
				\label{tab:sphTensorComp}
				\normalsize
			\end{table}
			
		\subsection{Spin-Field interactions}
			\subsubsection{The Zeeman Interaction}
				The Zeeman interaction is generally the largest of NMR interactions and describes the interaction of the spin with the static magnetic field $\vec{B}_0$, which is, for convenience, set to point along the z-axis in the lab frame. The expression in Eq. \ref{eq:zeeman_E} can be rewritten as a scalar product given in Eq. \ref{eq:genHamil}
				\begin{equation}
					\mathcal{H}_{\textrm{Z}} = \vec{I} \cdot (-\gamma \mathds{1}) \cdot \vec{B}_0.
				\end{equation}
				$\omega_0 = -\gamma |\vec{B}_0|$ is the Larmor frequency of the nuclear spin of interest and $\gamma$ is the nucleus specific gyromagnetic ratio, which is an observed ratio of magnetic moment to angular momentum, and its sign determines the sense of precession of the spin about the magnetic field.
			
			\subsubsection{The Chemical Shift Interaction}
				Each nucleus experiences an effective magnetic field slightly different from the external magnetic field $\vec{B}_0$. This is because of the magnetic field induced by the surrounding electrons at the location of each nucleus. This allows to distinguish nuclei of same species in different chemical environments. The induced magnetic field is dependent on the electronic charge distribution and is proportional to the external magnetic field $\vec{B}_0$. The effective magnetic field experienced by a nucleus is, therefore, given by
			\begin{equation}
				\vec{B}_{\tr{eff}} = \vec{B}_0 + \sigma \vec{B}_0,
			\end{equation}
			where $\sigma$ is the chemical shift tensor of the nucleus that quantifies the proportionality of the induced magnetic field to the external magnetic field. The chemical shift Hamiltonian in the lab frame is then given by,
			\begin{equation}
			\begin{alignedat}{2}
				\mathcal{H}_{\textrm{CS}} &= \vec{I} \cdot (-\gamma\sigma) \cdot \vec{B}_0\\
				&= \omega_0(\sigma_{xz}I_x + \sigma_{yz}I_y + \sigma_{zz}I_z)
			\end{alignedat}
			\label{eq:2-40}
			\end{equation}
			In high-field, the precession of the spin is considered fast enough to approximately average out $x$ and $y$ components. This simplifies Eq. \ref{eq:2-40} to $\mathcal{H}_{\tr{CS}} = \omega_0\sigma_{zz}I_z$.
			
			The chemical shift tensor can also be expressed in the PAS frame, where as given in Eq. \ref{eq:2-35},
			\begin{equation}
				\begin{alignedat}{2}
					\sigma^{\tr{CS}}_{iso} &= \frac{1}{3}(\sigma^P_{xx} + \sigma^P_{yy} + \sigma^P_{zz})\\
					\delta^{\tr{CS}} &= \sigma^P_{zz}-\sigma^{\tr{CS}}_{iso}\\
					\eta^{\tr{CS}} &= \frac{\sigma^P_{yy}-\sigma^P_{xx}}{\delta^{\tr{CS}}}.
				\end{alignedat}
			\label{eq:2-42}
			\end{equation}
			where the convention is $|\sigma^{P}_{zz} - \sigma^{\tr{CS}}_{iso}| \geq |\sigma^{P}_{xx} - \sigma^{\tr{CS}}_{iso}| \geq |\sigma^{P}_{yy} - \sigma^{\tr{CS}}_{iso}|$ so that $\eta^{\tr{CS}}$ is positive and smaller than 1, while $\delta^{\tr{CS}}$ is either positive or negative. Parameters defined in Eq. \ref{eq:2-42} are helpful as they determine the position, width and shape of the peak in the spectrum.\\
			
			\subsubsection{The Radio-frequency Interaction}
			\label{ssec:rot_frame}
				The rf Hamiltonian takes the form
				\begin{equation}
				\hat{\mathcal{H}}_{\textrm{rf}}(t) = \hat{\mathcal{H}}_{\textrm{rf}}(t) = 2\omega_{\textrm{rf}}(t)\cos(\omega_c t+\phi(t)) \hat{I}_x,
				\end{equation}
				where $\omega_0$ is the Larmor frequency of the nucleus of interest, $\omega_{\tr{rf}}(t)$ is the time dependent amplitude of the rf field, $\omega_c$ is the frequency of the rf field, called the \textit{carrier frequency} and $\phi(t)$ is the time-dependent phase of the rf field. It is often simpler to transform the entire description to the rotating frame described by the carrier frequency. The external Hamiltonian, which also includes the Zeeman interaction, in such a frame is given by,
				\begin{equation}
				\begin{alignedat}{2}
				\hat{\tilde{\mathcal{H}}}_{\textrm{ext}} 	& = \omega_0e^{i\omega_ct\hat{I}_z}\hat{I}_ze^{-i\omega_ct\hat{I}_z} +\\
				& \hspace*{30pt}  2\omega_{\textrm{rf}}\cos(\omega_0t+\phi(t))e^{i\omega_ct\hat{I}_z}\hat{I}_xe^{-i\omega_ct\hat{I}_z} - ie^{i\omega_ct\hat{I}_z}\frac{d(e^{-i\omega_ct\hat{I}_z})}{dt}\\
				& = \omega_0 \hat{I}_z +  2\omega_{\textrm{rf}}\cos(\omega_0t+\phi(t))(\hat{I}_x\cos(\omega_ct) - \hat{I}_y\sin(\omega_ct)) - \omega_c \hat{I}_z\\
				& = (\omega_0-\omega_c)\hat{I}_z + \omega_{\textrm{rf}}\Big(\hat{I}_x \cos\phi(t) + \hat{I}_y\sin\phi(t) \Big) + \\
				& \hspace*{30pt}\omega_{\textrm{rf}}\Big(\hat{I}_x\cos\big((\omega_0+\omega_c)t + \phi(t)\big) - \hat{I}_y\sin\big((\omega_0+\omega_c)t+\phi(t)\big)\Big) \label{eq:rf_zeeman}
				\end{alignedat}
				\end{equation}
				On averaging over the Larmor period ($\frac{2\pi}{\omega_0}$), which is orders of magnitude shorter than the cycle time of rf amplitude and phase, the average first-order Hamiltonian is
				\begin{equation}
				\begin{alignedat}{2}
				\hat{\overline{\tilde{\mathcal{H}}}}_{\tr{ext}} &= (\omega_0-\omega_c)\hat{I}_z + \omega_{\textrm{rf}}(\cos\phi(t) \hat{I}_x + \sin\phi(t) \hat{I}_y) + \\
				&\hspace{30pt}\int_{0}^{\frac{2\pi}{\omega_0}}\omega_{\textrm{rf}}\Big(\hat{I}_x\cos\big((\omega_0+\omega_c)t + \phi(t)\big) - \hat{I}_y\sin\big((\omega_0+\omega_c)t+\phi(t)\big)\Big)\\
				&= (\omega_0-\omega_c)\hat{I}_z + \omega_{\textrm{rf}}(\cos\phi(t) I_x + \sin\phi(t) I_y),
				\end{alignedat}
				\label{eq:rotframe_Hext}
				\end{equation}
				where $\omega_c \approx \omega_0$ i.e., the rf carrier frequency is set very close to the Larmor frequency of the nucleus, and $\phi(t)$ is considered piece-wise constant in the intervals defined by \big((n-1)$\frac{2\pi}{\omega_0}$,n$\frac{2\pi}{\omega_0}$\big), with integer n $\geq1$.\\
					
		\subsection{Spin-Spin interactions}
			\subsubsection{Dipolar-coupling interaction}
				The influence of magnetic dipole moment of one spin on another through space is termed the \textit{dipolar coupling} interaction and the Hamiltonian is given by
				\begin{equation}
					\hat{\mathcal{H}}_{\textrm{D}} = -\frac{\mu_0}{4\pi}\frac{\gamma_1\gamma_2}{|\vec{r}|^3} \Bigg(\frac{3}{|\vec{r}|^2}\Big(\vec{\hat{I}}_1\cdot\vec{r}\Big)\Big(\vec{\hat{I}}_1\cdot\vec{r}\Big) - \Big(\vec{\hat{I_1}}\cdot\vec{\hat{I}}_2\Big)\Bigg)
				\label{eq:2-43}
				\end{equation}
				where $\vec{r}$ is the distance vector connecting the two spins. This can be written in the form of Eq. \ref{eq:genHamil} as
				\begin{equation}
					\hat{\mathcal{H}}_{\textrm{D}} = \vec{\hat{I}}_1 \cdot D \cdot \vec{\hat{I}}_2
				\end{equation}
				where the elements of the traceless (and therefore purely anisotropic interaction) symmetric matrix $D$ are given by
				\begin{equation}
					\begin{alignedat}{2}
						D_{\alpha\beta} &= -\frac{\mu_0}{4\pi}\frac{\gamma_1\gamma_2}{|\vec{r}|^3} (-3e_\alpha e_\beta + \delta_{\alpha\beta})
					\end{alignedat}
				\end{equation}
				with $\alpha,\beta = x, y, z$ and $e_\alpha$ / $e_\beta$ are the components of a unit vector parallel to $\vec{r}$. In the PAS frame, since $D$ is symmetric and traceless,
				\begin{equation}
					D^{P} = \frac{-2\mu_0}{4\pi} \frac{\gamma_1\gamma_2}{|\vec{r}|^3}
					\begin{bmatrix}
						\frac{-1}{2} & 0 & 0\\
						0 & \frac{-1}{2} & 0\\
						0 & 0 & 1
					\end{bmatrix}.
				\end{equation}
				Representing the spin part of Eq. \ref{eq:2-43} in the rotating frame, as described in Sec. \ref{ssec:rot_frame}, and assuming high-field, the dipolar Hamiltonian term is given by
				\begin{equation}
					\hat{\mathcal{H}}_{\tr{D}} = -\frac{\mu_0}{4\pi} \frac{\gamma_1\gamma_2}{|\vec{r}^3|} \frac{3\cos^2\theta - 1}{2} (2\hat{I}_{1z}\hat{I}_{2z} - \hat{I}_{1x}\hat{I}_{2x} - \hat{I}_{1y}\hat{I}_{2y}),
				\end{equation}
				when the two nuclei are the same kind, homonuclear and
				\begin{equation}
					\hat{\mathcal{H}}_{\tr{D}} = -\frac{\mu_0}{4\pi} \frac{\gamma_1\gamma_2}{|\vec{r}^3|} \frac{3\cos^2\theta - 1}{2} 2\hat{I}_{1z}\hat{I}_{2z},
				\end{equation}
				when the two nuclei are of different kind, heteronuclear with $\theta$ being the angle the vector $\vec{r}$ makes with the external magnetic field. As can be seen here, at the magic angle, $3\cos^2\theta = 1$, making the dipolar coupling Hamiltonian null.\\
				
			\subsubsection{J-coupling interaction}
				The influence of magnetic dipole moment of one spin on another, mediated by the bond electrons is termed the J-coupling or the scalar coupling. The Hamiltonian for this interaction is given by,
				\begin{equation}
					\hat{\mathcal{H}}_{\tr{J}} = 2\pi \vec{\hat{I}}_1 \cdot J \cdot \vec{\hat{I}}_2,
				\end{equation}
				where the interaction matrix $J$, in principle, has both isotropic and anisotropic components. However only the isotropic component (the trace of $J$, denoted by $J_{\tr{iso}}$) is usually considered as the anisotropic component is experimentally indistinguishable from the dipolar coupling interaction and is included in the dipolar interaction tensor $D$. For light nuclei, the anisotropic component is known to be small and therefore neglected. Again with the high-field approximation, the Hamiltonian is given by,
				\begin{equation}
					\hat{\mathcal{H}}_{\tr{J}} = 2\pi J_{\tr{iso}} \vec{\hat{I}}_1 \cdot \vec{\hat{I}}_2
				\end{equation}
				for the homonuclear case, and
				\begin{equation}
					\hat{\mathcal{H}}_{\tr{J}} = 2\pi J_{\tr{iso}} \vec{\hat{I}}_{1z} \cdot \vec{\hat{I}}_{2z}
				\end{equation}
				for the heteronuclear case.\\
				
			The following chapters will use the concepts and mathematical description established in this chapter with focus on the chemical shift and dipolar coupling Hamiltonians. The average Hamiltonian theory to find effective Hamiltonian for any interaction frame Hamiltonian represented as product of complex exponentials with limited set of frequencies, will be discussed in the next chapter, followed by its application to dipolar recoupling pulse sequences in the subsequent chapters.

%% file: Chapters/chap3_DesignPrinciples.tex
\chapter{Design Principles}
\label{chap:DesignPrinciples}
	The knowledge on effective Hamiltonian is crucial in describing and understanding a solid-state NMR experiment. The performance of experiments is evaluated and design of novel experiments is guided, with the help of effective Hamiltonians. The time dependencies to the system Hamiltonian are brought in by MAS and rf irradiation. There are two major ways to deal with time-dependent Hamiltonians: the average Hamiltonian theory\cite{waugh1968js,haeberlen1968coherent,burum1981magnus,klarsfeld1989recursive} (AHT) and the Floquet theory\cite{floquet,shirley1965solution,maricq1982application,shimon1996}. AHT has been the most widely adopted approach in NMR, whereas Floquet theory is the preferred approach in most other fields\cite{chu2004beyond,kulander1991time}.\\
	
	In Floquet theory, the time-dependent finite-dimensional Hilbert space Hamiltonian is transformed into a time-independent infinite-dimensional Hamiltonian in the so called Floquet space, which is the Kronecker product of Fourier space and the Hilbert space\cite{augustine1995theoretical,boender1996stacking}. Strictly speaking, it is inappropriate to call the Floquet Hamiltonian time-independent, as the problem has been transformed out of time domain into frequency domain. Density operator evolution to arbitrary times can, however, be calculated directly in the Floquet space to obtain a spectrum\cite{schmidt1992floquet,kubo1990one}. This, for example, has aided in the analytical calculation of both the centrebands and the sidebands in MAS spectra\cite{schmidt1992floquet,nakai1992application,bain2001introduction}. In general this approach is of interest for numerical simulations\cite{levante1995formalized,baldus1995structure}. However, in practice, numerical simulations of the time-dependent Hilbert space Hamiltonian sliced to piece-wise constant Hamiltonians, are more efficient. To understand and gain insight into an experiment, an approximate analytical time-independent Hamiltonian in Hilbert space is sought. This can be achieved by diagonalising the Floquet Hamiltonian and projecting it back to the Hilbert space\cite{shimon1996}. More often than not, perturbation theory on the Floquet Hamiltonian, rather than full diagonalisation, is applied\cite{primas1963generalized}. The van Vleck transformation\cite{van1948dipolar}, which is a block diagonalisation method, is used to achieve this. The effective Hamiltonian so obtained has been shown to be identical to that obtained with AHT\cite{maricq1986application,llor1992equivalence} using Magnus expansion\cite{magnus1954exponential,salzman1986convergence,burum1981magnus} and the Hamiltonians are valid only at multiples of the periodicity (stroboscopic observation) defined by the time-dependent Hamiltonian. The details of the Floquet theory and van Vleck perturbation treatment is beyond the scope of this dissertation and can be found in the literature, in particular the reviews by S. Vega\cite{leskes2010floquet}, et al and M. Ernst\cite{scholz2010operator} are recommended.\\
	
	In this chapter, AHT is discussed and used to derive the effective Hamiltonian for a time-dependent Hamiltonian. This is followed by sequential discussion on situations with increasing complexity in the time dependency of the Hamiltonian. It will be shown that the interaction frame time modulations of single-spin operators can be described by using at most two frequencies per spin. One which reflects the periodicity of the pulse sequence acting on the spin in question and the other reflecting the overall rotation caused by a unit of the pulse sequence on the single-spin operators. This allows the total Hamiltonian for a solid-state MAS NMR experiment involving $N$ spin-$\frac{1}{2}$ nuclei to be described using at most $2N + 1$ fundamental frequencies, where one of the frequencies is a result of the time modulation of spatial part of the Hamiltonian. The overall rotation on a spin, caused by a pulse sequence can be numerically found using quaternions, an alternative description for rotations\cite{hamilton1853lectures,hamilton1969,klein1965}, and is described in the last section.

\section{Average Hamiltonian Theory}
	Consider a spin system under a time dependent Hamiltonian, $\hat{\mathcal{H}}(t)$, that can be written as a sum of two parts,
	\begin{equation}
		\hat{\mathcal{H}}(t) = \hat{\mathcal{H}}_{\tr{small}}(t) + \hat{\mathcal{H}}_{\tr{big}}(t).
	\end{equation}
	The effect of big Hamiltonian on the interesting small Hamiltonian, can easily be seen by transforming the problem to an interaction frame defined by $\hat{\mathcal{H}}_{\tr{big}}(t)$. This helps the series expansion of the effective Hamiltonian converge faster. As seen in Sec. \ref{sec:int_frame}, the interaction frame Hamiltonian $\hat{\tilde{\mathcal{H}}}(t)$ is given by,
	\begin{equation}
		\hat{\tilde{\mathcal{H}}}(t) = \hat{U}_{\tr{big}}^{\dagger}(t,0) \hat{\mathcal{H}}_{\tr{small}}(t)\hat{U}_{\tr{big}}(t,0),
		\label{eq:intFrHam}
	\end{equation}
	where
	\begin{equation}
		\hat{U}_{\tr{big}}(t,0) = \hat{\mathcal{T}}e^{-i\int_{0}^{t}\hat{\mathcal{H}}_{\tr{big}}(t')dt'}.
	\label{eq:3-3}
	\end{equation}
	In the interaction frame, the propagator for a time interval (0,$t$) is given by $\hat{\tilde{U}}(t,0) = \hat{\mathcal{T}}e^{-i\int_{0}^{t}\hat{\tilde{\mathcal{H}}}(t')dt'}$, as also given in Eq. \ref{eq:dyson}. As a product of unitary operators is again an unitary operator, $\hat{\tilde{U}}(\tau,0)$ for the interval 0 to $\tau$ can in principle be represented as a single operator,
	\begin{equation}
		\hat{\tilde{U}}(\tau,0) = e^{-i\hat{\overline{\tilde{\mathcal{H}}}}(\tau,0)\tau},
		\label{eq:prop_int_frame}
	\end{equation}
	where $\hat{\overline{\tilde{\mathcal{H}}}}$($\tau$,0) is the \textit{average} or \textit{effective} Hamiltonian with averaging done over the interval (0,$\tau$). The average Hamiltonian is given as a solution to the linear differential equation,
	\begin{equation}
		\frac{d\hat{\tilde{U}}(t)}{dt} = -i \hat{\tilde{\mathcal{H}}}(t)\hat{\tilde{U}}(t),
		\label{eq:dudt}
	\end{equation}
	through Magnus expansion\cite{magnus1954exponential,salzman1986convergence,burum1981magnus}. The solution is given by
	\begin{equation}
		\hat{\tilde{U}}(\tau,0) = e^{-i\sum\limits_{k = 1}^{\infty} \hat{\overline{\tilde{\mathcal{H}}}}^{(k)}(\tau,0)\tau}\hat{\tilde{U}}(0)
		\label{eq:ahtMagnus}
	\end{equation}
	where the initial value of the propagator $\hat{\tilde{U}}(0)$ is known to be $\mathds{1}$ and the average Hamiltonian $\hat{\overline{\tilde{H}}}(\tau,0)$ is given as a series $\sum_{k = 1}^{\infty} \hat{\overline{\tilde{\mathcal{H}}}}^{(k)}(\tau,0)$. The first two terms in the series are given by
	\begin{equation}
		\begin{aligned}
			\hat{\overline{\tilde{\mathcal{H}}}}^{(1)}(\tau,0) &= \frac{1}{\tau}\int_{0}^{\tau}\hat{\tilde{\mathcal{H}}}(t_1)dt_1 \\
			\hat{\overline{\tilde{\mathcal{H}}}}^{(2)}(\tau,0) &= \frac{1}{2i\tau}\int_{0}^{\tau}dt_1\int_{0}^{t_1}[\hat{\tilde{\mathcal{H}}}(t_1),\hat{\tilde{\mathcal{H}}}(t_2)]dt_2.
		\end{aligned}
		\label{eq:MagExpnsn}
	\end{equation}
	Note that from here on, the time interval $(t,0)$ will be dropped from the effective Hamiltonians and $(t,0)$ will be replaced with just (t) in the propagators for better readability. The Hamiltonian $\hat{\overline{\tilde{\mathcal{H}}}}$ is valid for any chosen fixed time $\tau$, where the averaging is done from 0 to $\tau$. However should the propagator given in Eq. \ref{eq:prop_int_frame} be valid for use at multiples of $\tau$, then it is understandably required that $\tau$ also represent the periodicity of the Hamiltonian $\hat{\tilde{\mathcal{H}}}$(t), i.e., $\hat{\tilde{\mathcal{H}}}(\tau+t) = \hat{\tilde{\mathcal{H}}}(t)$. The evolution at multiples of $\tau$ (i.e., $n\tau$, where $n$ is an integer) is then described by
	\begin{equation}
		\hat{\tilde{U}}(n\tau) = \Big(\hat{\tilde{U}}(\tau)\Big)^n = e^{-i\hat{\overline{\tilde{\mathcal{H}}}}(\tau)n\tau}.
		\label{eq:ahtU}
	\end{equation}

	In the following sections, the cases corresponding to the modulation of $\hat{\tilde{\mathcal{H}}}$(t) with single and multiple frequencies will be discussed.
	\subsection{Single frequency}
	\label{sect:3_SingleFrequency}
		Consider $\hat{\mathcal{H}}_{\tr{small}}(t)$ and $\hat{\mathcal{H}}_{\tr{big}}(t)$ which are periodic in time with $\tau_c$ and $\hat{U}_{\tr{big}}(t)$ is cyclic over the same time $\tau_c$, i.e., $\hat{U}_{\tr{big}}(\tau_c) = \mathds{1}$. The total Hamiltonian in interaction frame of $\hat{\mathcal{H}}_{\tr{big}}(t)$ can then be shown to also be periodic with $\tau_c$. i.e.,
			\begin{empheq}[right={\empheqrbrace \Rightarrow \hat{\tilde{\mathcal{H}}}(\tau_c + t) = \hat{\tilde{\mathcal{H}}}(t). }]{alignat=2}
			\begin{aligned}
				\hat{\mathcal{H}}_{\tr{small}}(\tau_c + t) &= \hat{\mathcal{H}}_{\tr{small}}(t)\\
				\hat{\mathcal{H}}_{\tr{big}}(\tau_c + t) &= \hat{\mathcal{H}}_{\text{big}}(t)\\
				\hat{U}_{\tr{big}}(\tau_c) &= \mathds{1} 
			\end{aligned}
		\label{eq:aht_assumptions}
		\end{empheq}
		In a MAS solid-state NMR experiment, this corresponds to a scenario where the internal Hamiltonians are periodic with $\tau_r$, the rf pulse sequence is rotor synchronised with period $\tau_r$ and $\hat{U}_{\tr{rf}}(\tau_r) = \mathds{1}$. The rf field interaction frame Hamiltonian is then periodic with $\tau_r$. In such a case, $\hat{\tilde{\mathcal{H}}}(t)$ can be Fourier transformed and written as,
		\begin{equation}
			\begin{alignedat}{2}
				\hat{\tilde{\mathcal{H}}}(t) = \sum_{n=-\infty}^{\infty}\hat{\tilde{\mathcal{H}}}^{(n)}e^{in\omega_rt},
			\end{alignedat}
		\end{equation}
		where $\hat{\tilde{\mathcal{H}}}^{(n)}$ are the Fourier components whose elements also contain spin operators. The calculation of these Fourier components are shown in Sec. \ref{sect:4_ff} for the case of amplitude-modulated pulse sequences and in Sec. \ref{sect:5_ff} for the case of amplitude- and phase-modulated pulse sequences.\\
		
		The effective Hamiltonian to first order, as given in Eq. \ref{eq:MagExpnsn}, is then simply,
		\begin{equation}
			\begin{alignedat}{2}
				\hat{\overline{\tilde{\mathcal{H}}}}^{(1)}(\tau_r,0) &= \frac{1}{\tau_r}\int_{0}^{\tau_r}\hat{\tilde{\mathcal{H}}}(t)dt\\
				&= \frac{1}{\tau_r}\sum_{n=-\infty}^{\infty}\hat{\tilde{\mathcal{H}}}^{(n)} \int_{0}^{\tau_r}e^{in\omega_rt} dt\\
				&= \hat{\tilde{\mathcal{H}}}^{(0)}.
			\end{alignedat}
		\end{equation}
		The second-order effective Hamiltonian is given by,
		\begin{gather}
			\begin{alignedat}{2}
				\hat{\overline{\tilde{\mathcal{H}}}}^{(2)} &= \frac{1}{2i\tau_r}\int_{0}^{\tau_r}dt_1\int_{0}^{t_1}[\hat{\tilde{\mathcal{H}}}(t_1),\hat{\tilde{\mathcal{H}}}(t_2)]dt_2.\\
				&= \frac{1}{2i\tau_r}\int_{0}^{\tau_r}\sum_{n_1,n_2}[\hat{\tilde{\mathcal{H}}}^{(n_1)},\hat{\tilde{\mathcal{H}}}^{(n_2)}]e^{in_1\omega_rt_1}dt_1\int_{0}^{t_1}e^{in_2\omega_rt_2}dt_2.
			\end{alignedat}
		\label{eq:irupa}
		\end{gather}
		There are four possible cases with the indices being summed over, depending on whether $n_1\omega_r = 0$ (resonance condition, abbreviated as $\texttt{res}$) or $n_1\omega_r \neq 0$ (non-resonance condition, abbreviated as $\texttt{non-res}$) and likewise for $n_2\omega_r$. The cases can be listed with corresponding commutators as,
		\begin{enumerate}
			\item $[\texttt{non-res}, \texttt{non-res}]$
			\item $[\texttt{res}, \texttt{non-res}]$
			\item $[\texttt{non-res}, \texttt{res}]$
			\item $[\texttt{res}, \texttt{res}]$.
		\end{enumerate}
		
		\subsubsection*{\underline{Cases 1 and 2:}}
			For the first two of these cases, where the second term in the commutator corresponds to a $\texttt{non-res}$, Eq. \ref{eq:irupa} simplifies to
			\begin{equation}
				= \frac{1}{2i\tau_r} \int_{0}^{\tau_r} \sum_{n_1,n_2}^{} [\hat{\tilde{\mathcal{H}}}^{(n_1)}, \hat{\tilde{\mathcal{H}}}^{(n_2)}] e^{in_1\omega_r t_1} \frac{(e^{in_2\omega_r t_1} - 1)}{i (n_2\omega_r)} dt_1.
			\label{eq:irupa2}
			\end{equation}
			This can further be seen as a sum of two terms, one involving the exponent of $n_2\omega_r$, while the other involving minus one. As the integration is over $\tau_r$, the first term gives a non-zero contribution only when $(n_1 + n_2)\omega_r = 0$ and allows the contribution to second-order effective Hamiltonian from $[\texttt{non-res}, \texttt{non-res}]$ to be reformulated as
			\begin{equation}
				= \frac{-1}{2} \sum_{n_0,n}^{} \frac{[\hat{\tilde{\mathcal{H}}}^{(n_0-n)}, \hat{\tilde{\mathcal{H}}}^{(n)}]}{n\omega_r},
			\end{equation}
			where the indices being summed over obey $n_0\omega_r = 0$ and $n\omega_r \neq 0$. Similarly, the second term results in a non-zero contribution only when $n_1\omega_r = 0$. This allows the contribution from $[\texttt{res}, \texttt{non-res}]$ to be reformulated as
			\begin{equation}
				= \frac{1}{2} \sum_{n_0,n}^{} \frac{[\hat{\tilde{\mathcal{H}}}^{(n_0)}, \hat{\tilde{\mathcal{H}}}^{(n)}]}{n\omega_r},
			\label{eq:3-16}
			\end{equation}
			where again, $n_0\omega_r = 0$ and $n\omega_r \neq 0$.
		
		\subsubsection*{\underline{Cases 3 and 4:}}
			For the last two of the listed cases, where the second term in the commutator corresponds to a $\texttt{res}$, Eq. \ref{eq:irupa} simplifies to
			\begin{equation}
				= \frac{1}{2i\tau_r} \int_{0}^{\tau_r} \sum_{n_1,n_2}^{} [\hat{\tilde{\mathcal{H}}}^{(n_1)}, \hat{\tilde{\mathcal{H}}}^{(n_2)}] t_1 e^{in_1\omega_r t_1} dt_1.
			\end{equation}
			Contribution from $[\texttt{non-res}, \texttt{res}]$ works out to be the same as given in Eq. \ref{eq:3-16}, while the contribution from $[\texttt{res}, \texttt{res}]$ works out to zero.\\
			
			The total second-order effective Hamiltonian can therefore be expressed as
			\begin{equation}
				\hat{\overline{\tilde{\mathcal{H}}}}^{(2)} = \frac{-1}{2} \sum_{n_0,n}^{} \frac{[\hat{\tilde{\mathcal{H}}}^{(n_0-n)}, \hat{\tilde{\mathcal{H}}}^{(n)}]}{n\omega_r} + \sum_{n_0,n}^{} \frac{[\hat{\tilde{\mathcal{H}}}^{(n_0)}, \hat{\tilde{\mathcal{H}}}^{(n)}]}{n\omega_r},
			\end{equation}
			where $n_0\omega_r = 0$ and $n\omega_r \neq 0$.
		
	\subsection{Multiple Frequencies}
	\label{sect:MultipleFrequencies}
		Consider $\hat{\mathcal{H}}_{\tr{small}}(t)$ that is periodic in time with $\tau_r$ ($=\frac{2\pi}{\omega_r}$) and $\hat{\mathcal{H}}_{\tr{big}}(t)$ that is periodic in time with a different time $\tau_m$ ($=\frac{2\pi}{\omega_m}$). Despite the fact that the sum of the two Hamiltonians, $\hat{\mathcal{H}}_{\tr{small}}(t) + \hat{\mathcal{H}}_{\tr{big}}(t)$, is necessarily periodic in time with $\tau'_c = \frac{2\pi}{\gcd(\omega_r,\omega_m)}$, the interaction frame Hamiltonian $\hat{\tilde{\mathcal{H}}}(t)$ is not necessarily periodic with $\tau'_c$, where $\gcd(a,b)$ refers to the greatest common divisor between $a$ and $b$. This is because, propagator of the big Hamiltonian at $\tau'_c$, $\hat{U}_{\tr{big}}(\tau'_c)$, may not necessarily be cyclic with time $\tau'_c$, i.e., $\hat{U}_{\tr{big}}(\tau'_c) \neq \mathds{1}$. But a multiple $p$ of $\tau'_c$ can always be found such that $\hat{U}_{\tr{big}}(p\cdot\tau'_c) = \mathds{1}$. Therefore, in general,
		\begin{empheq}[right={\empheqrbrace \Rightarrow \hat{\tilde{\mathcal{H}}}(\tau_c + t) = \hat{\tilde{\mathcal{H}}}(t),}]{alignat=2}
			\begin{aligned}
				\hat{\mathcal{H}}_{\tr{small}}\Bigg(\frac{2\pi}{\omega_r} + t\Bigg) &= \hat{\mathcal{H}}_{\tr{small}}(t)\\
				\hat{\mathcal{H}}_{\tr{big}}\Bigg(\frac{2\pi}{\omega_m} + t\Bigg) &= \hat{\mathcal{H}}_{\tr{big}}(t)\\
				\hat{U}_{\tr{big}}\big(\tau_c\big) &= \mathds{1}
			\end{aligned}
		\label{eq:318}
		\end{empheq}
		where $\tau_c = p \cdot \tau'_c = p \cdot \frac{2\pi}{\gcd(\omega_r,\omega_m)}$ where $p$ is an integer.\\
		
		In the special case, where the propagator of the big Hamiltonian is cyclic with time $\tau'_c$, i.e., $\hat{U}_{\tr{big}}(\tau'_c) = \mathds{1}$, the period of $\hat{\tilde{\mathcal{H}}}(t)$ is defined only by the frequencies $\omega_r$ and $\omega_m$ and is given by
		\begin{equation}
			\tau_c = \tau'_c = \frac{2\pi}{\gcd(\omega_r,\omega_m)}.
		\end{equation}
		The time dependency of the interaction frame Hamiltonian, in this special case, of $\tau_c = \tau'_c$, can therefore be expressed in the Fourier space with only these two frequencies as given by,
		\begin{equation}
			\hat{\tilde{\mathcal{H}}}(t) = \sum_{n,k}^{}\hat{\tilde{\mathcal{H}}}^{(n,k)} e^{in\omega_rt} e^{ik\omega_mt}
		\label{eq:3-20}
		\end{equation}
		where $\hat{\tilde{\mathcal{H}}}^{(n,k)}$ are the Fourier components, the calculation of which are shown later in Sec. \ref{sect:4_ff} and Sec. \ref{sect:5_ff}.	The effective or average Hamiltonian for the cycle time $\tau_c$ ($= \tau'_c$), to the first and second order are then given by,
		\begin{equation}
			\begin{alignedat}{2}
				\hat{\overline{\tilde{\mathcal{H}}}}^{(1)} &= \sum_{n_0,k_0} \hat{\tilde{\mathcal{H}}}^{(n_0,k_0)}
			\end{alignedat}
		\label{eq:321}
		\end{equation}
		and
		\begin{gather}
			\begin{alignedat}{2}
				\hat{\overline{\tilde{\mathcal{H}}}}^{(2)} &= \frac{-1}{2} \sum_{n,k}\frac{[\hat{\tilde{\mathcal{H}}}^{(n_0-n,k_0-k)},\hat{\tilde{\mathcal{H}}}^{(n,k)}]}{n\omega_r + k\omega_m} + \sum_{n_0,k_0}\sum_{n,k}\frac{[\hat{\tilde{\mathcal{H}}}^{(n_0,k_0)},\hat{\tilde{\mathcal{H}}}^{(n,k)}]}{n\omega_r + k\omega_m},
			\end{alignedat}
		\label{eq:322}
		\end{gather}
		where $n_0\omega_r + k_0\omega_m = 0$ and $n\omega_r + k\omega_m \neq 0$. The expressions in Eq. \ref{eq:321} and Eq. \ref{eq:322} can be derived by following the same procedure described in the previous section. It is noted here that even though pairs of $(n,k)$ will have different cycle times $\Big(\frac{2\pi}{n\omega_r + k\omega_m}\Big)$ in the integration, it will always be a factor of $\tau_c$ defined above and therefore the integration over $\tau_c$ will be valid.\\
		
		The expressions for the effective Hamiltonian given here for the first two orders, are exactly the same as obtained through Floquet theory that uses van Vleck transformation, and are available in literature reviews\cite{leskes2010floquet,scholz2010operator}.
		
		In the most general case, where the propagator of the big Hamiltonian, $\hat{U}_{\tr{big}}(t)$, is not cyclic with $\tau'_c$, but cyclic with some multiple $p$ of $\tau'_c$, i.e., $\hat{U}_{\tr{big}}(\tau'_c) \neq \mathds{1}$ but $\hat{U}_{\tr{big}}(p\cdot\tau'_c) = \mathds{1}$, the period of the interaction frame Hamiltonian, $\hat{\tilde{\mathcal{H}}}(t)$ is defined not only by the two frequencies $\omega_r$ and $\omega_m$, but also by a third frequency which reflects the multiple $p$. However, propagation for the interaction frame Hamiltonian, $\hat{\tilde{\mathcal{H}}}(t)$, can still be found at multiples of $\tau'_c$, using only time-independent averaged or effective Hamiltonians, by employing a trick.\\
		
		Say, $\hat{U}_{\tr{big}}(\tau'_c)$ can be found and expressed as,
		\begin{equation}
			\hat{U}_{\tr{big}}(\tau'_c) = e^{-i\hat{\mathcal{H}}_{\tr{eff}}\tau'_c},
		\label{eq:323}
		\end{equation}
		where $\hat{\mathcal{H}}_{\tr{eff}}$ is the time-independent average or effective of $\hat{\mathcal{H}}_{\tr{big}}(t)$ averaged over its periodic time $\tau'_c$. Since $\hat{\mathcal{H}}_{\tr{big}}$ is typically the rf field and/or isotropic chemical shift Hamiltonians, it only involves single-spin operators and so the average $\hat{\mathcal{H}}_{\tr{eff}}$ can be calculated using quaternions, which is discussed later in the chapter, in Sec. \ref{sect:quaternions}. Note that since $\hat{\mathcal{H}}_{\tr{big}}(t)$ is periodic in time with $\tau'_c$ (ref. Eq. \ref{eq:318}), the propagator given in Eq. \ref{eq:323} can be used to find the same at multiples of $\tau'_c$ as given by,
		\begin{equation}
			\hat{U}_{\tr{big}}(N\tau'_c) = \Big(\hat{U}_{\tr{big}}(\tau'_c)\Big)^N,
		\label{eq:324}
		\end{equation}
		where $N$ is an integer. This allows the different intervals of duration $\tau'_c$ of $\hat{\tilde{\mathcal{H}}}(t)$ to be related according to,
		\begin{equation}
			\hat{\tilde{\mathcal{H}}}(N\tau_c' + t) = \hat{U}_{\tr{big}}^{\dagger}(N\tau_c')\hat{\tilde{\mathcal{H}}}(t)\hat{U} _{\tr{big}}(N\tau_c').
		\label{eq:314}
		\end{equation}
		
		With the help of Eq. \ref{eq:314}, it can be shown that, propagators of the interaction frame Hamiltonian can be defined at multiples of $\tau'_c$, which is only a factor of the period $\tau_c$ ($= p\cdot\tau'_c$) of the interaction frame Hamiltonian $\hat{\tilde{\mathcal{H}}}(t)$, according to
		\begin{equation}
			\begin{aligned}
				\hat{\tilde{U}}(N\tau_c') &=e^{i\hat{\mathcal{H}}_{\tr{eff}}N\tau_c'} \cdot e^{-i\hat{\overline{\tilde{\mathcal{H}}}}N\tau_c'}.
			\end{aligned}
		\label{eq:modProp}
		\end{equation}
		Note that in Eq. \ref{eq:modProp}, $\hat{\overline{\tilde{\mathcal{H}}}}$ is the effective time-independent Hamiltonian of $\hat{\tilde{\mathcal{H}}}(t)$ averaged over the period $\tau_c$, while $\hat{\mathcal{H}}_{\tr{eff}}$ is the effective time-independent Hamiltonian of $\hat{\mathcal{H}}_{\tr{big}}(t)$ averaged over $\tau'_c$. This is advantageous as the propagators are now defined at time points ($\tau'_c$) shorter than the period ($\tau_c$) of $\hat{\tilde{\mathcal{H}}}(t)$, using only time-independent effective Hamiltonians.\\
		
		To verify Eq. \ref{eq:modProp}, consider $\tau_c$ sliced into $p$ intervals of duration $\tau'_c$ and each of those intervals further sliced into $m$ divisions of duration $\delta t$. The explicit expression for the propagator of the interaction frame Hamiltonian at $\tau_c$ can then be written as a product of exponentials, as given by,
		\begin{equation}
			\begin{alignedat}{2}
				\hat{\tilde{U}}^{\dagger}(\tau_c) &= \Bigg(\prod_{k=1}^{m}e^{i\hat{\tilde{\mathcal{H}}}(k\delta t)\delta t}\Bigg)\\
				& \hspace*{40pt} \hat{U}_{\tr{big}}(\tau'_c) \cdot \Bigg(\prod_{k=1}^{m}e^{i\hat{\tilde{\mathcal{H}}}(k\delta t)\delta t}\Bigg) \cdot \hat{U}_{\tr{big}}^{\dagger}(\tau'_c)\\
				& \hspace*{100pt} \cdots\\
				& \hspace*{60pt}
				\hat{U}_{\tr{big}}\big((p-1)\tau'_c\big) \cdot \Bigg(\prod_{k=1}^{m}e^{i\hat{\tilde{\mathcal{H}}}(k\delta t)\delta t}\Bigg) \cdot \hat{U}_{\tr{big}}^{\dagger}\big((p-1)\tau'_c\big),
			\end{alignedat}
		\label{eq:327}
		\end{equation}
		where Eq. \ref{eq:314} has been utilised. Note that in Eq.\ref{eq:327}, the first $q$ lines taken together on the RHS corresponds the propagator $\hat{\tilde{U}}^{\dagger}(q\cdot\tau'_c)$, with $q$ taking the values from 1 to $p$. By taking the exponentials corresponding to the $m$-th divisions out of the products and expressing them with original small and big Hamiltonians, as given by,
		\begin{equation}
			\begin{alignedat}{2}
				e^{i\hat{\tilde{\mathcal{H}}}(k\delta t)\delta t} &= \hat{U}_{\tr{big}}(\tau'_c) \cdot e^{i\hat{\mathcal{H}}_{\tr{small}}(m\delta t)\delta t} \cdot \hat{U}_{\tr{big}}^{\dagger}(\tau'_c),
			\end{alignedat}
		\end{equation}
		Eq. \ref{eq:327} can be rewritten as,
		\begin{equation}
			\begin{alignedat}{2}
				\hat{\tilde{U}}^{\dagger}(\tau_c) &= \Bigg(\prod_{k=1}^{m-1}e^{i\hat{\tilde{\mathcal{H}}}(k\delta t)\delta t}\Bigg) \hat{U}_{\tr{big}}(\tau'_c)e^{i\hat{\mathcal{H}}_{\tr{small}}(m\delta t)\delta t}\hat{U}_{\tr{big}}^{\dagger}(\tau'_c)\\
				& \hspace*{20pt} \hat{U}_{\tr{big}}(\tau'_c) \Bigg(\prod_{k=1}^{m-1}e^{i\hat{\tilde{\mathcal{H}}}(k\delta t)\delta t}\Bigg) \hat{U}_{\tr{big}}(\tau'_c)e^{i\hat{\mathcal{H}}_{\tr{small}}(m\delta t)\delta t}\hat{U}_{\tr{big}}^{\dagger}(2\tau'_c)\\
				& \hspace*{100pt} \cdots\\
				& \hspace*{40pt}
				\hat{U}_{\tr{big}}\big((p-1)\tau'_c\big) \Bigg(\prod_{k=1}^{m-1}e^{i\hat{\tilde{\mathcal{H}}}(k\delta t)\delta t}\Bigg) \hat{U}_{\tr{big}}(\tau'_c)e^{i\hat{\mathcal{H}}_{\tr{small}}(m\delta t)\delta t} \hat{U}_{\tr{big}}^{\dagger}(p\tau'_c).
			\end{alignedat}
		\label{eq:328}
		\end{equation}
		Two things can be inferred from Eq. \ref{eq:328}. One is that,
		\begin{equation}
			\begin{alignedat}{2}
				\hat{\tilde{U}}^{\dagger}(\tau_c) = e^{i\hat{\overline{\tilde{\mathcal{H}}}}\tau_c} &= \Big(e^{i\hat{\overline{\tilde{\mathcal{H}}}}\tau'_c}\Big)^p\\
				&= \Bigg(\Bigg(\prod_{k=1}^{m-1}e^{i\hat{\tilde{\mathcal{H}}}(k\delta t)\delta t}\Bigg) \hat{U}_{\tr{big}}(\tau'_c)e^{i\hat{\mathcal{H}}_{\tr{small}}(m\delta t)\delta t}\Bigg)^p.
			\end{alignedat}
		\label{eq:329}
		\end{equation}
		The second inference is that, the first $q$ lines taken together on the RHS of Eq. \ref{eq:328}, which corresponds to $\hat{\tilde{U}}^{\dagger}(q\tau'_c)$, can be rewritten using Eq. \ref{eq:329} to yield
		\begin{equation}
			\hat{\tilde{U}}^{\dagger}(q\tau'_c) = \Big(e^{i\hat{\overline{\tilde{\mathcal{H}}}}\tau'_c}\Big)^q \cdot \hat{U}_{\tr{big}}(q\tau'_c).
		\label{eq:330}
		\end{equation}
		By making use of Eqs. \ref{eq:323} and \ref{eq:324} in Eq. \ref{eq:330}, it is seen that Eq. \ref{eq:modProp} is true.\\
		
		It is worth reiterating that the effective Hamiltonian, $\hat{\overline{\tilde{\mathcal{H}}}}$, of the interaction frame Hamiltonian, $\hat{\tilde{\mathcal{H}}}(t)$, is an average over the period $\tau_c$, while the effective Hamiltonian $\hat{\mathcal{H}}_{\tr{eff}}$ of $\hat{\mathcal{H}}_{\tr{big}}(t)$ is averaged over the interval $\tau'_c$, where $\hat{\mathcal{H}}_{\tr{big}}(t)$ is periodic.\\
		
		Since $\hat{\mathcal{H}}_{\tr{big}}(t)$ is usually the rf field and/or isotropic chemical shift Hamiltonians, its average $\hat{\mathcal{H}}_{\tr{eff}}$ can be expressed as a linear combination of single-spin operators and the effective rotation given by Eq. \ref{eq:323} can be written for a single-spin as,
		\begin{equation}
			\hat{U}_{\tr{big}}(\tau'_c) = e^{-i\hat{\mathcal{H}}_{\tr{eff}}\tau'_c} = e^{-i \omega_{\tr{cw}}\tau'_c \hat{\mathcal{F}}}
		\end{equation}
		where $\hat{\mathcal{F}}$ is a linear combination of single-spin operators of the spin under consideration. The effective field $\omega_{\tr{cw}}$ and the rotation axis $\hat{\mathcal{F}}$ can numerically be found using a variety of methods, one of which is using \textit{quaternions}, described in Sec. \ref{sect:quaternions}. It can be seen that the periodicity of $\hat{\tilde{\mathcal{H}}}(t)$ is now given by $\tau_c = p\cdot\tau'_c$ such that $\omega_{\tr{cw}}\cdot p\cdot\tau'_c = 2\pi$. This implies that $\tau_c$ is defined not only by the two frequencies $\omega_r$ and $\omega_m$, but also by $\omega_{\tr{cw}}$ according to,
		\begin{equation}
			\tau_c = \frac{2\pi}{\gcd(\omega_r,\omega_m,\omega_{\tr{cw}})}.
		\end{equation}
		Therefore the time-dependent interaction frame Hamiltonian in this case is written in the Fourier space using the three frequencies, given by,
		\begin{equation}
			\begin{aligned}
				\hat{\tilde{\mathcal{H}}}(t) = \sum_{n,k,l}^{}\hat{\tilde{\mathcal{H}}}^{(n,k,l)} e^{in\omega_rt} e^{ik\omega_mt} e^{il\omega_{\tr{cw}}t}
			\end{aligned}
		\end{equation}
		The effective Hamiltonian, $\hat{\overline{\tilde{\mathcal{H}}}}$, subsequently can be derived to first and second-order as,
		\begin{equation}
			\begin{alignedat}{2}
				\hat{\overline{\tilde{\mathcal{H}}}}^{(1)} &= \sum_{n_0,k_0,l_0} \hat{\tilde{\mathcal{H}}}^{(n_0,k_0,l_0)}
			\end{alignedat}
		\end{equation}
		and
		\begin{gather}
			\begin{alignedat}{2}
				\hat{\overline{\tilde{\mathcal{H}}}}^{(2)} &= \frac{-1}{2} \sum_{n,k}\frac{[\hat{\tilde{\mathcal{H}}}^{(n_0-n,k_0-k,l_0-l)},\hat{\tilde{\mathcal{H}}}^{(n,k,l)}]}{n\omega_r + k\omega_m + l\omega_{\tr{cw}}} + \sum_{n_0,k_0,l_0}\sum_{n,k,l}\frac{[\hat{\tilde{\mathcal{H}}}^{(n_0,k_0,l_0)},\hat{\tilde{\mathcal{H}}}^{(n,k,l)}]}{n\omega_r + k\omega_m + l\omega_{\tr{cw}}},
			\end{alignedat}
		\end{gather}
		where $n_0\omega_r + k_0\omega_m + l_0\omega_{\tr{cw}}= 0$ and $n\omega_r + k\omega_m + l\omega_{\tr{cw}} \neq 0$.
	
\section{Quaternions}
\label{sect:quaternions}
	Quaternions are an alternative to describe rotations\cite{hamilton1853lectures,hamilton1969,klein1965}. Unlike unitary matrices and directional cosines that describe rotation axis of a pulse as functions of pulse parameters with possible discontinuities, quaternions describe the axis as continuous functions. Indeed, this favours the use of quaternions over directional cosines for optimisation procedures involving rotation axis\cite{emsley1992optimization}. This will be elucidated below.\\
	
\begin{figure}[!h]
		\centering
		\includegraphics[width=\linewidth]{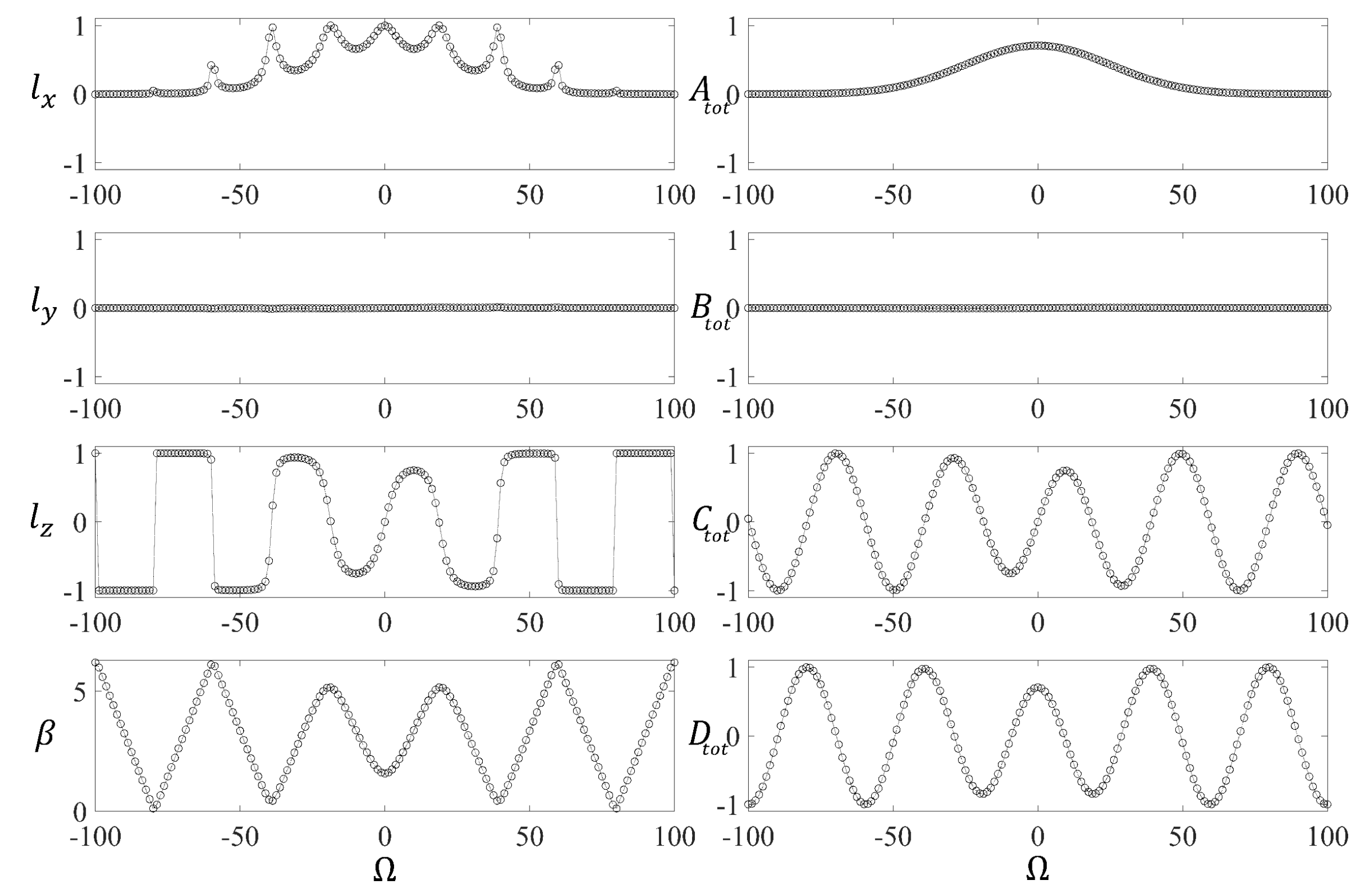}
		\caption{Overall propagator $\hat{U}(t,0)$ for a 50 ms $\pi/2$ Gaussian-shaped pulse with $\phi(t) = 0$ and $\omega_{\tr{rf}} =  a\cdot e^{-\frac{(t-b)^2}{2c^2}}$ where $a = \frac{250}{\sqrt{2\pi}}$, $b = 250$ms and $c = \frac{\pi}{2}\cdot\frac{1}{250}$, represented (left) as directional cosines and the flip angle and (right) as quaternion elements with varying offset $\Omega$ in Hz. Figure adapted from Ref. \cite{emsley1992optimization}.}
		\label{fig:3-1}
	\end{figure}

	Consider a periodic rf Hamiltonian, $\hat{\mathcal{H}}(t)$ with period $\tau_p$ given by
	\begin{equation}
		\hat{\mathcal{H}}(t) = \Omega \hat{I}_z + \omega_{\tr{rf}}(t) (\hat{I}_x \cos\phi(t) + \hat{I}_y \sin\phi(t))
	\end{equation}
	where offset from the rf carrier frequency is given by $\Omega = \omega_0 - \omega_c$, and $\omega_{\tr{rf}}(t)$ and $\phi(t)$ describe the time-dependent amplitude and phase of an arbitrary pulse. Dividing $\hat{\mathcal{H}}(t)$ into a finite number $N$ of intervals $\Delta t$, such that $\Delta t$ is short enough to assume $\hat{\mathcal{H}}$ to be constant in the interval and $N\Delta t = \tau_p$, the propagator defined in Eq. \ref{eq:3-3} is given by,
	\begin{equation}
		\hat{U}(\tau_p,0) = e^{-i \hat{\mathcal{H}}_N \Delta t} e^{-i \hat{\mathcal{H}}_{N-1} \Delta t} \cdots e^{-i \hat{\mathcal{H}}_1 \Delta t}
	\end{equation}
	where $\hat{\mathcal{H}}_j$ is the constant Hamiltonian for the $j$-th interval. For a spin-$\frac{1}{2}$, $I$, the propagator for the $j$-th pulse, $\hat{U}\big(j\Delta t, (j-1)\Delta t\big)$ can be expressed as a linear combination of Pauli matrices $\hat{I}_x$, $\hat{I}_y$, $\hat{I}_z$ and $\mathds{1}$ operators\cite{goldman1988quantum}. This can be expressed as
	\begin{equation}
		\hat{U}\big(j\Delta t, (j-1)\Delta t\big) = 
		\begin{bmatrix}
			U_{11}^{(j)} & U_{12}^{(j)}\\
			U_{21}^{(j)} & U_{22}^{(j)}\\
		\end{bmatrix},
	\end{equation}
	where
	\begin{equation}
		\begin{alignedat}{2}
			U_{11}^{(j)} & = \cos\frac{\beta_j}{2} - i\cos\theta_j\sin\frac{\beta_j}{2}\\
			U_{12}^{(j)} & = -i\sin\theta_j\sin\frac{\beta_j}{2}e^{-i\phi_j}\\
			U_{21}^{(j)} & = -i\sin\theta_j\sin\frac{\beta_j}{2}e^{i\phi_j}\\
			U_{22}^{(j)} & = \cos\frac{\beta_j}{2} + i\cos\theta_j\sin\frac{\beta_j}{2}\\
		\end{alignedat}
	\end{equation}
	with $\phi_j$ being the phase of the $j$-th pulse, while the polar angle $\theta_j$ and the flip angle $\beta_j$ are given by
	\begin{equation}
		\begin{alignedat}{2}
			\theta_j &= \tan^{-1}\big(\omega_{\tr{rf}}(j\Delta t)/\Omega\big)\\
			\beta_j &= \big(\Omega^2 + \omega^2_{\tr{rf}}(j\Delta t)\big)^{\frac{1}{2}} \Delta t.
		\end{alignedat}
	\end{equation}
	To visualise the position of the rotation axis of the $j$-th pulse, defined by phase and polar angle of the pulse, directional cosines are considered\cite{BLUMICH1985356}. They are simply the components of the axis of rotation along $x$, $y$ and $z$ directions and are given by
	\begin{equation}
		\begin{alignedat}{2}
			l_x^{(j)} &= \sin\theta_j\cos\phi_j\\
			l_y^{(j)} &= \sin\theta_j\sin\phi_j\\
			l_z^{(j)} &= \cos\theta_j.
		\end{alignedat}
	\end{equation}
	For a 50 ms Gaussian-shaped $\pi/2$ pulse, the directional cosines and the flip angle extracted from the overall propagator $\hat{U}(\tau_p,0)$ with varying offset is shown on the left side in Fig. \ref{fig:3-1}\footnote{Figure adapted from Ref. \cite{emsley1992optimization}. As the exact simulation parameters were not found in the said source, the author has been able to resimulate to only roughly match the original figure.}. Note the jump discontinuity in $l_z$, when the flip angle $\beta$ reaches 0 or $2\pi$ radians.
	
	The issue of discontinuity is avoided by describing rotations using quaternions, whose elements are related to directional cosines as,
	\begin{equation}
		\begin{alignedat}{2}
			A_{j} = l_x^{(n)} \sin\frac{\beta_n}{2},\\
			B_{j} = l_y^{(n)} \sin\frac{\beta_n}{2},\\
			C_{j} = l_z^{(n)} \sin\frac{\beta_n}{2},\\
			D_{j} = \cos\frac{\beta_n}{2},\\
		\end{alignedat}
	\end{equation}
	The overall quaternion is obtained by
	\begin{equation}
		\begin{alignedat}{2}
			\begin{bmatrix}
				A_{\tr{tot}}\\
				B_{\tr{tot}}\\
				C_{\tr{tot}}\\
				D_{\tr{tot}}
			\end{bmatrix}
			&=
			\begin{bmatrix}
				D_N & -C_N & B_N & A_N \\
				C_N & D_N & -A_N & B_N \\
				-B_N & A_N & D_N & C_N \\
				-A_N & -B_N & -C_N & D_N
			\end{bmatrix}
			\cdots
			\begin{bmatrix}
				D_2 & -C_2 & B_2 & A_2 \\
				C_2 & D_2 & -A_2 & B_2 \\
				-B_2 & A_2 & D_2 & C_2 \\
				-A_2 & -B_2 & -C_2 & D_2
			\end{bmatrix}
			\begin{bmatrix}
				A_1\\
				B_1\\
				C_1\\
				D_1
			\end{bmatrix}
		\end{alignedat}
	\end{equation}
	Elements of the overall quaternion calculated for the same 50 ms Gaussian shaped pulse is shown on the right side in Fig. \ref{fig:3-1}. Note that plots are free of any jump discontinuities associated with directional cosines.\\
	
	Euler angles is also a way to describe rotations. However it suffers from the phenomenon of gimbal-lock, where at particular configurations, two of the three degrees of freedom (corresponding to three Euler angles) become redundant. This constrains the motion, unless artificially moved out of those configurations. Quaternions also avoid the problem of gimbal-lock and are therefore the preferred description of rotations, as far as numerical optimisation, involving the axis of rotation, is concerned.\\
	
	Having established the expressions for the effective Hamiltonian in this chapter, the subsequent chapters will discuss the formalism developed to express the interaction frame Hamiltonian as product of complex exponentials with finite set of frequencies, a requirement for the validity of the expressions derived here. The next chapter will deal with pulse sequences that are only amplitude-modulated, while the subsequent chapter will deal with the formalism to handle pulse sequences that are both amplitude- and phase-modulated.

%% file: Chapters/chap4_AmpModSeq.tex
\chapter{Amplitude Modulated Pulse Sequences}
\label{chap:ampMod}
	In this chapter, a description that aids in expressing rf interaction frame time-dependent Hamiltonian, where the spin part is time-modulated only by an amplitude modulated rf pulse sequence, as a product of complex exponentials is introduced. This enables the application of formulas derived in chapter \ref{chap:DesignPrinciples} to obtain a time-independent effective Hamiltonian that describes the experiment. The idea is illustrated by first describing homonuclear radio-frequency driven recoupling (RFDR) \cite{rfdr_vega,rfdr_griffin_2008} experiment, where the spin part time modulation is rewritten with only one frequency (see Sec. \ref{sect:3_SingleFrequency}). The description helps with arguments in favour of introducing temporal displacement of $\pi$ pulses in the RFDR experiment for better transfer efficiencies.\\
	
	Later, the theoretical description for amplitude-modulated pulse sequences is applied to study transfer efficiency of $^{\tr{RESPIRATION}}$CP under chemical shift interaction, where the spin part is rewritten with two basic frequencies for every rf field channel (see Sec. \ref{sect:MultipleFrequencies}). The low polarisation transfer efficiency of $^{\tr{RESPIRATION}}$CP in spin systems with dominant chemical shift offset interactions is explained by the generation of second-order effective Hamiltonian term that is along the rf field axis, as this displaces the effective Hamiltonian axis from the plane transverse to the transfer axis, in the zero-quantum subspace. Effective Hamiltonian so calculated are compared with direct propagation numerical simulations. Validity of the description at large offset is questioned and explained with help of quaternion analysis. Aided by insights from the analysis, first a new variant of $^{\tr{RESPIRATION}}$CP is proposed. The variant, called Broadband-$^{\tr{RESPIRATION}}$CP is shown to perform better even under large isotropic chemical shift offsets. Secondly, insights offer arguments in favour of transforming into an interaction frame defined by both rf field and isotropic chemical shift offset, as against an interaction frame defined only by the rf field. Such a treatment would amount to handling phase-modulation in the pulse sequence. A full description to explain any amplitude- and phase-modulated pulse sequence is described in the next chapter.

	\section{Fundamental Frequencies}
	\label{sect:4_ff}
\begin{figure}[h]
			\centering
			\fbox{\includegraphics[width=\linewidth]{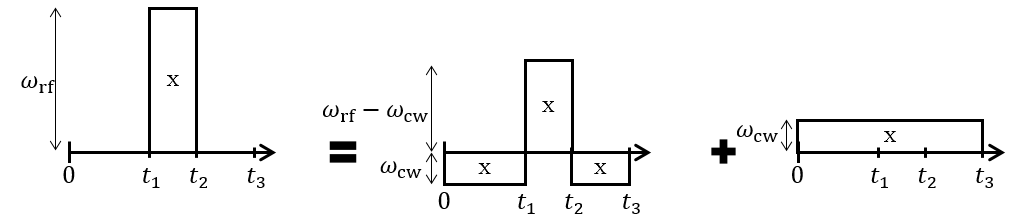}}
			\caption{Schematic representation of splitting a single pulse of amplitude $\omega_{\tr{rf}}$ as a sum of time-independent continuous wave component $\omega_{\tr{cw}}$ and a time-dependent amplitude modulated component with zero net rotation, $\omega_{\tr{rf}} - \omega_{\tr{cw}}$. Notice that in the first term of RHS, the area of $-X$ phase components with amplitude $\omega_{\tr{cw}}$ equals the area of $X$ phase component with amplitude $\omega_{\tr{rf}}-\omega_{\tr{cw}}$.}
			\label{fig:4-1}
		\end{figure}

		Consider a time-dependent rf field Hamiltonian, that is only modulated in amplitude, given by
		\begin{equation}
			\hat{\mathcal{H}}_{\tr{rf}}(t) = \omega_{\tr{rf}}(t) \hat{I}_x.
		\end{equation}
		This can be split into two components: a time-independent continuous wave component, $\omega_{\tr{cw}} = \overline{\omega_{\tr{rf}}(t)}$ and a time-dependent amplitude modulated component with zero net rotation, $\omega_{\tr{AM}}(t) = \omega_{\tr{rf}}(t) - \omega_{\tr{cw}}$. This is illustrated for a single pulse in Fig. \ref{fig:4-1} for better understanding and can be extended for multiple pulses, in a straightforward manner.
		\\
		
		The rf interaction frame transformation of single spin operators, say $\hat{I}_z$, can now be rewritten as
		\begin{equation}
			\begin{alignedat}{2}
				\hat{\tilde{I}}_z(t) &= \Big(\hat{\mathcal{T}}e^{-i\omega_{\tr{AM}}(t)\hat{I}_x t}\Big)^{\dagger} e^{i\omega_{\tr{cw}}\hat{I}_x t} \hat{I}_z e^{-i\omega_{\tr{cw}}\hat{I}_x t} \Big(\hat{\mathcal{T}}e^{-i\hat{I}_x\int_{0}^{t}\omega_{\tr{AM}}(t)t}\Big)\\
				&= \hat{I}_z \big(\cos\big(\omega_{\tr{cw}}t\big)\cos\big(\beta_m(t)\big)-\sin\big(\omega_{\tr{cw}}t\big)\sin\big(\beta_m(t)\big)\Big) +\\ 
				& \hspace*{30pt}\hat{I}_y \Big(\sin\big(\omega_{\tr{cw}}t\big)\cos\big(\beta_m(t)\big)+\cos\big(\omega_{\tr{cw}}t\big)\sin\big(\beta_m(t)\big)\Big)
			\end{alignedat}
			\label{eq:4_intFrame}
		\end{equation}
		where $\beta_m(t) = \int_{0}^{t}\omega_{\tr{AM}}(t')dt'$. This is valid as the two rf field components commute with each other at all times. As the flip angle imparted by the amplitude modulated component over the entire period defined by the pulse element is zero, the cosine and sine of it can be written as a Fourier series given by,
		\begin{equation}
			\begin{alignedat}{2}
				\cos\big(\beta_m(t)\big) &= \sum_{k = -\infty}^{\infty} a_{k}^z e^{ik\omega_m t}\\
				\sin\big(\beta_m(t)\big) &= \sum_{k = -\infty}^{\infty} a_{k}^y e^{ik\omega_m t}\\
			\end{alignedat}
			\label{eq:4_Fourier_Omegam}
		\end{equation}
		Eq. \ref{eq:4_Fourier_Omegam}, along with the complex exponential forms of $\cos(\omega_{\tr{cw}}t)$ and $\sin(\omega_{\tr{cw}}t)$ and the relations $\hat{I}^{+} = \hat{I}_z + i\hat{I}_y$ and $\hat{I}^{-} = \hat{I}_z - i\hat{I}_y$, can be used to rewrite Eq. \ref{eq:4_intFrame} as
		\begin{equation}
			\begin{alignedat}{2}
				\hat{\tilde{I}}_z(t) &= \sum_{k=-\infty}^{\infty}\sum_{\substack{l=-1\\l\neq0}}^{1}\hat{\mathcal{H}}^{(k,l)}e^{ik\omega_mt}e^{il\omega_{\tr{cw}t}}
			\end{alignedat}
			\label{eq:4_timeEq_FT}
		\end{equation}
		where the Fourier coefficients $\hat{\mathcal{H}}^{(k,\pm1)} = (a_{k,z} \pm ia_{k,y}) I^{\mp}$.\\
	
		The above procedure can be used to express all amplitude modulated rf field interaction frame time dependent Hamiltonian terms. It is now valid to use the expressions for effective Hamiltonian derived in Sec. \ref{sect:MultipleFrequencies}, for the case $\omega_{\tr{cw}} = 0$ and $\omega_{\tr{cw}} \neq 0$.
	
	\section{Adiabatic RFDR}
\begin{figure}[!h]
			\centering
			\includegraphics[width=\linewidth]{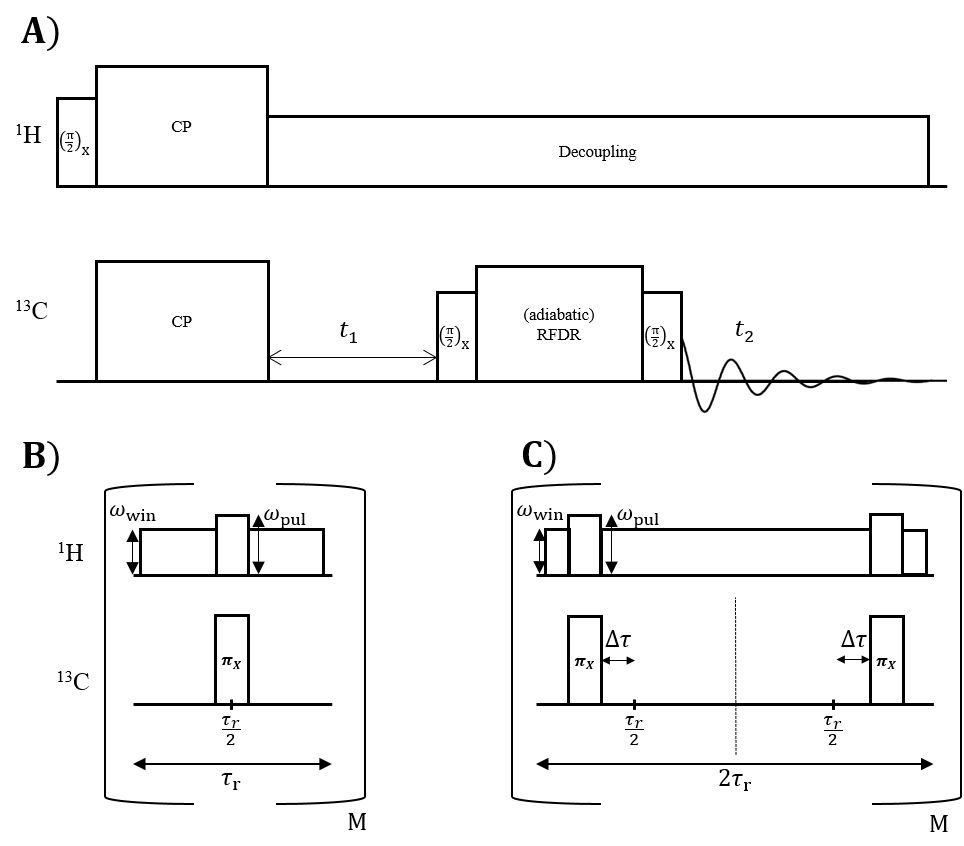}
			\caption{A $^{13}$C-$^{13}$C correlation experiment using either B) RFDR or C) adiabatic RFDR to mediate polarization transfer. The RFDR element consists of one $\pi$ pulse in the middle of every rotor period, which is repeated M times. In adiabatic RFDR, two rotor periods are considered where the temporal positions of the $\pi$ pulses are displaced in opposite directions with respect to the centre of the rotor period, by $\Delta \tau$. The element is repeated M times, with a new $\Delta \tau$ for every repetition. During the RFDR mixing element, $^1$H decoupling with constant x-phase is employed.}
			\label{fig:4-2}
		\end{figure}

\cite{straaso2016improved}
	\label{sect:4_adRFDR}
		In this section, a modification to the RFDR pulse sequence that improves the transfer efficiency, theoretically to 100\% is presented. The improvement is a result of adiabatically varying the temporal displacement of $\pi$ pulses in the pulse sequence from their original positions.	Consider two homonuclear coupled spin-$\frac{1}{2}$ nuclei, $I_1$ and $I_2$. In the rotating frame, the time dependent Hamiltonian is given by,
		\begin{equation}
			\hat{\mathcal{H}}(t) = \hat{\mathcal{H}}_I(t) + \hat{\mathcal{H}}_{II}(t) + \hat{\mathcal{H}}_{\tr{rf}}(t),
		\end{equation}
		with
		\begin{equation}
			\begin{aligned}
				\hat{\mathcal{H}}_I(t) &= \sum_{q=1}^{2} \sum_{n=-2}^{2}\omega_{I_q}^{(n)} e^{in\omega_rt}\hat{I}_{qz},
			\end{aligned}
		\label{eq:4_csH}
		\end{equation}
		\begin{equation}
			\begin{aligned}
				\hat{\mathcal{H}}_{II}(t) &= \sum_{n=-2}^{2} \omega_{I_1I_2}^{(n)} e^{in\omega_rt} (2\hat{I}_{1z}\hat{I}_{2z}-\hat{I}_{1x}\hat{I}_{2x}-\hat{I}_{1y}\hat{I}_{2y})
			\end{aligned}
		\label{eq:4_dipH}
		\end{equation}
		where the angular frequencies $\omega_{I_q}^{(n)}$ are the isotropic ($n = 0$) and anisotropic chemical shifts of the two nuclei, $\omega_{I_1I_2}^{(n)}$ the dipolar coupling and $\omega_r$ the spinning rate. The original RFDR and adiabatic RFDR pulse sequences are shown in a 2D correlation experiment depicted in Fig. \ref{fig:4-2}A. RFDR pulse sequence has a $\pi$ pulse in the middle of every rotor period, as shown in Fig. \ref{fig:4-2}B. To compensate for chemical shift offset and pulse imperfections by averaging certain higher-order terms in the Hamiltonian, XY-4 or XY-8 phase cycling scheme\cite{XY4_XY8_1,XY4_XY8_2} of the pulse train is employed. It will be shown in this section that by adiabatically varying the temporal placement of the $\pi$ pulses, polarisation transfer through the recoupled homonuclear dipolar coupling Hamiltonian can be improved. The modified pulse sequence, called the adiabatic RFDR pulse sequence is shown in Fig. \ref{fig:4-2}C. It is noted the sequence presented here is conceptually different from the variant found in literature, where the $\pi$ pulses are replaced with adiabatic inversion pulses, for improved stability of the experiment\cite{rfdr_adiabaticPi}. The basic unit, which is two rotor periods long in this case, has the two $\pi$ pulses shifted in time by $\Delta \tau$ in opposite directions from $\frac{\tau_r}{2}$ and $\frac{3\tau_r}{2}$. Clearly, $\Delta \tau = 0$ corresponds to the original RFDR pulse sequence. Convention here is that $\Delta \tau$ is negative when the first $\pi$ pulse is advanced in time with respect to $\frac{\tau_r}{2}$ (i.e., second $\pi$ is deferred from $\frac{3\tau_r}{2}$) and vice versa. A specific time-shift $\Delta \tau$ is repeated four times to accommodate XY-8 phase cycling scheme for the $\pi$ pulses. Ideally, in an adiabatic RFDR experiment, the time shift $\Delta \tau$ is changed gradually over repetitions.\\
		
\begin{figure}[h]
			\centering
			\fbox{\includegraphics[width=\linewidth]{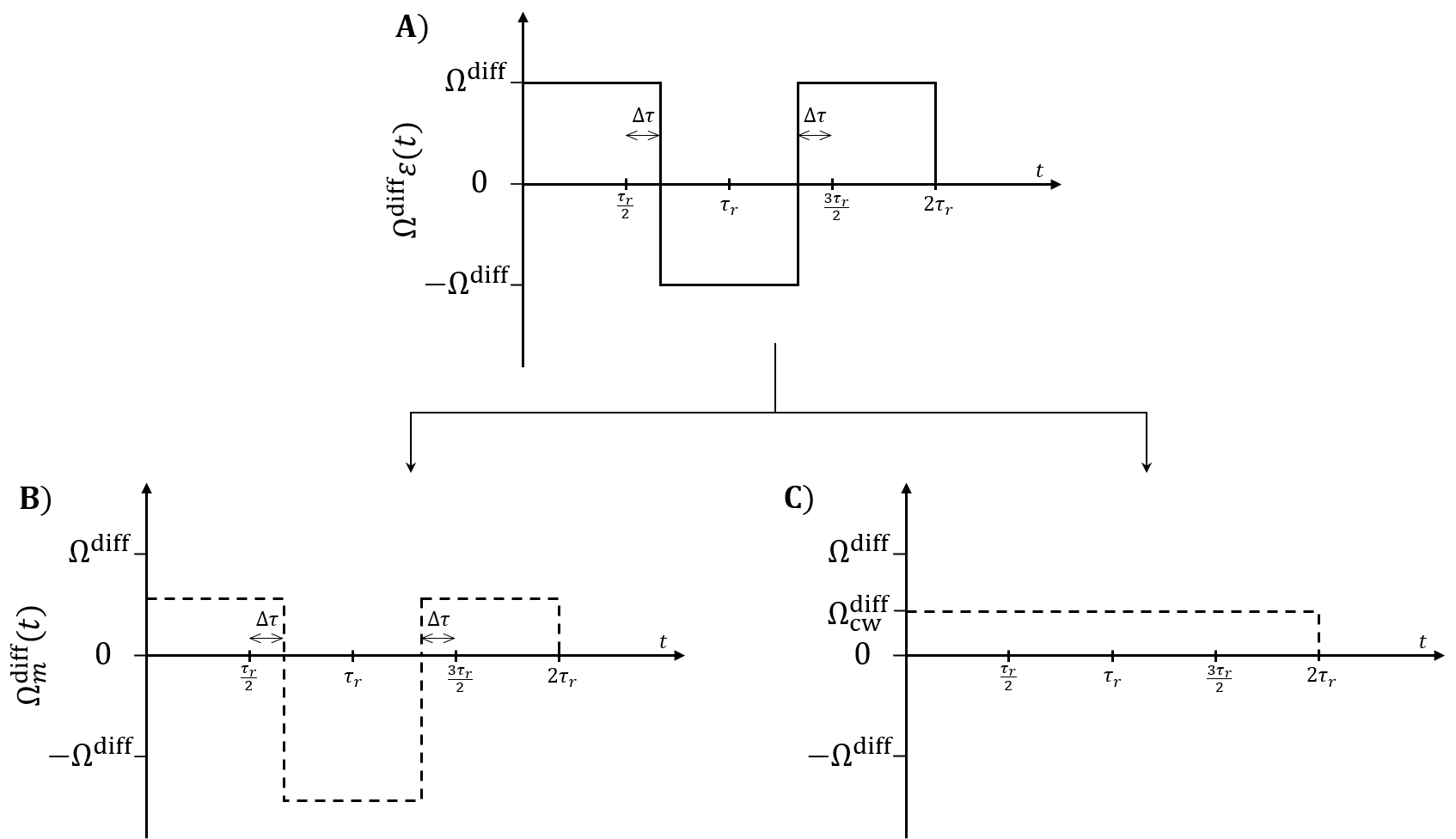}}
			\caption{A) Time-dependent chemical shift difference, $\Omega^{\tr{diff}}\varepsilon(t)$, between the two nuclei under adiabatic RFDR pulse sequence is split into two components: C) a time-independent part $\Omega_{\tr{cw}}^{\tr{diff}} = \overline{\Omega^{\tr{diff}}\varepsilon(t)}$ and B) a time-dependent part,$\Omega_m^{\tr{diff}}(t) = \Omega^{\tr{diff}}(t)-\Omega_{\tr{cw}}$, with zero net.}
			\label{fig:4-3}
		\end{figure}

		Assuming ideal $\pi$ pulses, the dipolar coupling Hamiltonian given in Eq. \ref{eq:4_dipH} remains unchanged in the rf field interaction frame. However the chemical shift Hamiltonian given in Eq. \ref{eq:4_csH} is modified and is given by,
		\begin{equation}
			\begin{aligned}
				\hat{\tilde{\mathcal{H}}}_I(t) &= \sum_{q=1}^{2} \sum_{n=-2}^{2}\omega_{I_q}^{(n)} e^{in\omega_rt} \varepsilon(t) \hat{I}_{qz},
			\end{aligned}
		\end{equation}
		where $\varepsilon(t)$ denotes the sign of spin part of the chemical shift Hamiltonian, at time $t$, and is the same for both spins. The isotropic chemical shift Hamiltonian can be rewritten as
		\begin{equation}
			\begin{alignedat}{2}
				\omega_{I_1}^{(0)} \varepsilon(t) \hat{I}_{1z} + \omega_{I_2}^{(0)} \varepsilon(t) \hat{I}_{2z} &= \Omega^{\tr{diff}} \varepsilon(t) \hat{I}_z^{ZQ} + \Omega^{\tr{sum}} \varepsilon(t) \hat{I}_z^{DQ}
			\end{alignedat}
		\label{eq:4_cs_zq_dq}
		\end{equation}
		where $\Omega^{\tr{diff}} = \omega_{I_1}^{(0)} - \omega_{I_2}^{(0)}$ and $\Omega^{\tr{sum}} = \omega_{I_1}^{(0)} + \omega_{I_2}^{(0)}$ are the difference and sum of the isotropic chemical shifts, respectively, while $\hat{I}_z^{ZQ} = \frac{1}{2}(\hat{I}_{1z}-\hat{I}_{2z})$ and $\hat{I}_z^{DQ} = \frac{1}{2}(\hat{I}_{1z}+\hat{I}_{2z})$ are the fictitious spin-$\frac{1}{2}$ zero-quantum and double-quantum operators\cite{fictitiousOps_Vega_1978}, respectively. As zero-quantum terms in the effective dipolar coupling Hamiltonian is what transfers in the RFDR experiment\cite{rfdr_Vega_1992,rfdr_griffin_1998}, it proves useful to transform the problem further into a frame defined by the zero quantum term in Eq. \ref{eq:4_cs_zq_dq}. To do so, a trick similar to the one discussed in Sec. \ref{sect:4_ff} is employed to split $\Omega^{\tr{diff}} \varepsilon(t)$, given in Fig. \ref{fig:4-3}A as a sum of time-independent component, $\Omega_{\tr{cw}}^{\tr{diff}} = \overline{\Omega^{\tr{diff}} \varepsilon(t)}$ shown in Fig. \ref{fig:4-3}C and a time-dependent component, $\Omega_m^{\tr{diff}}(t) = \Omega^{\tr{diff}} \varepsilon(t) - \Omega_{\tr{cw}}^{\tr{diff}}$ with zero mean, shown in Fig. \ref{fig:4-3}B.\\
		
		The total rf field interaction frame Hamiltonian of the system with all the components is now given by, 
		\begin{equation}
		 \begin{alignedat}{2}
			 \hat{\tilde{\mathcal{H}}}(t) &= \Omega^{\tr{diff}}_m(t) \hat{I}_z^{ZQ} + \Omega^{\tr{diff}}_{\tr{cw}} \hat{I}_z^{ZQ} + \Omega^{\tr{sum}}\varepsilon(t) \hat{I}_z^{DQ} + \sum_{q=1}^{2}\sum_{\substack{n=-2\\n\neq0}}^{2}\omega_{I_q}^{(n)} e^{in\omega_rt}\varepsilon(t) \hat{I}_{qz}\\
			 &\hspace{40pt} +\sum_{n=-2}^{2}\omega_{I_1I_2}^{(n)} e^{in\omega_rt} (2\hat{I}_{1z}\hat{I}_{2z}-\hat{I}_{1x}\hat{I}_{2x}-\hat{I}_{1y}\hat{I}_{2y}).
		 \end{alignedat}
		\label{eq:4_rfFrameH}
		\end{equation}
		By transforming the problem into a frame defined by the first term in Eq. \ref{eq:4_rfFrameH}, only the dipolar Hamiltonian term is changed, as all other terms commute with the first term. The transformation is given by,
		\begin{equation}
		 \begin{alignedat}{3}
			 \hat{\tilde{U}}(t) &= \hat{\mathcal{T}} e^{-i\hat{I}^{ZQ}_z\int_{0}^{t}\Omega^{\tr{diff}}_m(t')dt'} &= e^{-i\beta(t)\hat{I}^{ZQ}_z}.
		 \end{alignedat}
		\end{equation}
		Similar to Eq. \ref{eq:4_Fourier_Omegam}, cosine and sine of $\beta(t)$ can be written as $\cos(\beta(t)) = \sum_{k=-\infty}^{\infty}a_{k}^xe^{ik\omega_mt}$ and $\sin(\beta(t)) = \sum_{k=-\infty}^{\infty}a_{k}^ye^{ik\omega_mt}$, respectively. This leads to dipolar coupling Hamiltonian in the interaction frame of the first term of Eq. \ref{eq:4_rfFrameH} given by,
		\begin{equation}
		 \begin{alignedat}{2}
			 \hat{\tilde{\mathcal{H'}}}_{II}(t) = \sum_{n=-2}^{2}\sum_{k=-\infty}^{\infty}\hat{\tilde{\mathcal{H'}}}^{(n,k)} e^{in\omega_rt} e^{ik\omega_mt}
		 \end{alignedat}
		\end{equation}
		where the Fourier components work out to,
		\begin{equation}
		 \begin{aligned}
			 \hat{\tilde{\mathcal{H'}}}^{(n,k)} = \omega_{I_1I_2}^{(n)}(-\hat{I}_x^{ZQ}a_{k}^x + \hat{I}_y^{ZQ}a_{k}^y + 2\hat{I}_{1z}\hat{I}_{2z} \delta_{k,0}).
		 \end{aligned}
		\end{equation}
		The fictitious spin-$\frac{1}{2}$ zero-quantum operators are $\hat{I}_x^{ZQ} = \hat{I}_{1x}\hat{I}_{2x} + \hat{I}_{1y}\hat{I}_{2y}$ and $\hat{I}_y^{ZQ} = \hat{I}_{1y}\hat{I}_{2x} + \hat{I}_{1x}\hat{I}_{2y}$. The first-order effective dipolar coupling Hamiltonian is given by (see Sec. \ref{sect:MultipleFrequencies})
		\begin{equation}
			\hat{\overline{\tilde{\mathcal{H'}}}}^{(1)}_{II} = \sum_{n,k}\hat{\tilde{\mathcal{H'}}}^{(n,k)}
		\label{eq:4_1stOrdHomo}
		\end{equation}
		such that $n\omega_r + k\omega_m = 0$. As $2\omega_m = \omega_r$ for the problem here, $2n+k = 0$ and simplifies Eq. \ref{eq:4_1stOrdHomo} to
	 	\begin{equation}
		 	\hat{\overline{\tilde{\mathcal{H'}}}}^{(1)}_{II} = -\hat{I}_x^{ZQ}\omega_x^{\tr{eff}} + \hat{I}_y^{ZQ}\omega_y^{\tr{eff}}
		\label{eq:4_recDipHamil}
	 	\end{equation}
	 	where $\omega^{\tr{eff}}_x = \sum_{n,k}^{}\omega_{I_1I_2}^{(n)}a_{k}^x$ and $\omega^{\tr{eff}}_y = \sum_{n,k}^{}\omega_{I_1I_2}^{(n)}a_{k}^y$ with $n\omega_r + k\omega_m = 0$.\\

	 	The total first-order effective Hamiltonian is therefore given by,
	 	\begin{equation}
		 	\begin{alignedat}{2}
			 	\hat{\overline{\tilde{\mathcal{H'}}}}^{(1)} = \Omega^{\tr{diff}}_{\tr{cw}} \hat{I}_z^{ZQ} + \overline{\Omega^{\tr{sum}}\varepsilon(t)} \hat{I}_z^{DQ} + -\hat{I}_x^{ZQ}\omega_x^{\tr{eff}} + \hat{I}_y^{ZQ}\omega_y^{\tr{eff}} + \hat{\overline{\tilde{\mathcal{H'}}}}^{(1)}_{CSA}
		 	\end{alignedat}
		 \label{eq:4-16}
	 	\end{equation}
	 	where
	 	\begin{equation}
		 	\begin{alignedat}{2}
				\hat{\overline{\tilde{\mathcal{H'}}}}^{(1)}_{CSA} = \sum_{q=1}^{2}\sum_{\substack{n=-2\\n\neq0}}^{2} \omega_{I_q}^{(n)}\cdot 4 \cdot (-1)^{n+1} \frac{\sin(n\omega_r\Delta\tau)}{n\omega_r} \hat{I}_{qz}.
		 	\end{alignedat}
	 	\end{equation}
	 	It is seen from Eq. \ref{eq:4-16} that in the zero-quantum subspace, the effective dipolar coupling Hamiltonian, which is on the x,y-plane, adds with rest of the terms, including $\Omega^{\tr{diff}}_{\tr{cw}}\hat{I}_z^{ZQ}$, all along the longitudinal axis to give the total effective Hamiltonian. Therefore $\Omega^{\tr{diff}}_{\tr{cw}} = (\omega_{I_1}^{(0)}-\omega_{I_2}^{(0)})\frac{2\Delta\tau}{\tau_r}$, could in principle be varied slowly enough using $\Delta\tau$, such that the polarization is adiabatically changed from $\hat{I}_{1z}$ to $\hat{I}_{2z}$.\\
	 	
\begin{figure}[!h]
	 		\centering
	 		\includegraphics[width=\linewidth]{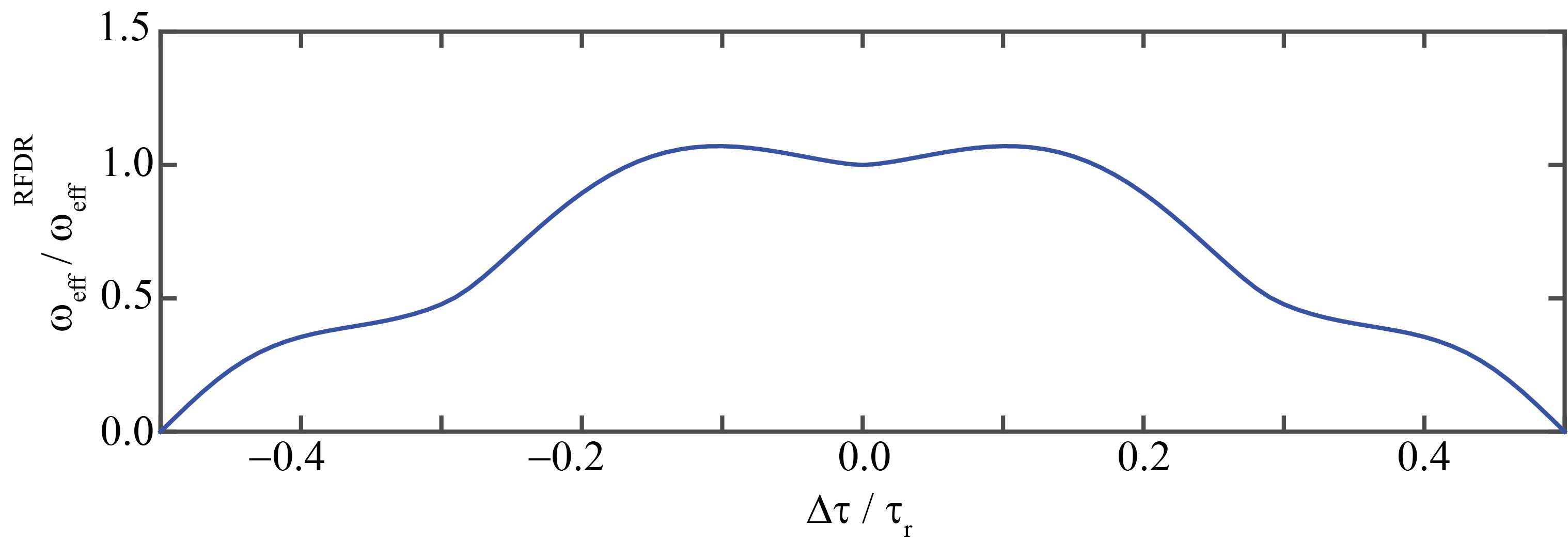}
	 		\caption{The powder averaged strength for the recoupled dipolar	Hamiltonian is shown with varying relative time-shift $\Delta\tau$ of the $\pi$-pulses. The strength is scaled to unity for $\Delta\tau = 0$. The calculation is done by setting $\Omega^{\tr{diff}}/2\pi = 1.2\omega_r/2\pi = 12$ kHz.}
	 		\label{fig:4-4}
	 	\end{figure}

	 	The powder-averaged strength of the effective dipolar coupling Hamiltonian given in Eq. \ref{eq:4_recDipHamil}, is given by
	 	\begin{equation}
		 	\begin{alignedat}{2}
			 	\omega_{\tr{eff}} = \frac{1}{8\pi^2}\int \sqrt{(\omega^{\tr{eff}}_x)^2 + (\omega^{\tr{eff}}_y)^2} d\alpha_{\tr{PR}} \sin\beta_{\tr{PR}} d\beta_{\tr{PR}} d\gamma_{\tr{PR}}.
		 	\end{alignedat}
	 	\end{equation}
	 	and is shown with varying $\Delta \tau$ in Fig. \ref{fig:4-4}. The calculation shown is done with $\Omega^{\tr{diff}}/2\pi = 1.2\omega_r/2\pi = 12$ kHz and MAS rate of 10 kHz. The extremes in the plot, where $\Delta\tau = \pm \tau_r$ corresponds to the two $\pi$ pulses being either at the same time or separated by two rotor periods, and leads to no recoupling. The centre of the plot, where $\Delta\tau = 0$ corresponds to the normal RFDR pulse sequence. Even though maximum transfer is achieved for a $\Delta$ that is slightly off zero, there is no significant gain in strength of the recoupled Hamiltonian by doing so compared to the normal RFDR pulse sequence. As the objective here is to adiabatically transfer, it is observed that strength of the recoupled dipolar interaction is not compromised much, for any of the $\Delta \tau$ used in the sweep. It is noted that the profile seen in Fig. \ref{fig:4-4} is highly dependent on the ratio of chemical shift different to the spinning rate and presence of CSA term further complicates the overall transfer prediction.
	 	
	 	To find the optimal sweep that adiabatically performs $I_{1z} \rightarrow I_{2z}$ transfer for a real system with a significant CSA, numerical simulations were performed. For a fixed $N$ XY-8 blocks of the pulse sequence, $\Delta\tau$ for each of the N blocks can be found using\cite{vinod2008swept}
		\begin{equation}
			\tr{x}_i = \tr{x}_{\tr{co}} \cdot \bigg(-1 + 2 \frac{i}{N-1}\bigg)\\
			\Delta\tau_i = \frac{\tau_{\tr{sweep}}}{2} \frac{\tan \tr{x}_i}{\tan \tr{x}_{\tr{co}}}
		\end{equation}
		where $\tau_{\tr{sweep}}$ is the sweep size, $\tr{x}_{\tr{co}}$ is the tangential cut-off angle and $i$ denotes the block number, taking either of the values $0, 1, \cdots N-1$. For $N\leq3$, it is noted that $\tr{x}_\tr{co}$ is inconsequential. If $N$ = 1, $\Delta\tau$ is chosen to be zero as in the normal RFDR pulse sequence. Simulations were done on open-source SIMPSON\cite{bak2000simpson,tosner2014computer} software and based on a representative $^{13}$CO to $^{13}$C$_{\alpha}$ spin pair in a polypeptide with the chemical shift parameters ($\delta_{\tr{iso}}^{\tr{CS}}, \delta_{\tr{aniso}}^{\tr{CS}}, \eta^{\tr{CS}}, \alpha_{\tr{PC}}^{\tr{CS}}, \beta_{\tr{PC}}^{\tr{CS}}, \gamma_{\tr{PC}}^{\tr{CS}}$) set to (170 ppm, -76 ppm, 0.90, 0$^{\circ}$, 0$^{\circ}$, 90$^{\circ}$) and (50 ppm, -20 ppm, 0.43, 90$^{\circ}$, 90$^{\circ}$, 0$^{\circ}$) for $^{13}$CO and $^{13}$C$_{\alpha}$ respectively\cite{bak2002specification}. The dipole interaction parameters ($b_{I_1I_2}, \beta_{\tr{PE}}^{\tr{D}}, \gamma_{\tr{PE}}^{\tr{D}}$) were (-2142 Hz, 90$^{\circ}$, 120.8$^{\circ}$), MAS rate was 10 kHz and the $^1$H Larmor frequency was set to 400 MHz. Powder averaging was achieved using the REPULSION\cite{bak1997repulsion} scheme with 66 $\alpha_{\tr{CR}}$, $\beta_{\tr{CR}}$ pairs and 9 $\gamma_{\tr{CR}}$ angles. The duration of the $\pi$ pulses was set to $5\mu$s and ideal $^1$H heteronuclear decoupling was assumed. The results for grid search optimisations of $\tau_{\tr{sweep}}$ and $\tr{x}_\tr{co}$ for a fixed $N$ are shown in Tab. \ref{tab:adrfdr_opt}.
		\begin{table}
			\centering
			\begin{tabular}{|l|cccccccccc|}
				\hline
				\hline
				N & 1 & 2 & 3 & 4 & 5 & 6 & 7 & 8 & 9 & 10\\
				\hline
				$\tau_{\tr{sweep}}$ & 0 & 2.9 & 2.5 & 3.2 & 3.3 & 3.4 & 3.7 & 3.5 & 3.7 & 3.6 \\
				$\tr{x}_\tr{co}$ & - & - & - & 89 & 80 & 80 & 79 & 79 & 81 & 81\\
				\hline
			\end{tabular}
		\caption{The optimal adiabatic parameters ($\tau_{\tr{sweep}}, \tr{x}_\tr{co}$), obtained using a grid search, for various numbers of XY-8 element blocks (N). Note that the sweep form, i.e., $\tr{x}_\tr{co}$, is immaterial for $N =$ 1, 2, and 3.}
		\label{tab:adrfdr_opt}
		\end{table}
		
\begin{figure}[!h]
	 		\centering
	 		\includegraphics[width=\linewidth]{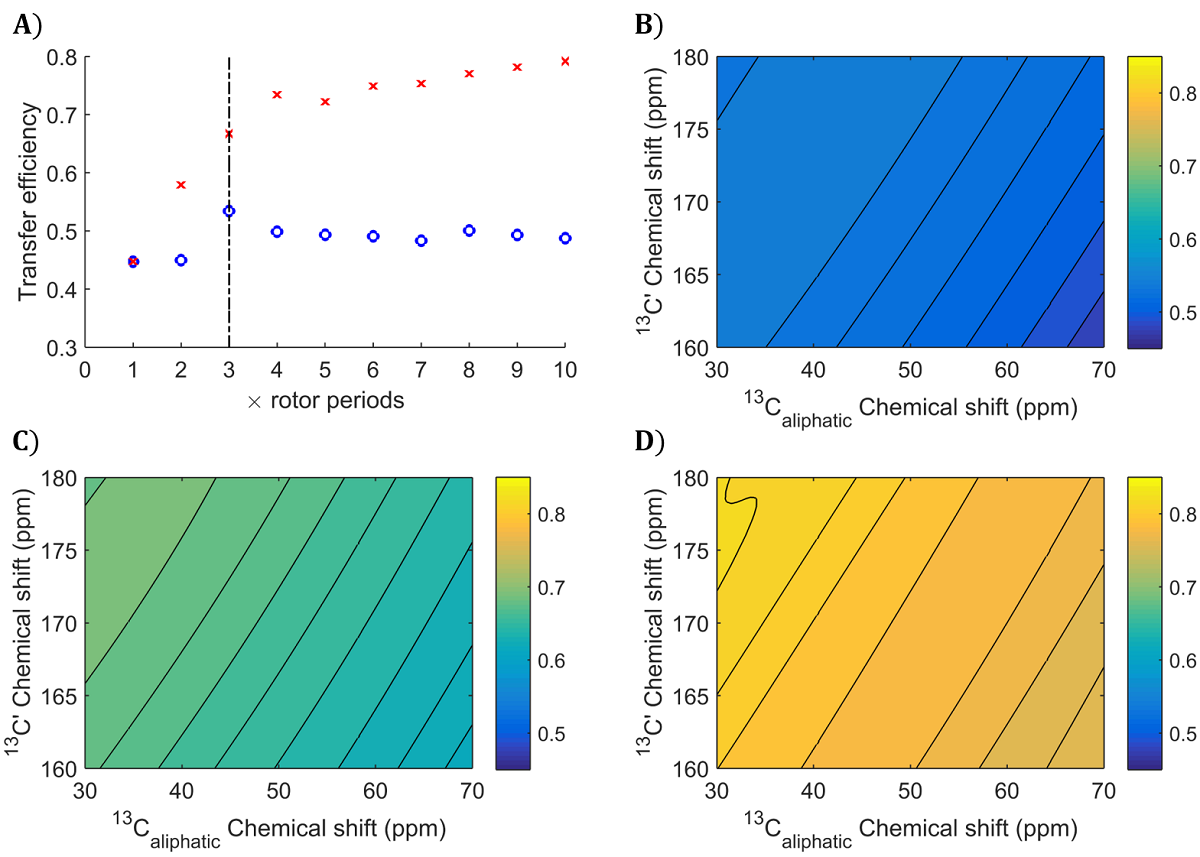}
	 		\caption{Simulations for the transfer efficiencies for RFDR (blue squares) and adiabatic RFDR (red squares) with varying mixing time. For a given number of blocks, $N$, the corresponding time-shifts $\Delta\tau_i$ of the most efficient adiabatic RFDR sequence has been selected. Dashed line at N = 3	indicates the time point of maximum transfer efficiency of standard RFDR. The transfer efficiency as function of chemical-shifts for both involved nuclei are numerically calculated for (b) RFDR (N = 3), (c) adiabatic RFDR (N = 3), and (d) adiabatic RFDR (N = 10). All simulations were done for a 400 MHz spectrometer at 10.0 kHz MAS. The rf field strength of the $\pi$-pulses was set to 100 kHz and a XY-8 phase cycling was employed.}
	 		\label{fig:4-5}
	 	\end{figure}

		In Fig. \ref{fig:4-5}A, simulated transfer efficiencies for RFDR and the different optimal adiabatic RFDR pulse sequences are compared. Normal RFDR reaches a maximum transfer of 53.3\% after 24 rotor period ($N = 3$), while adiabatic RFDR reaches 66.7\% for the same mixing time. Adiabatic RFDR achieves about 79\% transfer efficiency with a mixing time of 8ms ($N = 10$). The robustness towards chemical shift offset variations are seen in the plots given in Fig. \ref{fig:4-5}B-D, where RFDR (Fig. \ref{fig:4-5}B) with a mixing time of 24 rotor periods ($N = 3$), adiabatic RFDR (Fig. \ref{fig:4-5}C) with a mixing time of 24 rotor periods ($N = 3$) and adiabatic RFDR (Fig. \ref{fig:4-5}D) with a mixing time of 80 rotor periods ($N = 10$) are shown. It is evident that the polarisation transfer is higher over the entire chemical shift region for adiabatic RFDR pulse sequence compared to the normal RFDR pulse sequence.
	 	
\begin{figure}[!h]
	 		\centering
	 		\includegraphics[width=\linewidth]{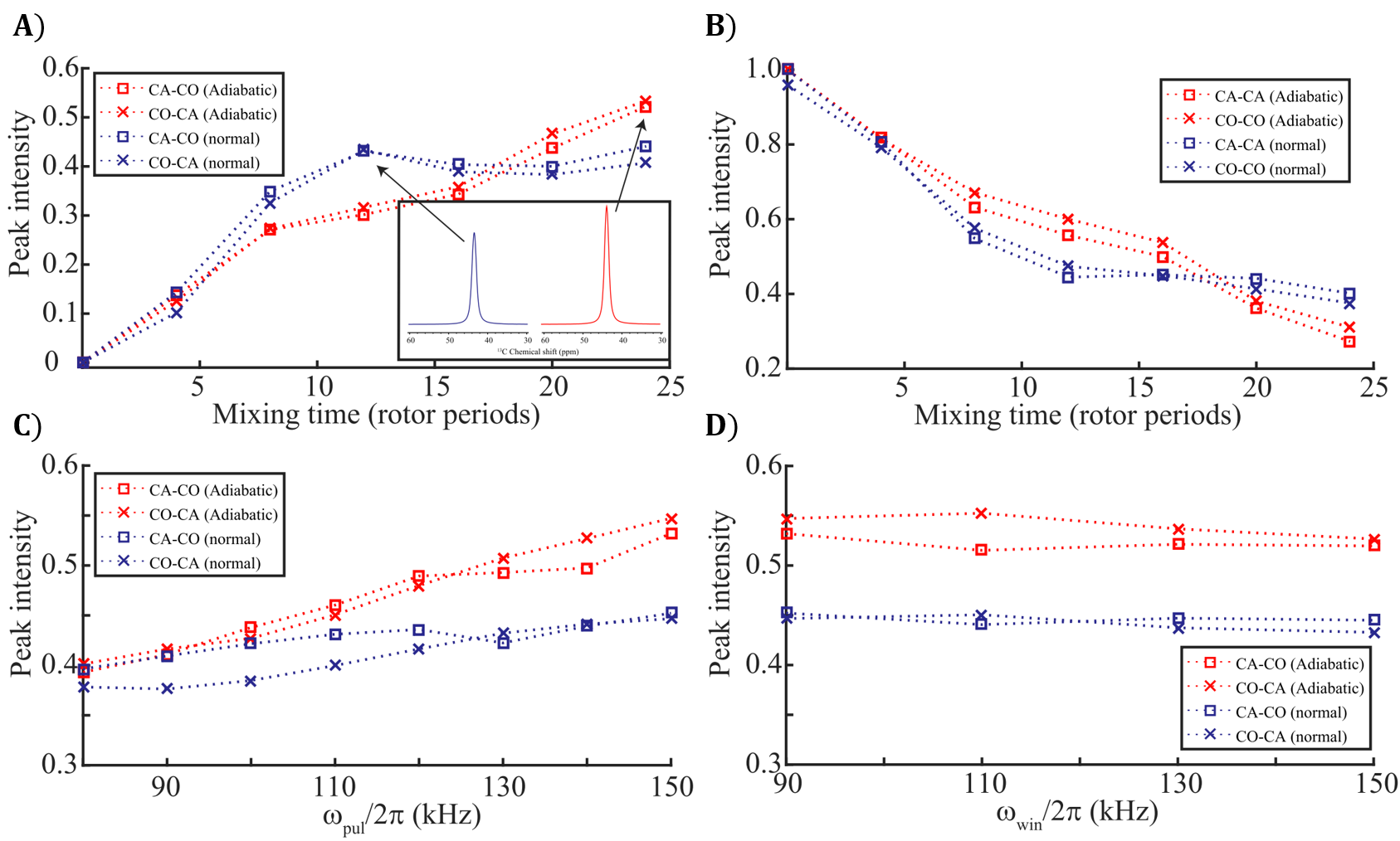}
	 		\caption{Extracted experimental peak intensities from 2D spectra on uniformly $^{13}$C,$^{15}$N-labelled glycine using RFDR (blue crosses and squares) and adiabatic RFDR (red crosses and squares). In (a) signal intensities of cross peaks and (b) diagonal peaks as function of mixing time are presented. $^1$H decoupling was employed during the mixing elements using $\omega_{\tr{pul}}/2\pi = 150$ kHz and $\omega_{\tr{win}}/2\pi = 90$ kHz. Note, for the adiabatic RFDR points, the entire sweep is only executed at a mixing time of 24 rotor periods (N = 3). The inset in right lower corner of (a) shows spectrum slices for highest $^{13}$CO to $^{13}$C$_{\alpha}$ cross peaks for RFDR (blue) and adiabatic RFDR (red). In (c) cross peaks intensities as function of 1H decoupling strength during the $\pi$-pulses with $\omega_{\tr{win}}/2\pi = 90$ kHz and (d) cross peaks intensities as function of 1H decoupling strength during the windows between the $\pi$-pulses with $\omega_{\tr{pul}}/2\pi = 150$ kHz is presented using a total mixing time of 12 rotor periods for RFDR and 24 rotor periods for the adiabatic version. All peak intensities have been scaled relative to the diagonal peak from a 2D spectrum without any mixing element and recorded on a 400 MHz spectrometer at 10.0 kHz MAS. The rf field strength of the $\pi$-pulses was set to 100 kHz, a XY-8 phase cycling was employed in all experimental approaches.}
	 		\label{fig:4-6}
	 	\end{figure}

	 	Fig. \ref{fig:4-6} presents a comparison of the experimental transfer efficiencies for normal RFDR and the adiabatic RFDR ($N = 3$) pulse sequences. The transfer efficiencies have been extracted from slices of 2D spectra recorded on uniformly $^{13}$C, $^{15}$N labelled glycine at 10 kHz MAS on a 400 MHz spectrometer. Fig. \ref{fig:4-6}A and B show the peak intensities of cross peaks and diagonal peaks with varying mixing time. The peak intensities have been integrated and normalised to the diagonal peak intensity from a spectrum without any mixing element. The adiabatic RFDR is recorded consistently with $\tau_{\tr{sweep}} = 2.5\mu$s. Hence, only the last measured data point (24 rotor periods with $N = 3$) exploits the entire sweep of the given pulse sequence and matches with the simulated data presented in Fig. \ref{fig:4-5}. The adiabatic RFDR is seen to reach a maximum transfer efficiency of about 55\% which is a gain of more than 20\% over the normal RFDR pulse sequence that reached a maximum transfer efficiency of 45\% at 12 rotor periods. The inset in Fig. \ref{fig:4-6}A shows spectrum slices extracted from the highest $^{13}$CO to $^{13}$C$_{\alpha}$ cross peaks for RFDR and adiabatic RFDR pulse sequences. The signal-to-noise ratio was determined from the reference spectrum to be about 6000. In Fig. \ref{fig:4-6}B, the diagonal peak intensities can be seen dropping continuously for the adiabatic RFDR, whereas for the normal RFDR sequence, it seems like the polarisation is equilibrating between the two carbon nuclei. The equilibration after 12 rotor periods can be attributed to the different effective dipolar coupling strengths for a powder sample where polarisation will be transferred either forward or backward for certain crystallites.
	 	
	 	The observed experimental transfer efficiency for $N = 3$ was found to be lower than the one found using numerical simulation in Fig. \ref{fig:4-5} (55\% compared to 66.7\%). This may be explained by several aspects, as follows. The numerical simulations were performed for an isolated two-spin system without relaxation which does not describe the full spin dynamics in a multi-spin system. In particular $^1$H decoupling, which was assumed to be ideal for the numerical simulations are not valid in the experiments. It has been discussed that better decoupling performance may be achieved using moderate CW irradiation during the windows between the $\pi$ pulses and strong CW irradiation during the $\pi$ pulses\cite{bayro2008radio}. Fig. \ref{fig:4-6}C presents experimental data for the peak intensities of cross peaks with varying $^1$H rf field strength $\omega_{\tr{pul}}/2\pi$, that is applied during the $\pi$ pulses on $^{13}$C. The performance is seen to improve for both version of RFDR, however the improvement is higher for adiabatic RFDR pulse sequence. Fig. \ref{fig:4-6}D presents the cross peak intensities with varying strength of CW irradiation $\omega_{\tr{win}}/2\pi$ that is applied between the $\pi$ pulses on $^{13}$C, while $\omega_{\tr{pul}} = 150$ kHz. It is observed that the transfer efficiencies are insensitive to changes in $^1$H decoupling CW irradiation between the $\pi$ pulses.\\
	 	
	 	In summary, the adiabatic variant of the RFDR experiment, that gradually varies temporal positions of $\pi$ pulses during the mixing time, significantly improves polarisation transfer. Theoretically, the chemical shift difference, which is dependent on the temporal placements of the $\pi$ pulses, is understood to adiabatically vary the total effective Hamiltonian such that the polarisation is transferred from one nucleus to the other.
	 	
	\section{$^{\textrm{RESPIRATION}}$CP}\cite{nielsen2016theoretical}
	\label{sect:resp}
		A highly approved and endorsed heteronuclear dipolar recoupling pulse sequence is the Hartmann-Hahn cross-polarisation (CP)\cite{hartmann1962,stejskal1977,Stejskal1979}. It comprises of simultaneous CW rf irradiations on two heteronuclear spins $I$ and $S$ under a MAS rate $\omega_r$, with the two rf amplitudes $\omega_{\tr{rf}}^{\tr{(q)}}$ satisfying $\omega_{\tr{rf}}^{\tr{(I)}}+\omega_{\tr{rf}}^{\tr{(S)}} = n\omega_r$, with $\tr{q} = \tr{I or S}$. Several modifications of the CP experiment have been proposed over the years, to ensure good performance even under inhomogeneous rf field or chemical shift offset\cite{khaneja2006composite,hediger1993cross,kolbert1995broadband,bjerring2003solid}. One such modification is the Rotor Echo Short Pulse IRaAdiaTION mediated cross polarization ($^{\tr{RESPIRATION}}$CP)\cite{jain2012efficient,nielsen2013adiabatic,jain2014low,basse2014efficient}. In this section, a theoretical description for any amplitude modulated pulse sequence is detailed by illustrating $^{\tr{RESPIRATION}}$CP. It turns out that the rf field interaction frame Hamiltonian in such cases is described by two basic frequencies for every rf pulse channel, as discussed in Sec. \ref{sect:MultipleFrequencies}.\\
	
		Consider two heteronuclear coupled spin-$\frac{1}{2}$ nuclei, $I$ and $S$. In the rotating frame, the time-dependent Hamiltonian is given by,
		\begin{equation}
			\hat{\mathcal{H}}(t) = \hat{\mathcal{H}}_I(t) + \hat{\mathcal{H}}_S(t) + \hat{\mathcal{H}}_{IS}(t) + \hat{\mathcal{H}}_{\tr{rf}}
		\end{equation}
		with
		\begin{equation}
			\begin{alignedat}{2}
				\hat{\mathcal{H}}_I(t) &= \sum_{n=-2}^{2} \omega_{I}^{(n)} e^{in\omega_rt} \hat{I}_{z}\\
				\hat{\mathcal{H}}_S(t) &= \sum_{n=-2}^{2} \omega_{S}^{(n)} e^{in\omega_rt} \hat{S}_{z}\\
				\hat{\mathcal{H}}_{IS}(t) &= \sum_{n=-2}^{2} \omega_{IS}^{(n)} e^{in\omega_rt} 2\hat{I}_{z}\hat{S}_z\\
				\hat{\mathcal{H}}_{\tr{rf}}(t) &= \omega_{\tr{rf}}^{\tr{(I)}}(t)\hat{I}_x + \omega_{\tr{rf}}^{\tr{(S)}}(t)\hat{S}_x
			\end{alignedat}
		\label{eq:resp_H}
		\end{equation}
		The $^{\tr{RESPIRATION}}$CP pulse sequence is shown on the left side in Fig. \ref{fig:4-6}, with $\tau_m^{(\tr{S})} = \tau_m^{\tr{(I)}} = \tau_r$. The pulse sequence is repeated M times to accomplish transfer. In order to write the time dependency of $\hat{\mathcal{H}}_{\tr{rf}}(t)$ also as complex exponential, like rest of the Hamiltonians in Eq. \ref{eq:resp_H}, splitting of the rf field explained in Sec. \ref{sect:4_ff} is employed. The rf field on each channel is split into two components, a time dependent amplitude modulated component $\omega_{\tr{AM}}^{\tr{(q)}}(t)$ with zero net rotation on single spin operators and a time-independent continuous wave component $\omega_{\tr{cw}}^{\tr{(q)}}$. The splitting is shown on the right side in Fig. \ref{fig:4-6}.
		\begin{figure}[!h]
			\centering
			\fbox{\includegraphics[width=\linewidth]{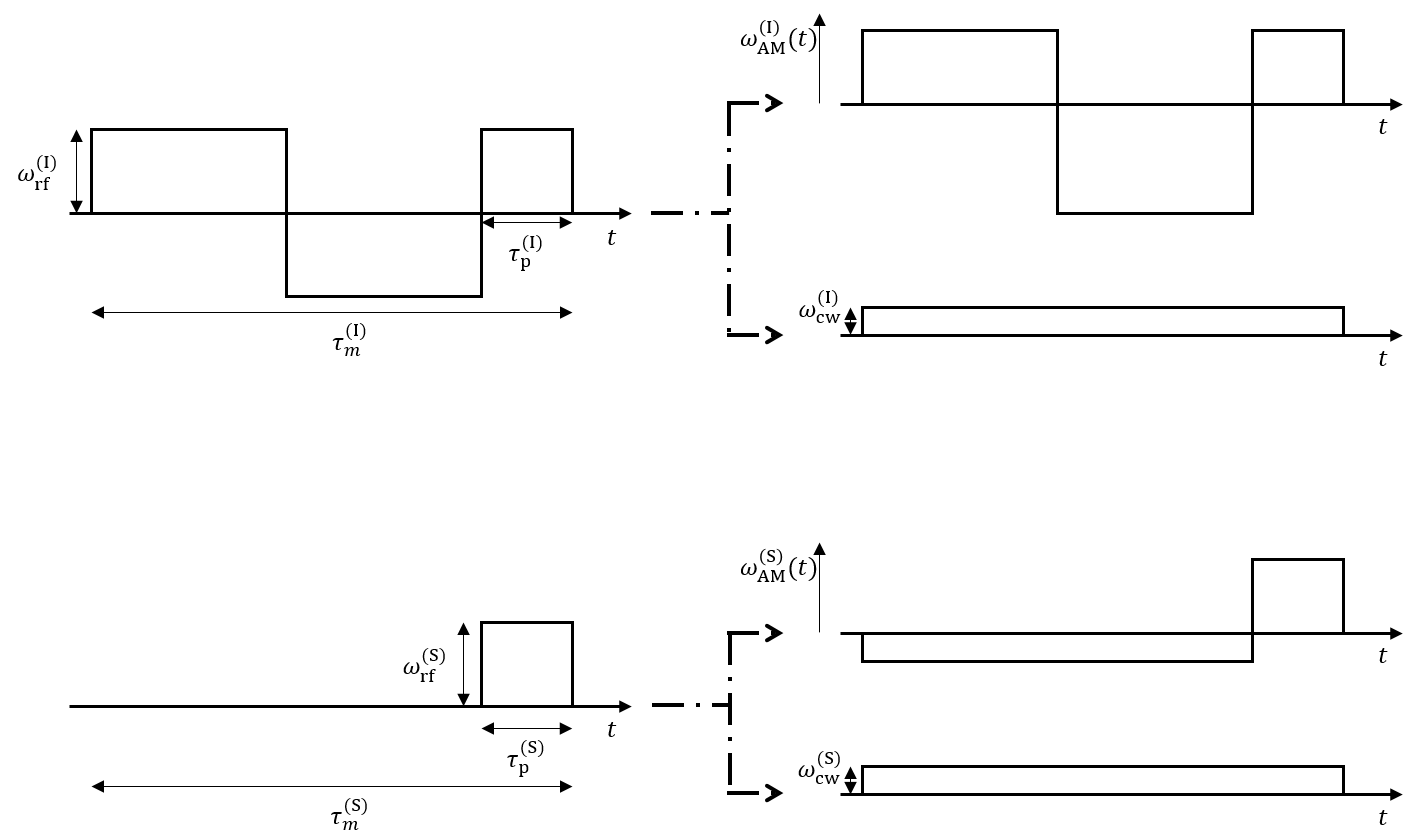}}
			\caption{Fig7}
			\label{fig:4-7}
		\end{figure}
		The rf field Hamiltonian given in Eq. \ref{eq:resp_H} can therefore be rewritten as
		\begin{equation}
			\begin{alignedat}{2}
				\hat{\mathcal{H}}_{\tr{rf}}(t) &= \big(\omega_{\tr{AM}}^{\tr{(I)}}(t) + \omega_{\tr{cw}}^{\tr{(I)}}\big) \hat{I}_x + \big(\omega_{\tr{AM}}^{\tr{(S)}}(t) + \omega_{\tr{cw}}^{\tr{(S)}}\big) \hat{S}_x
			\end{alignedat}
		\end{equation}
		where $\omega_{\tr{cw}}^{\tr{(q)}} = \overline{\omega_{\tr{rf}}^{\tr{(q)}}(t)}$ and $\omega_{\tr{AM}}^{\tr{(q)}}(t) = \omega_{\tr{rf}}^{\tr{(q)}}(t) - \omega_{\tr{cw}}^{\tr{(q)}}$. As explained in Sec. \ref{sect:4_ff}, the rf field interaction frame Hamiltonian can therefore be written as
		\begin{equation}
			\begin{alignedat}{2}
				\hat{\tilde{\mathcal{H}}}(t) = \sum_{n=-2}^{2}\sum_{k_{\tr{I}}=-\infty}^{\infty}\sum_{l_{\tr{I}}=-1}^{1}\sum_{k_{\tr{S}}=-\infty}^{\infty}\sum_{l_{\tr{S}}=-1}^{1} \hat{\tilde{\mathcal{H}}}^{(n,k_{\tr{I}},l_{\tr{I}},k_{\tr{S}},l_{\tr{S}})} e^{in\omega_rt} e^{ik_{\tr{I}}\omega_m^{\tr{(I)}}t} e^{il_{\tr{I}}\omega_{\tr{cw}}^{\tr{(I)}}t} e^{ik_{\tr{S}}\omega_m^{\tr{(S)}}t} e^{il_{\tr{S}}\omega_{\tr{cw}}^{\tr{(S)}}t}
			\end{alignedat}
		\label{eq:4_respTFH}
		\end{equation}
		with the Fourier components given by
		\begin{equation}
			\begin{alignedat}{2}
				\hat{\tilde{\mathcal{H}}}^{(n,k_{\tr{I}},\pm1,0,0)} &= \omega_{I}^{(n)} (a_{k_{\tr{I}}}^z \pm il a_{k_{\tr{I}}}^y) \hat{I}^{\mp}\\
				\hat{\tilde{\mathcal{H}}}^{(n,0,0,k_{\tr{S}},\pm1)} &= \omega_{S}^{(n)} (a_{k_{\tr{S}}}^z \pm il a_{k_{\tr{S}}}^y) \hat{S}^{\mp}\\
				\hat{\tilde{\mathcal{H}}}^{(n,k_{\tr{I}},\pm1,k_{\tr{S}},\pm1)} &= \omega_{IS}^{(n)} (a_{k_{\tr{I}}}^z \pm il a_{k_{\tr{I}}}^y) (a_{k_{\tr{S}}}^z \pm il a_{k_{\tr{S}}}^y) \hat{I}^{\mp}\hat{S}^{\mp}\\
				\hat{\tilde{\mathcal{H}}}^{(n,k_{\tr{I}},\pm1,k_{\tr{S}},\mp1)} &= \omega_{IS}^{(n)} (a_{k_{\tr{I}}}^z \pm il a_{k_{\tr{I}}}^y) (a_{k_{\tr{S}}}^z \mp il a_{k_{\tr{S}}}^y) \hat{I}^{\mp}\hat{S}^{\pm}\\
			\end{alignedat}
		\label{eq:4_23}
		\end{equation}
		
		In this section, only the dipolar coupling Hamiltonian in the effective Hamiltonian is discussed. The influence of chemical shift Hamiltonian in is discussed later in Chapter \ref{chap:genSeq}.\\
		
\begin{figure}[!h]
			\centering
			\includegraphics[width=\linewidth]{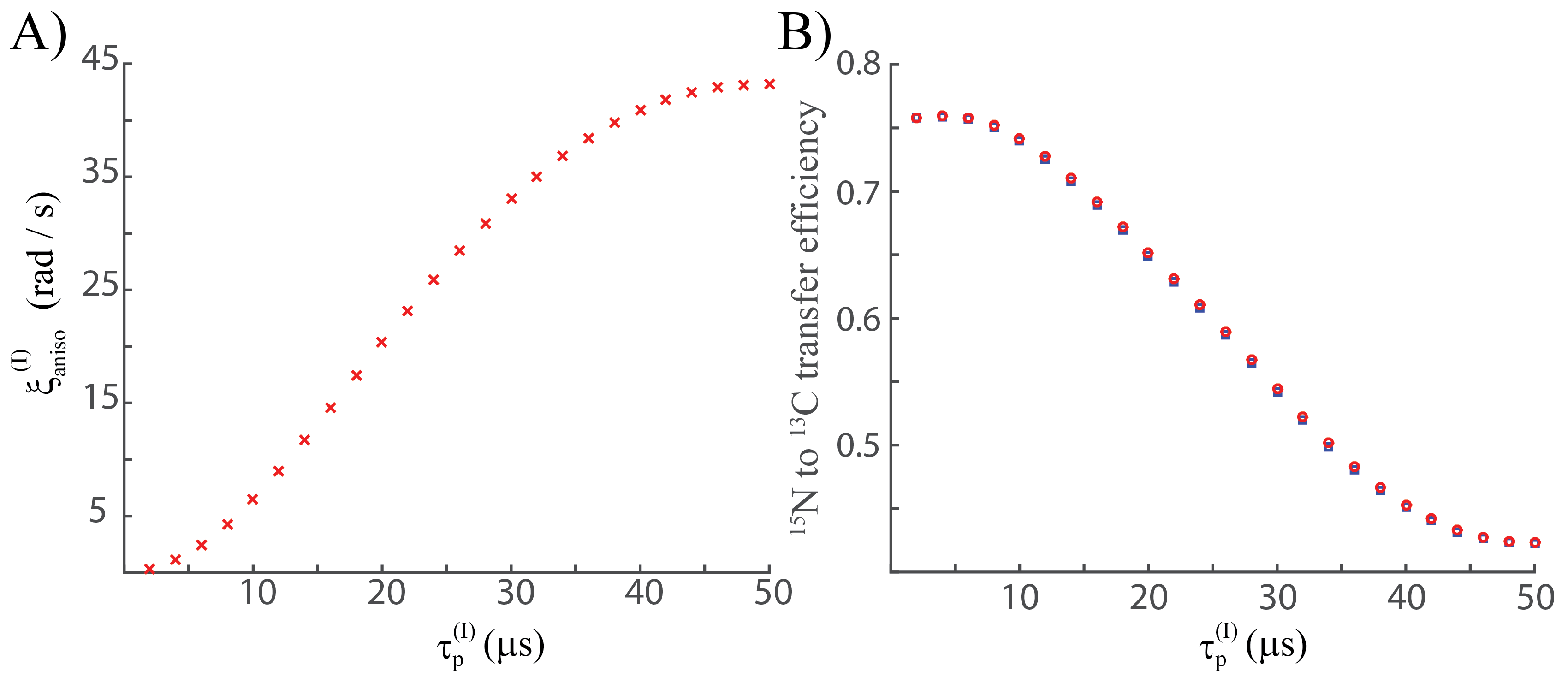}
			\caption{A) The strength $\xi_{\tr{aniso}}^{(\tr{S})}$ of effective second-order anisotropic chemical shift Hamiltonian given in Eq. \ref{eq:5_secord_aniso}, averaged over powder, expressed with varying short-pulse length $\tau_p^{\tr{(S)}}$ of $^{\tr{RESPIRATION}}$CP. B) Transfer efficiencies for $\hat{I}_x \rightarrow \hat{S}_x$ transfer against varying short pulse length, $\tau_p^{(\tr{S})}$. Red circles show the transfer performed with the calculated total effective Hamiltonian, while the blue boxes show the transfer obtained through direct-propagation numerical simulations.}
			\label{fig:5-1}
		\end{figure}

		The first-order effective Hamiltonian for the time-dependent Hamiltonian given in Eq. \ref{eq:4_respTFH}, in line with discussions in Chapter \ref{chap:DesignPrinciples}, is given by
		\begin{equation}
			\hat{\overline{\tilde{\mathcal{H}}}}^{(1)} = \sum_{n,k_{\tr{I}},l_{\tr{I}},k_{\tr{S}},l_{\tr{S}}}^{} \hat{\tilde{\mathcal{H}}}^{(n,k_{\tr{I}},l_{\tr{I}},k_{\tr{S}},l_{\tr{S}})}
		\label{eq:4_resp1stord}
		\end{equation}
		where the sum is over the quintuples $(n,k_{\tr{I}},l_{\tr{I}},k_{\tr{S}},l_{\tr{S}})$ that satisfy the resonance condition
		\begin{equation}
			n\omega_r + k_{\tr{I}} \omega_m^{\tr{(I)}} + l_{\tr{I}} \omega_{\tr{cw}}^{\tr{(I)}} + k_{\tr{S}} \omega_m^{\tr{(S)}} + l_{\tr{S}} \omega_{\tr{cw}}^{\tr{(S)}} = 0.
		\end{equation}
		For $^{\tr{RESPIRATION}}$CP, as the pulse sequence is rotor synchronised, the relation $\omega_r = \omega_m^{\tr{(I)}} = \omega_m^{\tr{(S)}}$ holds true and as the short pulses on both channels are identical, the relation $\omega_{\tr{cw}}^{\tr{(I)}} = \omega_{\tr{cw}}^{\tr{(S)}}$ holds true. Eq. \ref{eq:4_resp1stord} is therefore greatly simplified to
		\begin{equation}
			\hat{\overline{\tilde{\mathcal{H}}}}^{(1)} = \sum_{n=-2}^{2}\sum_{k_{\tr{I}}=-\infty}^{\infty} \hat{\tilde{\mathcal{H}}}^{(n,k_{\tr{I}},\pm1,-(k_{\tr{I}}+n),\mp1)}.
			\label{eq:4_resp1stordSimpl}
		\end{equation}
		The observation that $l_{\tr{I}} = -l_{\tr{S}}$ is true in Eq. \ref{eq:4_resp1stordSimpl} suggests that the recoupled dipolar Hamiltonian terms are zero-quantum. Therefore the form of effective first-order dipolar coupling Hamiltonian can be written as a linear combination of fictitious spin-$\frac{1}{2}$ operators $\hat{F}_{z}^{ZQ} (= 2\hat{I}_z\hat{S}_z + 2\hat{I}_y\hat{S}_y)$ and $\hat{F}_y^{ZQ} (= 2\hat{I}_z\hat{S}_y - 2\hat{I}_y\hat{S}_z)$. This Hamiltonian enables the transfer,
		\begin{equation}
		\hat{I}_x = \hat{F}^{ZQ}_x + \hat{F}^{DQ}_x \xrightarrow{(c\hat{F}^{ZQ}_z + d\hat{F}^{ZQ}_y)t} -\hat{F}^{ZQ}_x + \hat{F}^{DQ}_x = \hat{S}_x,
		\end{equation}
		where $c$ and $d$ define the Hamiltonian vector in the zero-quantum operator subspace and $\sqrt{c^2+d^2} = \frac{\pi}{t}$.	Illustration of this is given in Fig. \ref{fig:5-1}A, where the red arrow describes the effective first-order dipolar coupling Hamiltonian.\\
		
\begin{figure}[!h]
			\centering
			\includegraphics[width=0.7\linewidth]{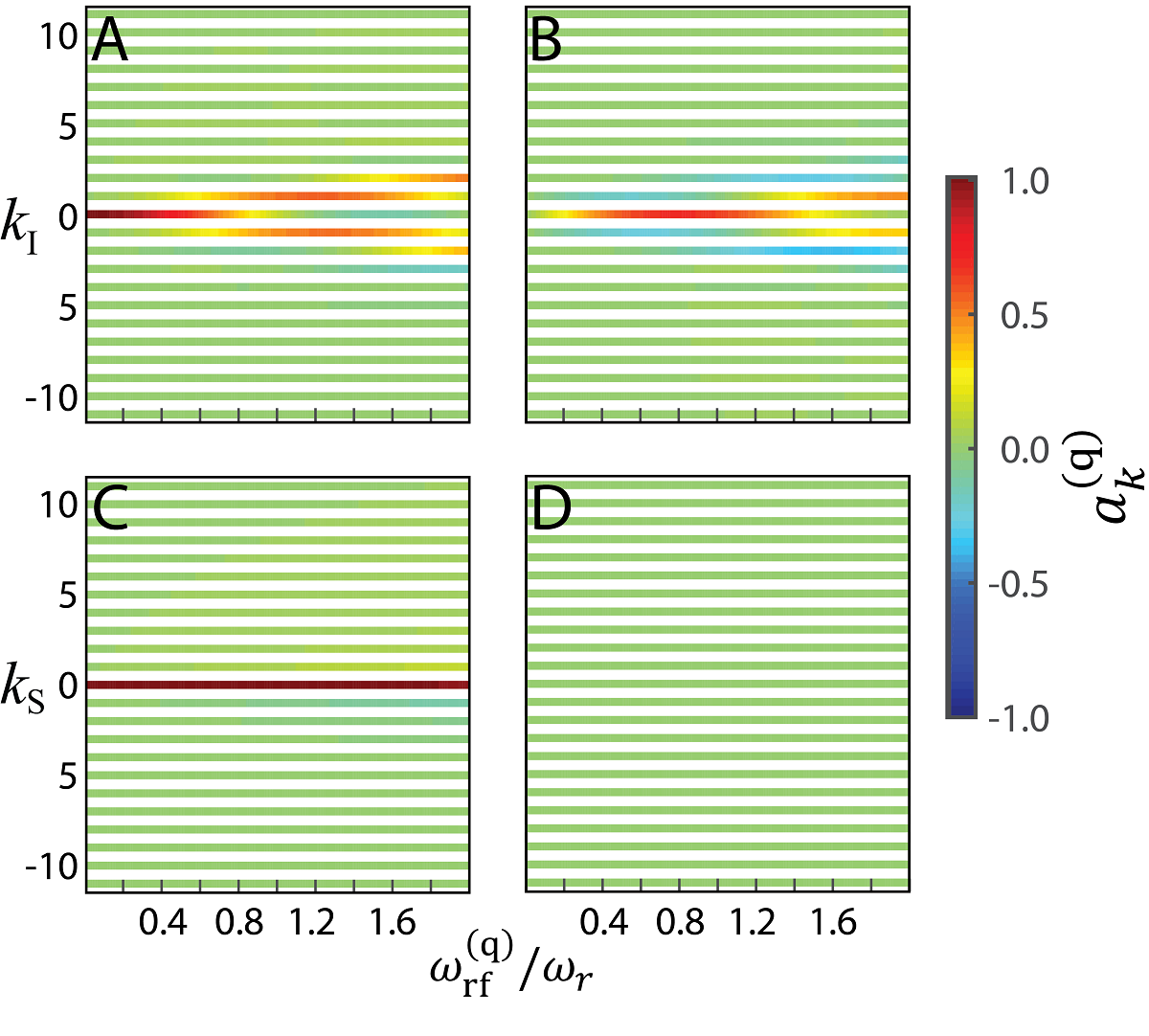}
			\caption{Real (A and C) and Imaginary parts (B and D) of the magnitude of Fourier coefficients, $a_k^{(\tr{q})} = a_{k_{\tr{q}}}^z + ia_{k_{\tr{q}}}^y$ with varying $k_{\tr{q}}$ and rf field amplitude, $\omega_{\tr{rf}}^{\tr{(q)}}$ scaled to $\omega_r$, that was kept constant for all pulses in the sequence. The duration of the short pulse was set to $\tau_p^{\tr{(q)}} = \frac{4}{60}\tau_r$. Plots A and B correspond to $I$-spin (q = $I$), while plots C and D correspond to $S$-spin (q = $S$).}
		\label{fig:4-8}
		\end{figure}

		To determine the size of the effective dipolar coupling Hamiltonian, the Fourier components according to the last of Eq. \ref{eq:4_23} have to be calculated. In Fig. \ref{fig:4-8}, the Fourier coefficients, $a_k^{(\tr{q})} = a_{k_{\tr{q}}}^z + ia_{k_{\tr{q}}}^y$, are shown with varying $|\omega_{\tr{rf}}^{\tr{(q)}}|$ which is set as the amplitude for both the amplitude modulated (for the interval 0 to $\tau_m^{\tr{(q)}} - \tau_p^{\tr{(q)}}$) and the short pulse (of duration $\tau_p^{\tr{(q)}}$). The duration $\tau_p^{\tr{(q)}}$ was set to $\tau_m^{\tr{(q)}}/15$. It is evident that for experimentally realistic values, the coefficients are non-zero only in the interval given by $\pm10$, i.e., $k_{\tr{q}} \in [-10,10]$.
		
\begin{figure}[!h]
			\centering
			\includegraphics[width=\linewidth]{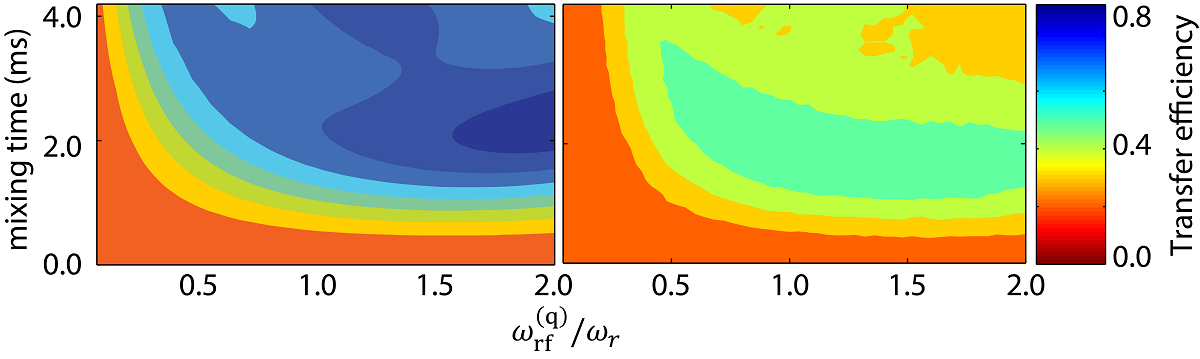}
			\caption{Analytical and experimental $^{15}$N $\rightarrow ^{13}$C transfer efficiencies with varying rf field amplitude, $\omega_{\tr{rf}}^{\tr{(q)}}$. The MAS rate was set to $\omega_r$ = 16.67 kHz, while the duration of the short pulses set to $\tau_p^{\tr{(q)}} = 4\mu$s. Data in (B) was recorded on uniformly $^{13}$C, $^{15}$N-labelled alanine using a 600 MHz spectrometer.}
			\label{fig:4-9}
		\end{figure}

		The effective dipolar Hamiltonian so found is utilised to calculate $\hat{I}_x \rightarrow \hat{S}_x$ transfer efficiency. The efficiency with varying rf field amplitude, which is set constant for the entire pulse sequence, and mixing time is shown in Fig. \ref{fig:4-9}A for a MAS rate of 16.7 kHz and short pulse length $\tau_p^{\tr{(q)}} = 4\mu$s. Fig. \ref{fig:4-9}B shows the experimental data for $^{15}$N $\rightarrow ^{13}$C$_\alpha$ transfer recorded on a 600 MHz spectrometer with the same MAS rate and other pulse parameters. The experimental plots are qualitatively comparable to the theoretical plots, however the 20-25\% difference in transfer efficiency between the plots can be attributed to the ignored CSA interaction in the theoretical calculation.
	
		As it has already been shown in previous studies that $^{\tr{RESPIRATION}}$CP is intrinsically broad-banded in $I$ spin, and owing to similar form of chemical shift analysis for the two spins, focus will be on the effect of chemical shift interaction of $S$-spin in this work.\\
		
		The effective first-order chemical shift Hamiltonian is given by
		\begin{equation}
			\hat{\bar{\tilde{\mathcal{H}}}}_{\tr{I}}^{(1)} = \sum_{n_0,k_\tr{I},l_\tr{I}} \hat{\tilde{\mathcal{H}}}^{(n_0,0,0,k_\tr{S},l_\tr{S})},
		\label{eq:5_firstord}
		\end{equation}
		where the triplet $(n_0,k_\tr{S},l_\tr{S})$ satisfy the resonance condition, $(n_0+k_{\tr{S}})\omega_r + l_\tr{S} \omega_{\tr{cw}}^{(\tr{I})} = 0$. Since $\omega_{\tr{cw}}^{(\tr{S})}$ is not an integral multiple of $\omega_r$, there is no set $(n,k_\tr{S},l_\tr{S})$ that satisfy the resonance condition and therefore, the effective first-order chemical shift Hamiltonian for $^{\tr{RESPIRATION}}$CP pulse sequence is not present.\\
		
\begin{figure}[!h]
			\centering
			\includegraphics[width=.5\linewidth]{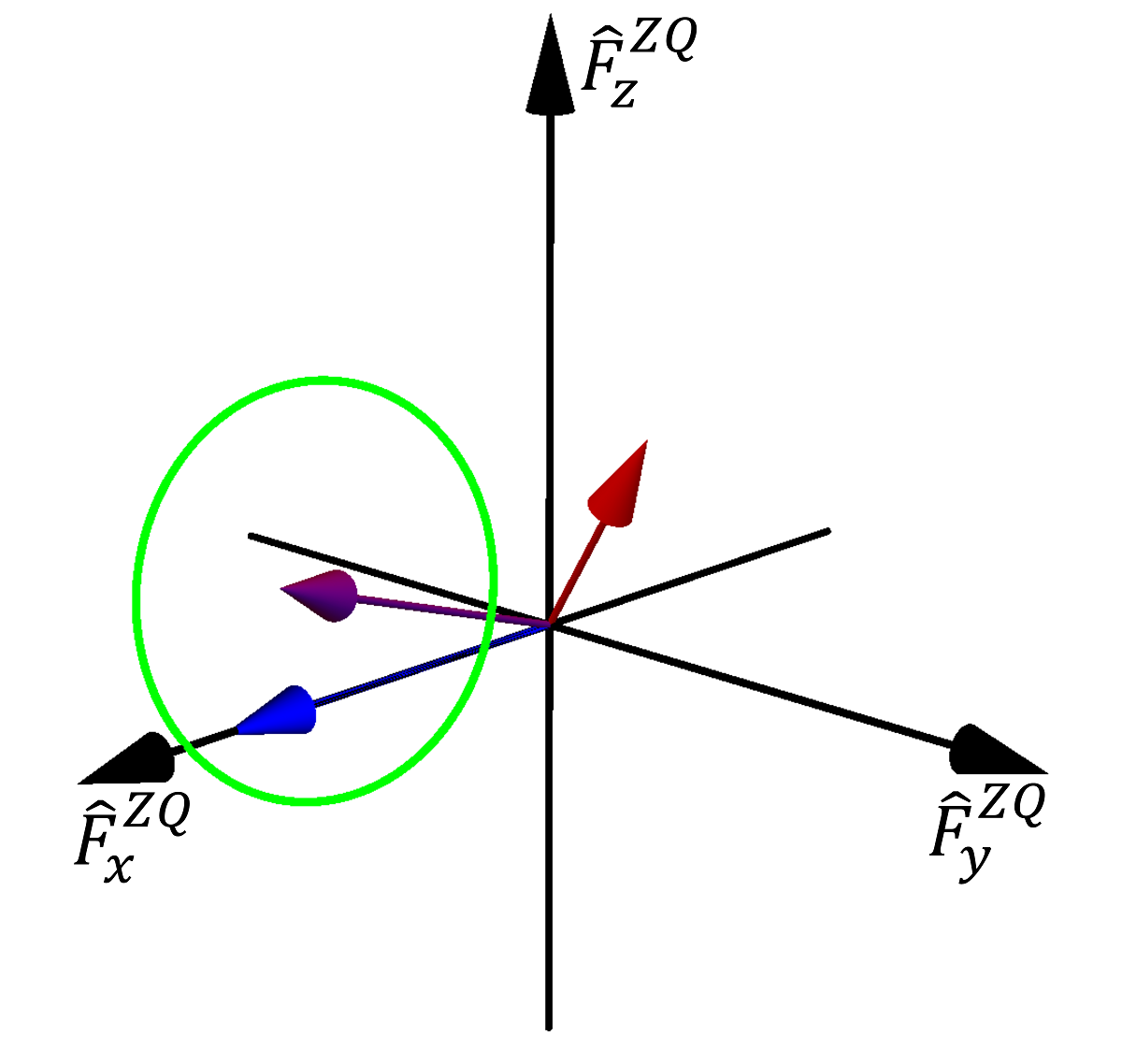}
			\caption{Illustration of time-evolution of the density operator (green line) under the effective Hamiltonian (purple arrow) for $^\tr{{RESPIRATION}}$CP. The effective Hamiltonian depicted here is the sum of effective first-order dipolar coupling Hamiltonian (red arrow) and effective second-order I-spin chemical shift Hamiltonian (blue arrow) in the ZQ subspace.}
		\label{fig:5-0}
		\end{figure}

		The second-order terms in the effective Hamiltonian that could potentially result in single-spin operators are the ones where the commutator is between chemical shift interaction and itself. Therefore the effective second-order Hamiltonian of interest takes the form
		\begin{equation}
			\hat{\bar{\tilde{\mathcal{H}}}}_{\tr{S}}^{(2)} = -\frac{1}{2}\sum_{n_0,k_\tr{S},l_\tr{S}}\sum_{n,\kappa_\tr{S},\lambda_\tr{S}} \frac{[\hat{\tilde{\mathcal{H}}}^{(n_0-n,0,0,k_\tr{S}-\kappa_\tr{S},l_\tr{S}-\lambda_\tr{S})}, \hat{\tilde{\mathcal{H}}}^{(n,0,0,\kappa_\tr{S},\lambda_\tr{S})}]}{(n + \kappa_\tr{S})\omega_r + \lambda_\tr{S}\omega_{\tr{cw}}^{(\tr{S})}},
		\label{eq:5_secord}
		\end{equation}
		where $(n+ \kappa_\tr{S})\omega_r + \lambda_\tr{S}\omega_{\tr{cw}}^{(\tr{S})} \neq 0$. It is noted here that in principle, all three contributions, i.e., commutator of isotropic chemical shift with itself, anisotropic chemical shift with itself, and isotropic with anisotropic chemical shift are present. However for easier visualisation of the effects of isotropic and anisotropic chemical shift interactions on the transfer, only the cases where either of these interactions is present and not both together, are considered. In the case where only the anisotropic chemical shift interaction is present, the effective second-order chemical shift Hamiltonian is given by,
		\begin{equation}
			\begin{alignedat}{2}
				\hat{\bar{\tilde{\mathcal{H}}}}_{\tr{S},\tr{aniso}}^{(2)} &= -\frac{1}{2}\sum_{\substack{n_0\\n_0\neq n}}^{}\sum_{\substack{n,\kappa_\tr{S}\\ n\neq0}} \frac{[\hat{\tilde{\mathcal{H}}}^{(n_0-n,0,0,-n_0-\kappa_\tr{S},\mp1)}, \hat{\tilde{\mathcal{H}}}^{(n,0,0,\kappa_\tr{S},\pm1)}]}{(n+\kappa_\tr{S})\omega_r \pm \omega_{\tr{cw}}^{(\tr{S})}}\\
				&= \xi_{\tr{aniso}}^{(\tr{S})} \hat{S}_x,
			\end{alignedat}
		\label{eq:5_secord_aniso}
		\end{equation}
		while for the case where only the isotropic chemical shift interaction is present, the same is given by,
		\begin{equation}
			\begin{alignedat}{2}
				\hat{\bar{\tilde{\mathcal{H}}}}_{\tr{S},\tr{iso}}^{(2)} &= -\frac{1}{2}\sum_{\kappa_\tr{S}} \frac{[\hat{\tilde{\mathcal{H}}}^{(0,0,0,-\kappa_\tr{S},\mp1)}, \hat{\tilde{\mathcal{H}}}^{(0,0,0,\kappa_\tr{S},\pm1)}]}{\kappa_\tr{S}\omega_m^{(\tr{S})} \pm \lambda_\tr{I}\omega_{\tr{cw}}^{(\tr{S})}}\\
				&= 4\pi^2\delta_{\tr{iso}}^{(\tr{S})^2}\xi_{\tr{iso}}^{(\tr{S})} \hat{S}_x,
			\end{alignedat}
		\label{eq:5_secord_iso}
		\end{equation}
		where $\delta_{\tr{iso}}^{(\tr{S})} = \frac{\omega_{\tr{S}}^{(0)}-\omega_{\tr{S}}^{\tr{rf}}}{2\pi}$ with $\omega_{\tr{S}}^{\tr{rf}}$ being the rf carrier frequency for the $S$ spin channel. As the contributions to effective second-order chemical shift Hamiltonian is only of the form $\hat{S}_x$, which can be written as $\hat{S}_x = -\hat{F}_x^{ZQ} + \hat{F}_x^{DQ}$, the resultant total effective Hamiltonian in the zero-quantum subspace in shifted away from the $\hat{F}_{z,y}^{ZQ}$ plane. This is illustrated in Fig. \ref{fig:5-0}, where the blue arrow describes the effective second-order chemical shift term, and the total effective Hamiltonian is shown by the purple arrow. Trajectory of the initial density operator ($\hat{F}_x^{ZQ}$) under the total effective Hamiltonian follows the green curve shown in the figure and results in diminished transfer efficiency.\\
		
		The powder-averaged strength of effective second-order chemical shift Hamiltonian given in Eq. \ref{eq:5_secord_aniso} was computed and is presented in Fig. \ref{fig:5-1}B as $\xi_{\tr{aniso}}^{(\tr{S})}$ with varying $\tau_p^{(\tr{S})}$, length of the short pulse. The calculations were performed for 20 kHz MAS and with anisotropic chemical shift, $\delta_{\tr{aniso}}^{(\tr{S})} = 5.0$ kHz. $\tau_p^{(\tr{S})}$ was varied, while also ensuring that $\omega_{\tr{cw}}^{(\tr{S})}$ remained constant by relating $\omega^{(\tr{S})} = \frac{2\pi}{25\tau_p^{(\tr{S})}}$. This is done to ensure constant dipolar recoupling conditions for the entire considered case space, that spans from an ideal pulse ($\tau_p^{(\tr{S})} = 0$) to a continuous-wave irradiation ($\tau_p^{(\tr{S})} = \tau_r = 50\mu$s). The strength can be seen proportional to the length of short pulse. Impact of the above calculated chemical shift Hamiltonian on the transfer efficiency is studied by also calculating the effective first-order dipolar coupling Hamiltonian with a dipolar coupling constant of $\frac{b_{IS}}{2\pi} = 50$ Hz. The transfer efficiency calculated with the total effective Hamiltonian, sum first-order dipolar coupling and second-order anisotropic chemical shift Hamiltonians, for 46 ms of mixing time is shown in Fig. \ref{fig:5-1}C as red circles and is seen to correspond well with Fig. \ref{fig:5-1}B. Additionally, the propagation is verified by direct-propagation numerical simulations, shown in Fig \ref{fig:5-1}C as blue boxes.
		
\begin{figure}[!h]
			\centering
			\includegraphics[scale=1.3]{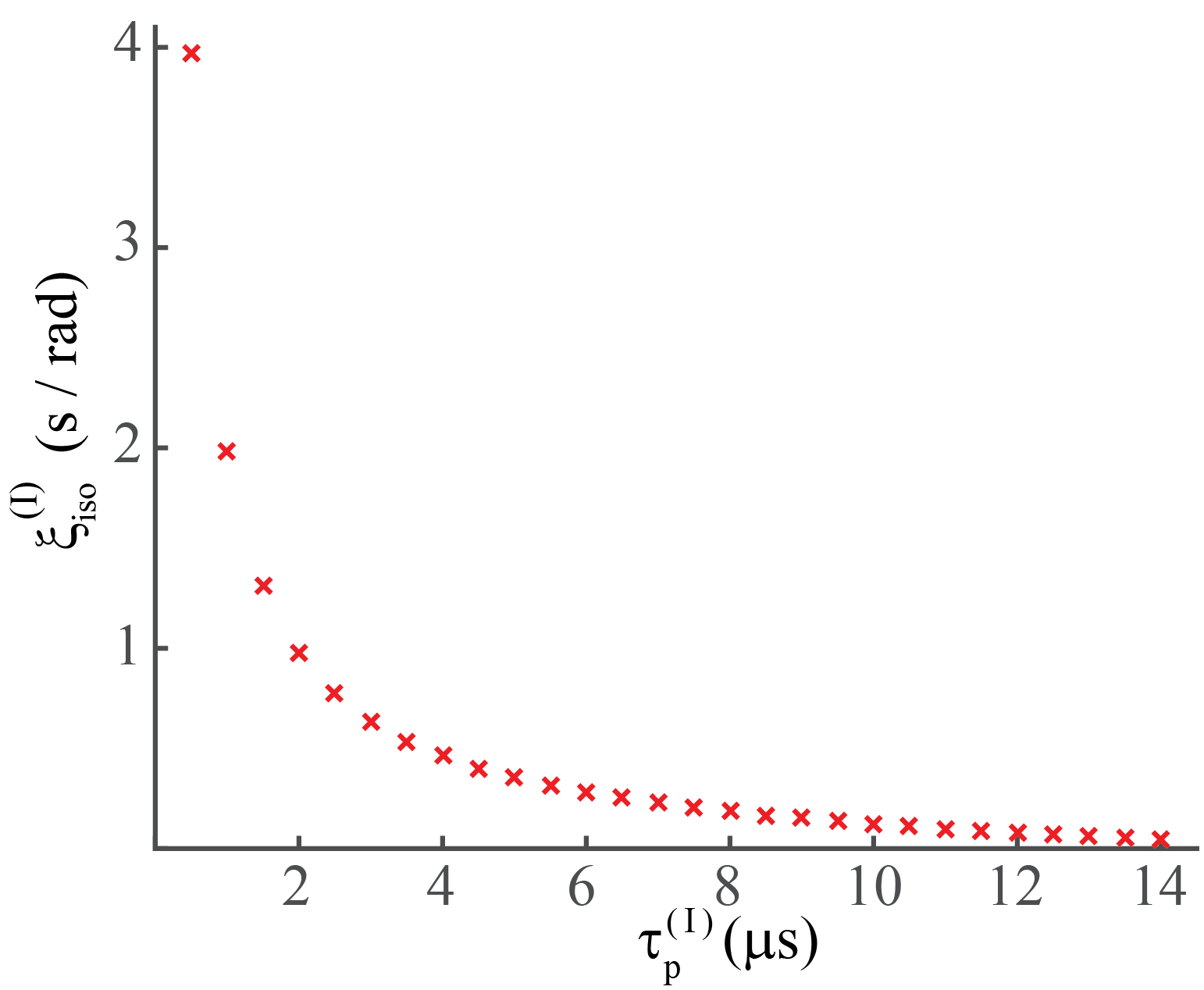}\\
			\caption{Offset-independent effective second-order isotropic chemical shift Hamiltonian, $\xi_{\tr{iso}}^{(\tr{S})}$, calculated for varying short pulse length, while $\omega^{(\tr{S})}$ was fixed at $2\omega_r$ and MAS rate was set to 20 kHz.}
		\label{fig:5-2}
		\end{figure}

		For the case where only isotropic chemical shift is present, the strength of $\hat{S}_z$ term given in Eq. \ref{eq:5_secord_iso} is $4\pi^2\delta_{\tr{iso}}^{(\tr{S})^2}\xi_{\tr{iso}}^{(\tr{S})}$. As the strength of the effective second-order chemical shift Hamiltonian depends on square of the offset, convergence of the Magnus series defining the effective Hamiltonian will be slower at large isotropic chemical shift offset. The strength also depends on length of the short pulse through $a_{\kappa_{\tr{S}}}^{(\tr{S})}$ coefficients and this is shown in Fig. \ref{fig:5-2}, where $\omega^{(\tr{S})} = 2\omega_r$ and the MAS rate was set to 20 kHz. Increasing the short pulse length (thereby increasing $\omega_{\tr{cw}}^{(\tr{S})}$), can be seen to suppress the second-order isotropic chemical shift Hamiltonian term. Of course, varying the short pulse length also affects the effective dipolar coupling Hamiltonian and so it is important to compare the transfer profiles with varying short pulse lengths, against isotropic chemical shift values.
		
\begin{figure}[!h]
			\centering
			\includegraphics[scale=0.95]{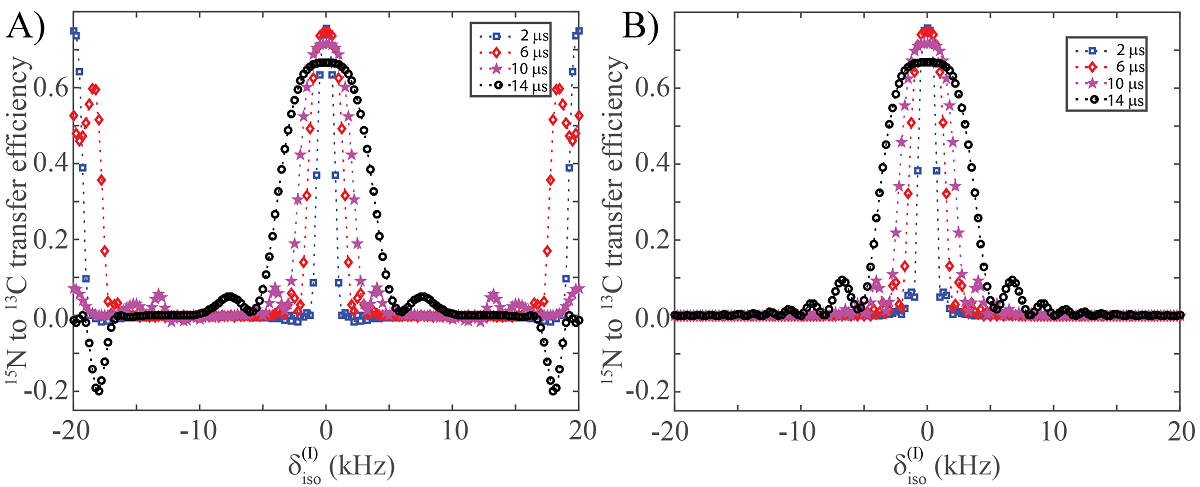}\\
			\caption{Transfer efficiencies for A) direct-propagation numerical simulations and B) evolution under total effective Hamiltonian (sum of effective first-order dipolar coupling and second-order isotropic chemical shift Hamiltonians), for $\hat{I}_x \rightarrow \hat{S}_x$ transfer with varying isotropic I-spin chemical shift. Calculations were done for a $^{15}N$, $^{13}C$ spin-system at 20 kHz MAS and rf amplitudes satisfying $\omega_{\tr{rf}}^{(\tr{S})} = \omega_{\tr{rf}}^{(\tr{S})} = 2\omega_{\tr{rf}}$. The four curves in both figures correspond to short pulse lengths of 2$\mu$s (blue squares), 6$\mu$s (red diamonds), 10$\mu$s (magenta stars) and 14$\mu$s (black circles) and mixing times of 2.4 ms, 2.3 ms, 2.2 ms and 2.2 ms, respectively.}
		\label{fig:5-3}
		\end{figure}

		Transfer efficiencies at varying isotropic chemical shift offset for $\hat{I}_x \rightarrow \hat{S}_x$, were obtained through direct-propagation simulations for short pulse lengths of $2\mu$s, $6\mu$s, $10\mu$s and $14\mu$s and are shown in Fig. \ref{fig:5-3}A. The MAS rate was set to 20 kHz, the rf field strengths satisfied, $\omega_{\tr{rf}}^{(\tr{S})} = \omega_{\tr{rf}}^{(\tr{I})} = 2\omega_r$, dipolar coupling constant $\frac{b_{\tr{IS}}}{2\pi}$ was set to 1 kHz and the isotropic S-spin chemical shift was zero. Even though longer short pulse lengths correspond to broader efficient transfer profiles around zero chemical shift offset, the broadest ($\tau_p^{(\tr{S})} = 14\mu$s) corresponds a range of only $\pm$5 kHz. The same set of simulations were repeated using the total effective Hamiltonian as calculated above using Eqs. \ref{eq:5_firstord} and \ref{eq:5_secord}. The results are shown in Fig. \ref{fig:5-3}B, and are seen to be consistent with the direct-propagation simulations shown in Fig. \ref{fig:5-3}A for isotropic chemical shift offset values in the range of $\pm5$ kHz. This justifies our choice to ignore the effective second-order Hamiltonian terms that involve commutators between the isotropic chemical shift term and the dipolar coupling term. However the second-order isotropic chemical shift Hamiltonian is too large at higher offset values, and this is evident as the Figs \ref{fig:5-3}A and B do not match at the regime. To overcome this problem, it is better to transform the Hamiltonian into an interaction frame defined by both the rf field and isotropic chemical shift interactions, rather than to transform just into the rf field interaction frame.\\
		
\begin{figure}[!h]
			\centering
			\begin{minipage}[c]{0.6\textwidth}
				\includegraphics[width=\linewidth]{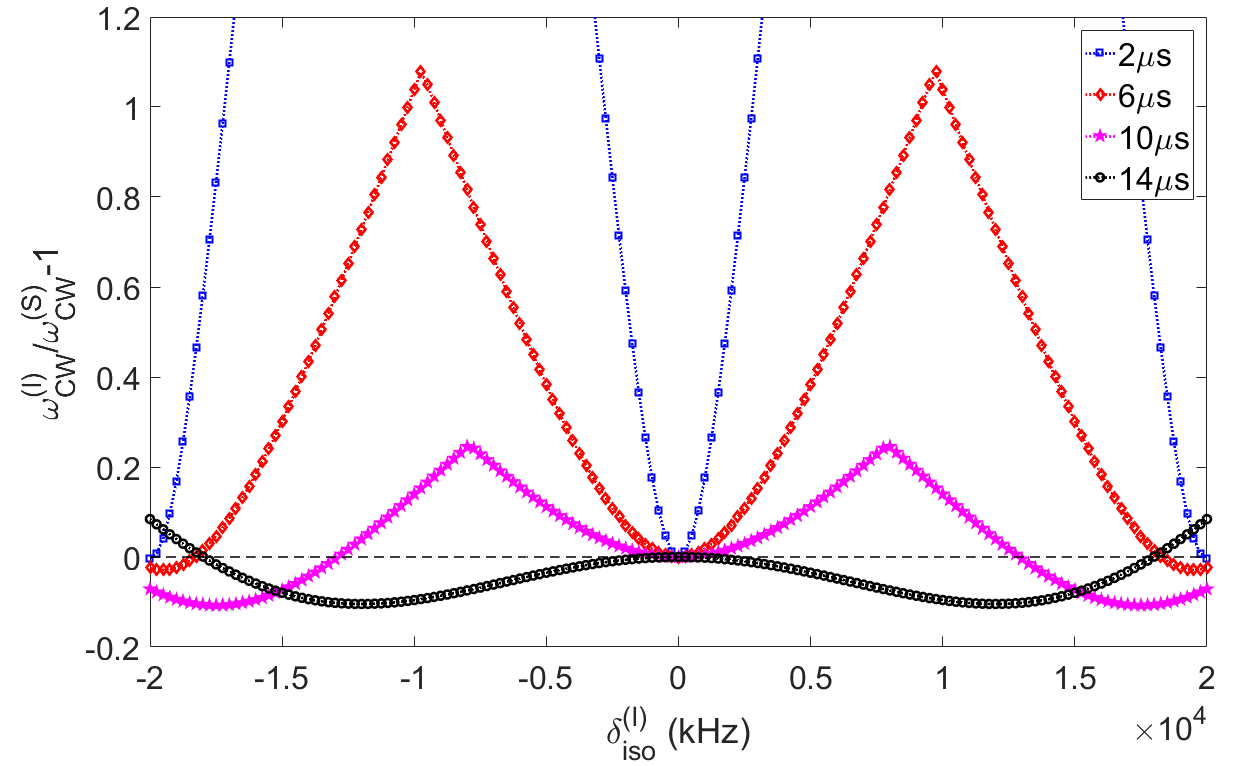}
			\end{minipage}
			\begin{minipage}[c]{0.35\textwidth}
				\centering
				\begin{tabular}[b]{|c|c|c|}
					\hline
					$\tau_p^{(\tr{S})}/\mu$s & $\delta_{\tr{iso}}^{(\tr{S})}/$kHz & $\hat{S}_x$\\
					\hline
					\hline
					$2$ & $\pm20$ & 0.99\\
					\hline
					$6$ & $\pm20$ & 0.80\\
					\hline
					$10$ & $\pm12.5$ & 0.09\\
					\hline
					$14$ & $\pm18.5$ & -0.38\\
					\hline
				\end{tabular}
			\end{minipage}
			\caption{Effective fields $\omega_{\tr{cw}}^{(\tr{q})}$ (left) for combined rf field and isotropic chemical shift interactions found using quaternions, and represented as a quantity which if zero, corresponds to $\omega_{\tr{cw}}^{\tr{(I)}} = \omega_{\tr{cw}}^{\tr{(S)}}$ and therefore effective first-order dipolar coupling resonance conditions being satisfied. The unit-scale x-component of the calculated effective fields are tabulated (right).}
		\label{fig:5-4}
		\end{figure}

		By transforming into a frame defined by the combined rf field and isotropic chemical shift interactions, the only two frequencies of the five fundamental frequencies that change are $\omega_{\tr{cw}}^{(\tr{q})}$. However the trick of splitting the amplitude-modulated rf field can not be applied to this case, as the operators no longer commute at different times. However, as shown in Sec. \ref{sect:quaternions}, the effective frequency $\omega_{\tr{cw}}^{(\tr{q})}$ and the effective axis can be found using quaternions. The effective fields, $\omega_{\tr{cw}}^{(\tr{q})}$, were calculated for parameters used in Fig. \ref{fig:5-3} and are shown in Fig. \ref{fig:5-4}. It is noted here that $\omega_{\tr{cw}}^{(\tr{S})} = \omega_{\tr{cw}}^{(\tr{I})}$ corresponds to presence of effective first-order dipolar coupling resonance conditions, and in Fig. \ref{fig:5-4}, this is where the y-axis is zero. Evidently that the broader efficient transfer profiles seen in Fig. \ref{fig:5-3} can be explained by corresponding shallower profiles in Fig. \ref{fig:5-4}, around $\delta_{\tr{iso}}^{(\tr{S})} = 0$ kHz. This is because shallower (or flatter) zero-values correspond to resonance conditions being valid over larger ranges. It can be further seen that efficient transfers seen at higher offsets in Fig. \ref{fig:5-3} are also explained by Fig. \ref{fig:5-4}. It is noted here since the transfer to $\hat{S}_x$ from $\hat{I}_x$ is desired and that $\omega_{\tr{cw}}^{(\tr{I})}$ is already along $x$, $\omega_{\tr{cw}}^{(\tr{S})}$ should also be directionally along $x$ to satisfy the resonance conditions. $\hat{S}_x$ component of the effective fields for different short pulse lengths at $\delta_{\tr{iso}}^{(\tr{S})} \neq 0$ are tabulated in Fig. \ref{fig:5-4}. This explains the transfers seen at higher offset values seen in Fig. \ref{fig:5-3}A, including the negative transfer efficiency for $\tau_p^{(\tr{S})} = 14\mu$s at $\pm18.5$ kHz. From this, it is clear that a broad-banded pulse sequence should satisfy the condition, $\omega_{\tr{cw}}^{(\tr{S})} = \omega_{\tr{cw}}^{(\tr{I})}$, for larger range of offsets. As generation of effective field plots as shown in Fig. \ref{fig:5-4} is straightforward and takes little to no time, it is faster and easier to test new ideas this way, rather than to do a full direct-propagation or effective-Hamiltonian-driven-propagation simulations.
		\\
		
	\section{Broadband-$^{\tr{RESPIRATION}}$CP}
\begin{figure}[!h]
			\centering
			\includegraphics[scale=0.45]{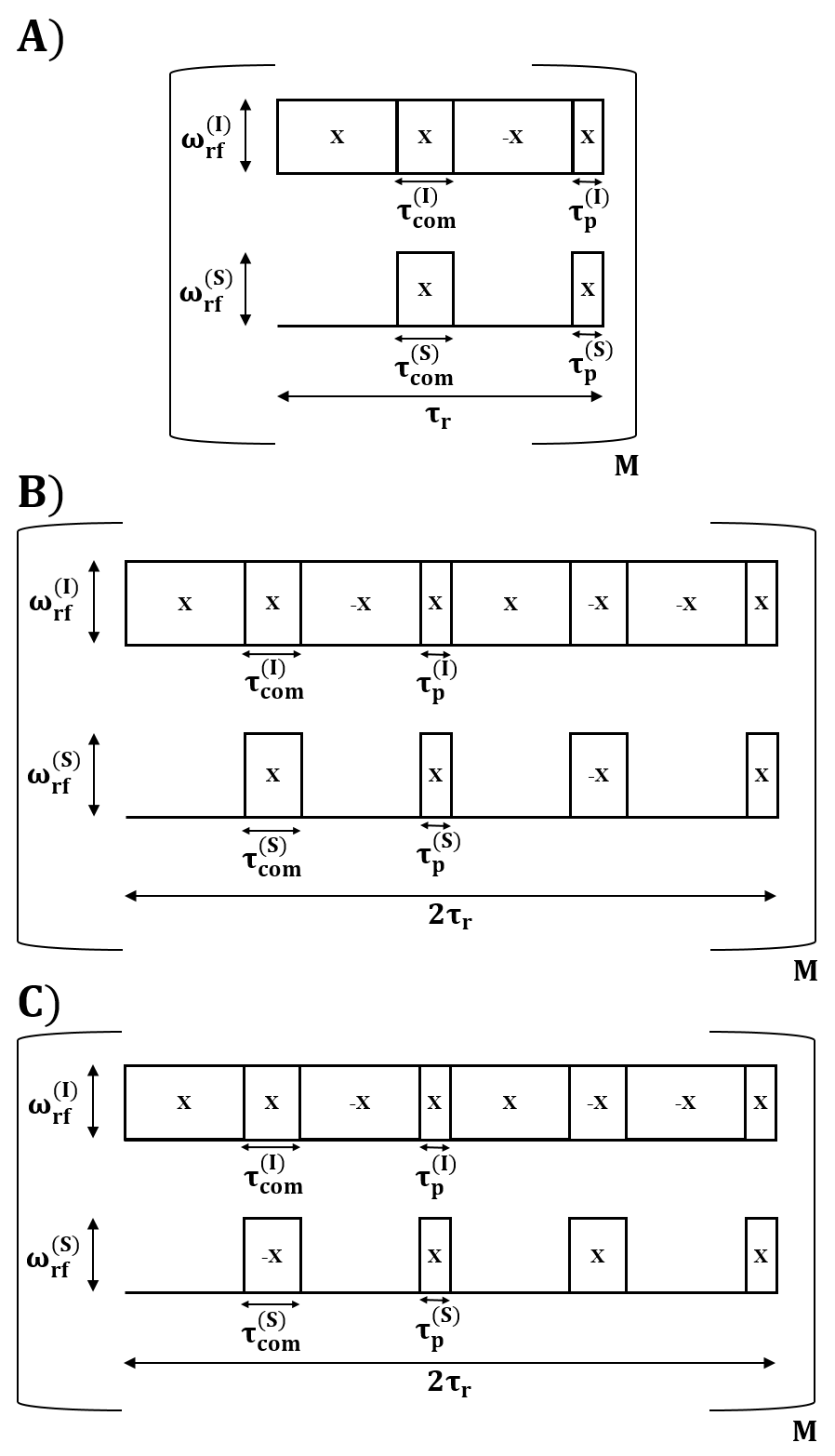}
			\caption{$^{\tr{RESPIRATION}}$CP pulse sequence given in Fig. \ref{fig:4-7}, but with additional offset-compensating pulses of duration $\tau_{\tr{com}}^{(\tr{q})}$ inserted in the middle of every rotor period. A) Phase of the offset-compensating pulses are set to $X$. B) Phases of the offset-compensating pulses are phase-cycled with $X$ and $-X$ synchronously on both channels. C) Phases of the offset-compensating pulses are phase-cycled with $X$ and $-X$ asynchronously on both channels.}
			\label{fig:5-5}
		\end{figure}

\begin{figure}[!h]
			\centering
			\includegraphics[scale=0.45]{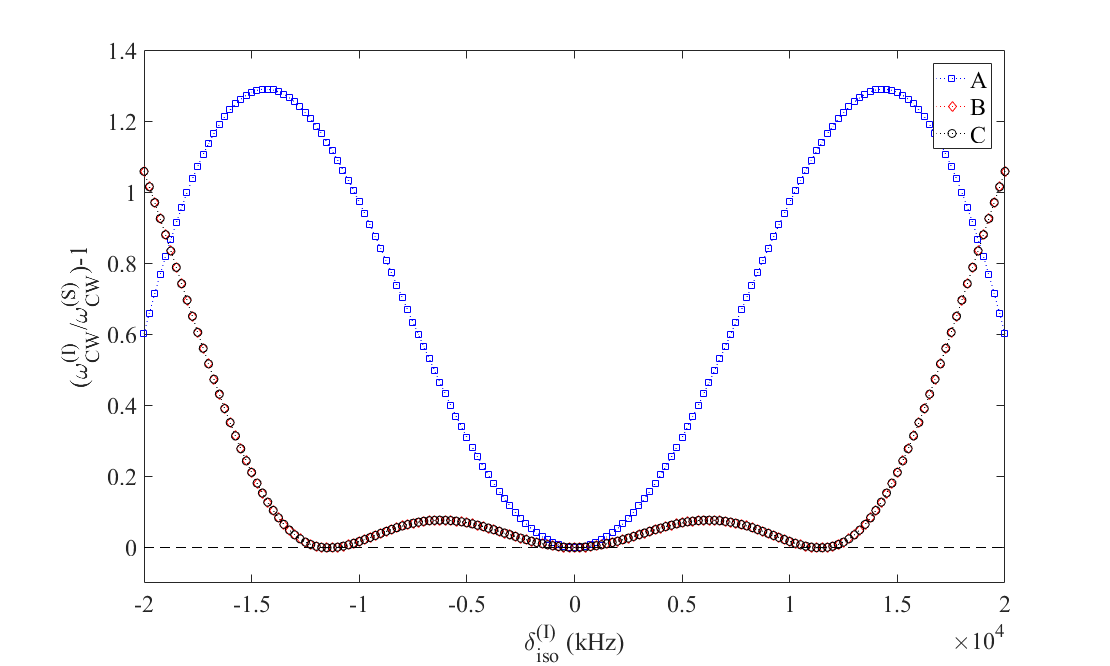}
			\caption{Effective field frequencies of the combined rf field and isotropic chemical shift. The rf pulse sequences are given in Fig. \ref{fig:5-5}. The effective fields are represented with varying isotropic chemical shift, and represented as a quantity, which if zero corresponds to presence of resonance conditions.}
		\label{fig:5-6}
		\end{figure}

\cite{shankar2016handling}
	\label{sect:bb_resp}
		To make the I-spin channel better offset-compensated, it was natural to insert a single pulse of length $\tau_{\tr{com}}^{(\tr{S})}$ in the middle of each free evolution period. It is important to include the compensation pulse on the S-spin channel as well, to avoid nullifying the recoupled dipolar Hamiltonian. The question of optimized flip angle of the compensation pulses and whether they should be phase cycled are now simply answered by the effective fields plots they generate. The pulse sequence with compensation pulses is shown in Fig. \ref{fig:5-5}A and two variants with phase cycling of the compensation pulses are shown in Fig. \ref{fig:5-5}B and \ref{fig:5-5}C.
		The effective field frequency plots for the above pulse sequences, when the flip angle of the compensation pulses are $\pi$ are shown in Fig. \ref{fig:5-6}, from which it is evident that pulse sequences where the compensation pulses are phase cycled perform better with respect to offset-compensation.\\
		
\begin{figure}[!h]
			\centering
			\includegraphics[width=\linewidth]{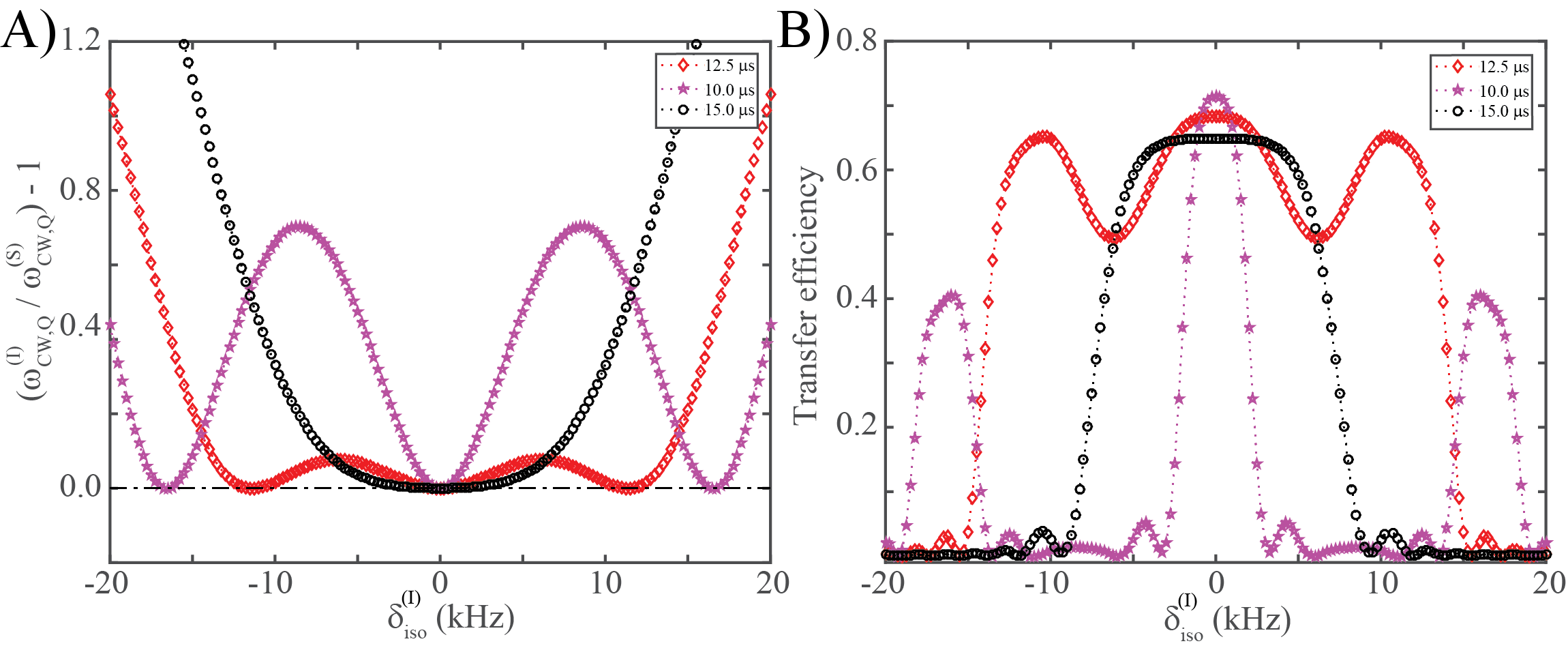}
			\caption{A) Effective field frequencies of combined rf field, defined by pulse sequence given in Fig. \ref{fig:5-5}B, and isotropic chemical shift. The offset-compensating pulses in the pulse sequence was set to produce a flip angle of $\pi$ ($\tau_{\tr{com}}^{(\tr{q})} = 12.5\mu$s), $0.8\pi$ ($\tau_{\tr{com}}^{(\tr{q})} = 10\mu$s) and $1.2\pi$ ($\tau_{\tr{com}}^{(\tr{q})} = 15\mu$s), where the spinning rate $\omega_r = 20$ kHz and $\omega_{\tr{rf}}^{(\tr{q})} = 2\omega_r$. B) Transfer efficiencies corresponding to A).}
		\label{fig:5-7}
		\end{figure}

		To verify if the flip angle of $\pi$ radians is best for the compensation pulses, the effective field frequency plot for the pulse sequence given in Fig. \ref{fig:5-6} was calculated for a flip angle of less than $\pi$ ($\tau_{\tr{com}}^{\tr{(q)}} = 10\mu$s) radians and greater than $\pi$ ($\tau_{\tr{com}}^{\tr{(q)}} = 15\mu$s) radians and are shown along with that for the pulse sequence where the flip angle is exactly $\pi$ ($\tau_{\tr{com}}^{\tr{(q)}} = 12.5\mu$s) radians in Fig. \ref{fig:5-7} (left). The corresponding transfer efficiencies obtained through direct-propagation simulations are also shown in Fig. \ref{fig:5-7} (right), which substantiate the claims made using only the effective field plots.\\
		
\begin{figure}[!h]
			\centering
			\includegraphics[width=\textwidth]{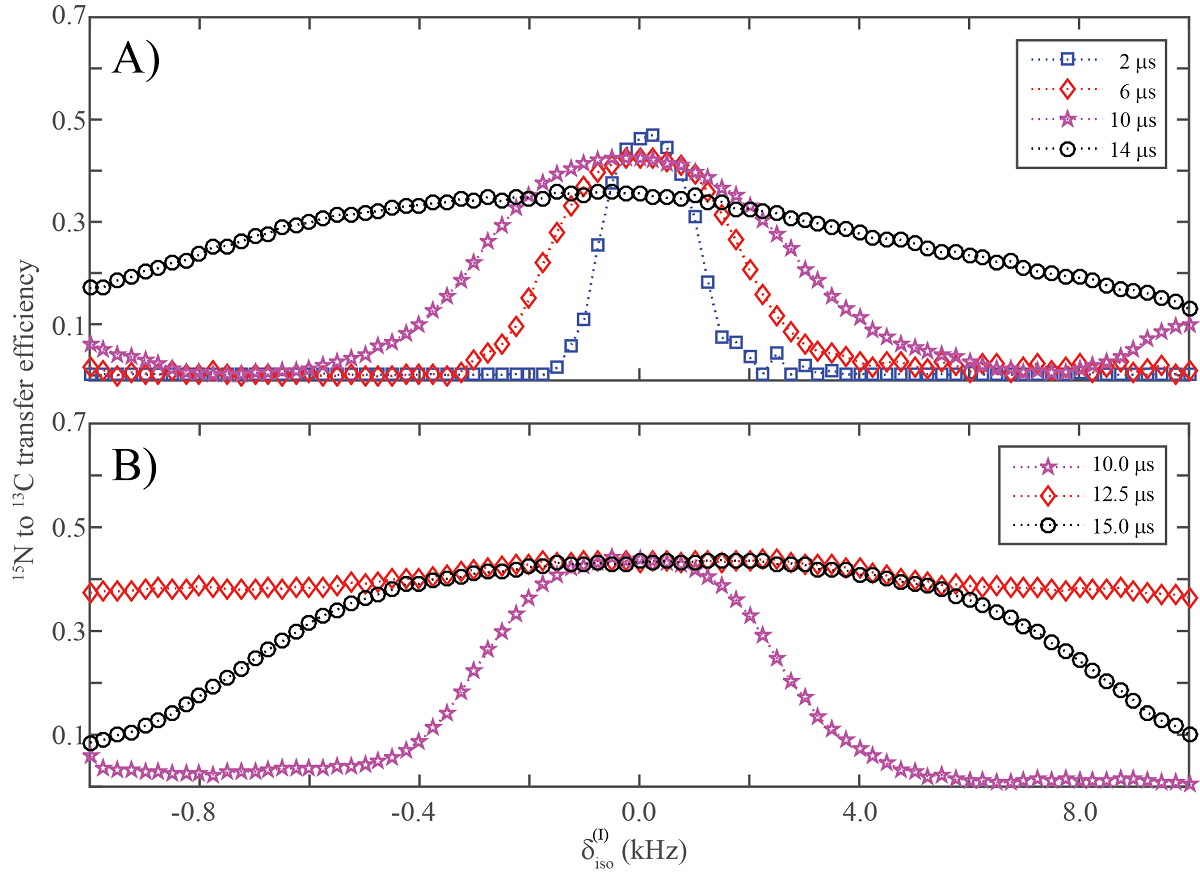}
			\caption{Extracted experimental $^{15}$N to	$^{13}$C$_{\alpha}$ transfer efficiencies obtained for a powder sample of $^{15}$N,$^{13}$C$_{\alpha}$-labelled glycine using A) $^{\tr{RESPIRATION}}$CP and
				B) BB-$^{\tr{RESPIRATION}}$CP (b) with varying isotropic chemical shift offset on the $S$-spin rf channel. The data were recorded on a 400 MHz spectrometer at 20 kHz MAS. In A) $\tau_p^{\tr{(q)}} = 2$ $\mu$s (blue boxes) with a total mixing time of 2.2 ms (M = 44), 6 $\mu$s (red diamonds) with M = 40, 10 $\mu$s (purple stars) with M = 40, and 14 $\mu$s (black circles) with M = 40. In B) $\tau_p^{\tr{(q)}} = 2$ $\mu$s while $\tau_{\tr{com}}^{\tr{(q)}} = 10$ $\mu$s (purple stars), 12.5 $\mu$s(red diamonds), 15 $\mu$s (black circles), all with total mixing time equal to = 2.0 ms (M = 20).}
			\label{fig:5-10}
		\end{figure}

		To experimentally validate the results, $^{15}$N ($\hat{I}_x$) $\rightarrow ^{13}$C$_{\alpha}$ ($\hat{S}_x$) transfer efficiency with varying $S$-spin isotropic chemical shift was measured in selectively $^{15}$N,$^{13}$C$_{\alpha}$-labelled glycine on a 400 MHz spectrometer using both $^{\tr{RESPIRATION}}$CP and its broadband variant. The MAS rate was set to 20 kHz. The measured transfer efficiency is shown in Fig. \ref{fig:5-10}A for $^{\tr{RESPIRATION}}$CP where the duration of the short pulse was varied to take the values 2$\mu$s, 6$\mu$s, 10$\mu$s and 14$\mu$s. The same for BB-$^{\tr{RESPIRATION}}$CP is shown in Fig. \ref{fig:5-10}B where the duration of the short pulse was constant at 2$\mu$s, while the duration of the offset-compensating pulse was varied to take the values $10\mu$s ($<\pi$), $12.5\mu$s (=$\pi$) and $15\mu$s ($>\pi$). The experimentally observed values correspond well with theoretical prediction. The understanding is that offset compensation can be tailored by altering the effective rotation of the offset-compensating pulses without much affecting the recoupled dipolar coupling Hamiltonian.
		
\begin{figure}[!h]
			\centering
			\includegraphics[width=\textwidth]{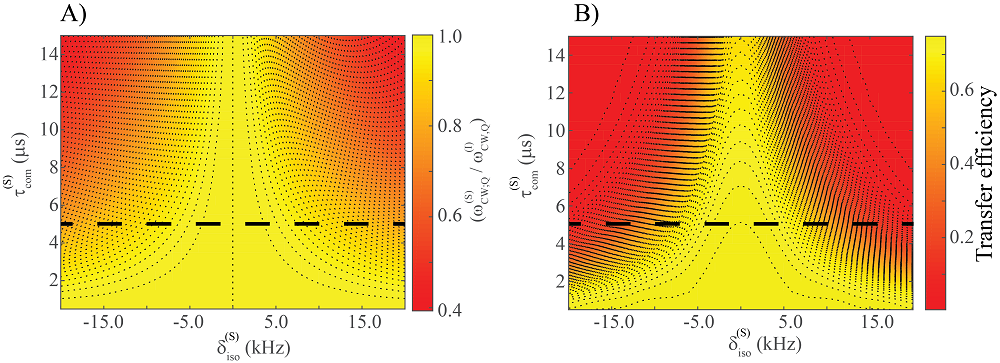}
			\caption{A) Ratio of effective field frequencies $\omega_{\tr{cw}}^{(\tr{I})}/\omega_{\tr{cw}}^{(\tr{S})}$ for combined rf field and isotropic chemical shift, represented with respect to varying S-spin isotropic chemical shift and length of offset-compensating pulse of $\pi$ radians on S-spin rf channel. B) Transfer efficiency corresponding to A).}
			\label{fig:5-8}
		\end{figure}

		As the offset-compensating pulses are also inserted on the $I$-spin rf channel, effect of the $\pi$ pulses in this channel, in particular the length $\tau_{\tr{com}}^{(\tr{I})}$ is studied using effective field frequencies. For the pulse sequence given in Fig \ref{fig:5-5}B, $\tau_{\tr{com}}^{(\tr{I})}$ was varied while keeping the flip angle of $\pi$ radians constant through corresponding variation in $\omega_{\tr{rf}}^{(\tr{I})}$ (only for the offset-compensating pulse), and the effective field frequencies obtained are shown in Fig. \ref{fig:5-8}. The rf field amplitude for rest of the pulses was set at $\omega_{\tr{rf}}^{(\tr{q})} = 2\omega_r$, the spinning rate was set at $\omega_r = 20$ kHz, short pulse lengths were $\tau_{p}^{(\tr{q})} = 2\mu$s. The finding is that, the shorter the offset-compensating pulse, better the transfer. This is substantiated by the transfer efficiency offset plots obtained through direct-propagation numerical simulations shown in Fig. \ref{fig:5-8}B.\\
		
\begin{figure}[!h]
			\centering
			\includegraphics[width=\textwidth]{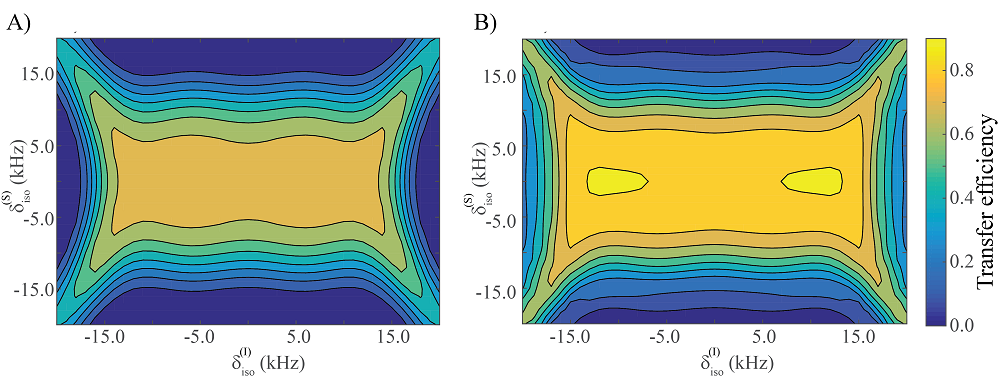}
			\caption{Numerical simulations for the transfer efficiencies with varying $I$- and $S$-spin chemical shift offset for A) BB-$^{\tr{RESPIRATION}}$CP and B) adiabatic BB-$^{\tr{RESPIRATION}}$CP with $\tau_{\tr{com}}^{\tr{(q)}} = 2$ $\mu$s, $\tau_p^{\tr{(q)}} = 2$ $\mu$s and $\omega_{\tr{rf}}^{(q)} = 2\omega_r$. The total mixing time was 2.2 ms (M = 22) for BB-$^{\tr{RESPIRATION}}$CP, and 7.0 ms (M=70) for adiabatic BB-$^{\tr{RESPIRATION}}$CP.}
		\label{fig:5-9}
		\end{figure}

		In Fig. \ref{fig:5-9}A, transfer efficiency with respect to isotropic chemical shift of both spins, obtained through direct-propagation simulation, is shown and it is seen that the offset-compensation is wider ($\pm13$ kHz) with efficiencies above 73\% than for $^{\tr{RESPIRATION}}$CP with respect to the $S$-spin rf field channel. The parameters used here correspond to the hyphenated black line in Fig. \ref{fig:5-8}, where $\tau_{\tr{com}}^{(\tr{I})} = 5\mu$s ($\implies \omega_{\tr{rf}}^{(\tr{I})} = 100$ kHz for $I$-spin channel compensation pulses), $\omega_r = 20$ kHz and $\omega_{\tr{rf}}^{(\tr{S})} = \omega_{\tr{rf}}^{(\tr{I})} = 2\omega_r$ for rest of the pulses.\\
		
		To further improve the transfer efficiency of the BB-$^{\tr{RESPIRATION}}$CP experiment, an additional amplitude sweep was added to the phase-alternating pulses in the $I$-spin rf channel. Similar to the discussion in Sec. \ref{sect:4_adRFDR}, the additional $\hat{F}^{ZQ}_z$ in the effective Hamiltonian helps drag the polarisation adiabatically from $\hat{I}_z$ to $\hat{S}_z$. Transfer efficiency so calculated with respect to varying isotropic chemical shift of both spins is shown in Fig. \ref{fig:5-9}B. The simulation used the same parameters as used in Fig. \ref{fig:5-9}A, but with a total mixing time of 7.0 ms (M=70). The sweep parameters were $\Delta = 800$ Hz and $d_{\tr{est}} = 160$ Hz. The maximum transfer efficiency obtained is about 90\% and the offset compensation is also slightly broader on both the rf channels.
		
\begin{figure}[!h]
			\centering
			\includegraphics[width=\textwidth]{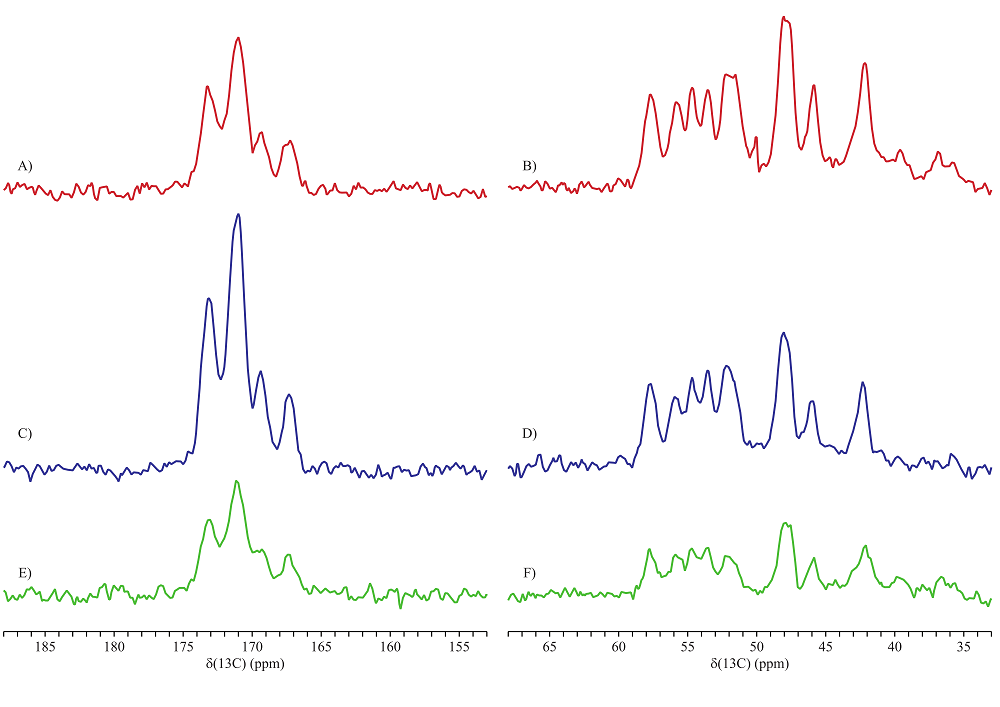}
			\caption{A comparison of 1D spectra for $^{15}$N $\rightarrow ^{13}$C$_{\alpha}$ (A, C and E) and $^{15}$N $\rightarrow ^{13}$CO (B, D, and F) transfer for SNFGAILSS amyloid fibrils with uniform $^{13}C$, $^{15}$N labeling on the FGAIL residues recorded on a 950 MHz spectrometer at 22.2 kHz MAS.
			The transfer is mediated by adiabatic versions of DCP (A and B) and $^{\tr{RESPIRATION}}$CP (C and D) with a carrier frequency of 170 ppm (A and C) and 50 ppm (B and D), while it is mediated by adiabatic BB-$^{\tr{RESPIRATION}}$CP with a carrier frequency of 100 ppm (E and F).}
			\label{fig:5-11}
		\end{figure}

		To demonstrate the performance of BB-$^{\tr{RESPIRATION}}$CP transfer element on a more a complicated system of biological relevance, 1D and 2D experiments were carried out on a sample of SNNFGAILSS amyloid fibrils, and compared with 1D $^{\tr{RESPIRATION}}$CP and DCP experiments for $^{15}$N $\rightarrow ^{13}$C transfer. The experiments were carried out on a 950 MHz spectrometer at 22.2 kHz MAS. For the comparisons adiabatic versions of all three pulse sequences were used. For the DCP experiment, the adiabatic amplitude sweep was added to the $^{13}$C channel, while linear sweeps were added to the phase-alternating pulses of both $^{\tr{RESPIRATION}}$CP experiments. The spectra obtained from 1D experiments recorded with adiabatic DCP, adiabatic $^{\tr{RESPIRATION}}$CP and adiabatic BB-$^{\tr{RESPIRATION}}$CP are shown in three rows respectively, in Fig. \ref{fig:5-11}, where the first column depicts the transfer to $^{13}$C$_{\alpha}$ and the second column depicts the transfer to $^{13}$CO. The two spectra each for the band-selective adiabatic DCP and adiabatic $^{\tr{RESPIRATION}}$CP were recorded individually by setting the rf carrier frequency at 170 ppm ($^{13}$C$_{\alpha}$) and 50 ppm ($^{13}$CO), while the two spectra for adiabatic BB-$^{\tr{RESPIRATION}}$CP were recorded simultaneously by setting the rf carrier frequency at 100 ppm. 
		The transfer efficiencies, $\varepsilon_{\tr{NC}}$, are shown in Tab. \ref{tab:bb_trans} and were calculated according to $\varepsilon_{\tr{NC}} = \frac{\eta_{\tr{HNC}}}{4\cdot\eta_{\tr{C}}\cdot \varepsilon_{\tr{HN}}}$, where $\eta_{\tr{HNC}}$ is the $^{13}$C signal intensity in the $^1$H $\rightarrow^{15}$ N$\rightarrow^{13}$C transfer, $\eta_{\tr{C}}$ is the direct excitation $^{13}$C signal intensity and $\eta_{\tr{HN}}$ is the transfer efficiency of $^1$H $\rightarrow ^{15}$N transfer. Admittedly, the transfers obtained for adiabatic BB-$^{\tr{RESPIRATION}}$CP are lower than that obtained for the other two pulse sequences. However, since BB-$^{\tr{RESPIRATION}}$CP transfers to the two $^{13}$C sites in a single experiment from the same number of $^{15}$N nuclei, as the other two sequences that transfer to the two $^{13}$C sites in two individual experiments, fair comparison between the pulse sequences would be to compare the weighted total transfers, shown in the last column of the table.
		\begin{table}[!h]
			\centering
			\begin{tabular}[b]{|c|c|c|c|}
				\hline
				& $^{15}$N $\rightarrow ^{13}$C$_{\alpha}$ & $^{15}$N $\rightarrow ^{13}$CO & weighted total\\
				\hline
				\hline
				 DCP & 27\% & 14\% & 20.5\%\\
				\hline
				$^{\tr{RESPIRATION}}$CP & 18\% & 27\% & 22.5\%\\
				\hline
				BB-$^{\tr{RESPIRATION}}$CP & 10\% & 14\% & 24\%\\
				\hline
			\end{tabular}
			\caption{Transfer efficiencies obtained for $^{15}$N $\rightarrow ^{13}$C transfers using adiabatic versions of DCP, $^{\tr{RESPIRATION}}$CP and BB-$^{\tr{RESPIRATION}}$CP pulse sequences, on a sample of SNNFGAILSS amyloid fibrils.}
		\label{tab:bb_trans}
		\end{table}
		
		In summary, the effect of isotropic chemical shift offset on heteronuclear transfer efficiency for $^{\tr{RESPIRATION}}$CP has been understood theoretically and verified experimentally. The theoretical analysis has been put to use in designing a variant that is broadbanded and its validity had been verified by experimental data recorded on 400 MHz and 950 MHz spectrometers. Though the description has been applied to analyse only the RFDR and $^{\tr{RESPIRATION}}$CP pulse sequences, the description could be applied to any amplitude-modulated pulse sequences to find the effective Hamiltonian and gain insights into the experiments.

%% file: Chapters/chap5_GenSeq.tex
\chapter{General Pulse Sequences}
\label{chap:genSeq}
	The previous chapter detailed the formalism to find the effective Hamiltonian in problems where the rf field interactions were only amplitude-modulated. The formalism can not be used to treat problems where the rf field interactions are modulated in both amplitude and phase, since the underlying assumption, that the rf field Hamiltonian at different times commute, is not valid here. In this chapter, a general formalism that enables the representation of a time-dependent rf field interaction frame Hamiltonian, where the rf field is both amplitude- and phase-modulated, in the Fourier space. This in turn enables the use of formulae derived in chapter \ref{chap:DesignPrinciples} for calculating the time-independent effective Hamiltonian. Sec. \ref{sect:5_ff} deals with the formalism, while Sec. \ref{sect:5-c7} illustrates the same by describing C-symmetry based homonuclear dipolar recoupling pulse sequences.\\
	
	\section{Fundamental Frequencies}
	\label{sect:5_ff}
		Similar to the formalism for the amplitude-modulated pulse sequences, given in Sec. \ref{sect:4_ff}, to find the interaction frame Hamiltonian, it is sufficient to find the interaction frame transformation of single-spin operators of individual spins. This is valid as long as the interaction frame transformation is defined by only single-spin operators, and is typically the case with them defined by rf field and/or isotropic chemical shift interactions. Another reason is that two-spin coupling terms in the Hamiltonian can be written as a Kronecker product of two single-spin operators - \textit{separable states}. It was discussed in Sec. \ref{sect:bb_resp} that it is better to transform into a frame defined by both the rf field and the isotropic chemical shift interaction together, rather than a frame defined by just the rf field interaction. Therefore, consider a combined rf field and isotropic chemical shift interaction experienced by a spin $I$, given by,
		\begin{equation}
			\hat{\mathcal{H}}_{\tr{rf+iso}}(t) = \omega_{\tr{I}}^{(0)} \hat{I}_z + \omega_{\tr{rf}}(t)(\hat{I}_x\cos \phi(t) + \hat{I}_y\sin \phi(t)),
		\end{equation}
		where $\omega_I^{(0)} = \omega_0^{\tr{(I)}} - \omega_c$ represents the isotropic chemical shift of spin $I$, with $\omega_0^{(I)}$ denoting the Larmor frequency of spin $I$ and $\omega_c$ denoting the rf carrier frequency of the rf channel. Say the rf field Hamiltonian, and therefore the combined rf and isotropic chemical shift Hamiltonian, $\hat{\mathcal{H}}_{\tr{rf+iso}}(t)$, is periodic with $\tau_m = \frac{2\pi}{\omega_m}$ and under MAS, the internal Hamiltonian is periodic with $\tau_r = \frac{2\pi}{\omega_r}$. The independency of the spatial and spin parts of the internal Hamiltonian can be exploited here. Since the spatial part of the internal Hamiltonian is already represented as a Fourier series, and is unaffected by interaction frame transformation, it can kept out of the discussion concerning spin-part transformations and later appended to interaction frame spin-parts to represent the total interaction frame Hamiltonian. The spin-parts of internal Hamiltonian are not time-modulated as yet.\\
		
		The propagator for the combined rf and isotropic chemical shift Hamiltonian for the interval (0,$\tau_m$), for each spin can be expressed,
		\begin{equation}
			\hat{U}_{\tr{rf+iso}}(\tau_m) = e^{-i\omega_{\tr{cw}} \tau_m \hat{\mathcal{F}}},
		\label{eq:5_2}
		\end{equation}
		where $\hat{\mathcal{F}}$ is a linear combination of single-spin operators of the spin ($I$) under consideration and $\omega_{\tr{cw}}$ is the effective field frequency that produces an average flip of $\omega_{\tr{cw}}t$ on the single-spin operators. The periodicity of interaction frame single-spin operators is therefore $\frac{2\pi}{\gcd(\omega_m,\omega_{\tr{cw}})}$, if $\omega_{\tr{cw}}$ is non-zero and simply $\frac{2\pi}{\omega_m} = \tau_m$ if $\omega_{\tr{cw}}$ is zero. The Fourier space representation of the interaction frame single-spin operators accordingly can be represented with respect to either one or two frequencies and is be shown below.\\
		
		\subsubsection*{\underline{Case: $\omega_{\tr{cw}} = 0$}}
			In the scenario where the propagator given in Eq. \ref{eq:5_2} is identity, i.e., $\omega_{\tr{cw}} = 0$, then the interaction frame single-spin operators are periodic with the same period of the combined rf and isotropic chemical shift Hamiltonian, $\tau_m$. The interaction frame transformation of single-spin operators is given by,
			\begin{equation}
				\hat{I}_{j} \xrightarrow{\hat{\mathcal{H}}_{\tr{rf+iso}}(t)} \sum_{j' = x,y,z}^{} c_{j,j'}(t) \hat{I}_{j'},
			\end{equation}
			where the time-dependent components $c_{j,j'}(t)$ are given by,
			\begin{equation}
				c_{j,j'}(t) = \langle \hat{I}_{j'} | \hat{U}^{\dagger}_{\tr{rf+iso}}(t) \cdot \hat{I}_j \cdot \hat{U}_{\tr{rf+iso}}(t) \rangle
			\label{eq:5-4}
			\end{equation}
			with $j$ and $j'$ each taking any of the three values $x, y$ or $z$. The components $c_{j,j'}(t)$ that are periodic in time with $\tau_m$ (because $\hat{U}_{\tr{rf+iso}}(\tau_m) = \mathds{1}$) can be expressed as a Fourier series given by,
			\begin{equation}
				c_{j,j'}(t) = \sum_{k_{\tr{I}} = -\infty}^{\infty} a_{j,j'}(k) \cdot e^{ik_{\tr{I}}\omega_m t},
			\label{eq:5-5}
			\end{equation}
			where $a_{j,j'}(k)$ are the complex Fourier coefficients. The interaction frame single-spin operators are then explicitly given by
			\begin{equation}
				\hat{I}_j \xrightarrow{\hat{\mathcal{H}}_{\tr{rf+iso}}(t)} \sum_{k_{\tr{I}} = -\infty}^{\infty} \Big(a_{j,x}(k) e^{ik_{\tr{I}}\omega_m t} \cdot \hat{I}_x +a_{j,y}(k) e^{ik_{\tr{I}}\omega_m t} \cdot \hat{I}_y + a_{j,z}(k) e^{ik_{\tr{I}}\omega_m t} \cdot \hat{I}_z\Big),
			\label{eq:5-6}
			\end{equation}
			where $j$ can take any of the values $x, y$ or $z$.
			
		\subsubsection*{\underline{Case: $\omega_{\tr{cw}} \neq 0$}}
			In the scenario where the propagator given in Eq. \ref{eq:5_2} is valid and $\omega_{\tr{cw}} \neq 0$, then the interaction frame single-spin operators are periodic with $\frac{2\pi}{\gcd\big(\omega_m,\omega_{\tr{cw}}\big)}$. Therefore the coefficients $c_{j,j'}(t)$ in Eq. \ref{eq:5-4}, can not be written as a Fourier series with just $\omega_m$ as was done for the case of $\omega_{\tr{cw}} = 0$ in Eq. \ref{eq:5-5}. To overcome this, interaction frame transformation of single-spin operators are rewritten as,
			\begin{equation}
				\begin{alignedat}{2}
					\hat{I}_j \xrightarrow{\hat{\mathcal{H}}_{\tr{rf+iso}}(t)} & \hspace*{20pt}	\hat{U}^{\dagger}_{\tr{rf+iso}}(t) \cdot \hat{I}_j \cdot \hat{U}_{\tr{rf+iso}}(t)\\
					&=  e^{i\omega_{\tr{cw}}\hat{\mathcal{F}}t} \underbrace{e^{-i\omega_{\tr{cw}}\hat{\mathcal{F}}t} \hat{U}^{\dagger}_{\tr{rf+iso}}(t) \cdot \hat{I}_j \cdot \hat{U}_{\tr{rf+iso}}(t) e^{i\omega_{\tr{cw}}\hat{\mathcal{F}}t}}_{\text{periodic with $\tau_m = \frac{2\pi}{\omega_m}$}} e^{-i\omega_{\tr{cw}}\hat{\mathcal{F}}t}
				\end{alignedat}
			\label{eq:5-7}
			\end{equation}
			where a positive and negative rotation about $\hat{\mathcal{F}}$ has been padded. The reason for this is that the term bracketed in Eq. \ref{eq:5-7} is now periodic with $\tau_m$ and therefore its projection can be rewritten as a Fourier series. For mathematical convenience that will be apparent by Eq. \ref{eq:5-12}, a rotated basis \{$x',y',z'$\} is defined such that $\hat{\mathcal{F}}$ is along the $z'$. Projection of the bracketed time-dependent term in Eq. \ref{eq:5-7} on the rotated basis allows the transformation given in Eq. \ref{eq:5-7} to be rewritten as 
			\begin{equation}
				\begin{alignedat}{2}
					\hat{I}_j \xrightarrow{\hat{\mathcal{H}}_{\tr{rf+iso}}(t)} & \hspace*{20pt} e^{i\omega_{\tr{cw}}\hat{\mathcal{F}}t} \Big(\sum_{j'}^{}c^{rot}_{j,j'}(t) \cdot \hat{I}_{j'}\Big) e^{-i\omega_{\tr{cw}}\hat{\mathcal{F}}t}
				\end{alignedat}
			\label{eq:5-9}
			\end{equation}
			where
			\begin{equation}
				c^{rot}_{j,j'}(t) = \langle \hat{I}_{j'} | e^{-i\omega_{\tr{cw}}\hat{\mathcal{F}}t} \hat{U}^{\dagger}_{\tr{rf+iso}}(t) \cdot \hat{I}_j \cdot \hat{U}_{\tr{rf+iso}}(t) e^{i\omega_{\tr{cw}}\hat{\mathcal{F}}t} \rangle,
			\label{eq:5-8}
			\end{equation}
			and the index $j$ can take any of the values ${x,y,z}$ while $j'$ can take any of the values ${x',y',z'}$. Note that only the projection operators $\hat{I}_{j'}$ are in the rotated basis, while the initial operators $\hat{I}_j$ are in the conventional basis. The components $c^{rot}_{j,j'}(t)$ given in Eq. \ref{eq:5-8}, being periodic with $\tau_m$, can now be written as a Fourier series given by,
			\begin{equation}
				c^{rot}_{j,j'}(t) = \sum_{k_{\tr{I}} = -\infty}^{\infty} a_{j,j'}(k) \cdot e^{ik_{\tr{I}}\omega_m t},
			\label{eq:5-10}
			\end{equation}
			where $a_{j,j'}(k)$ are the complex Fourier coefficients. The Fourier series expression in Eq. \ref{eq:5-10} substituted in Eq. \ref{eq:5-9} leads to,
			\begin{equation}
				\begin{alignedat}{2}
				\hat{I}_j \xrightarrow{\hat{\mathcal{H}}_{\tr{rf+iso}}(t)} & \hspace*{20pt} e^{i\omega_{\tr{cw}}\hat{\mathcal{F}}t} \Big(\sum_{j'}^{}\sum_{k_{\tr{I}} = -\infty}^{\infty} a_{j,j'}(k) \cdot e^{ik_{\tr{I}}\omega_m t} \cdot \hat{I}_{j'}\Big) e^{-i\omega_{\tr{cw}}\hat{\mathcal{F}}t}
				\end{alignedat}
			\label{eq:5-11}
			\end{equation}
			Now the rotation defined by $e^{-i\omega_{\tr{cw}}\hat{\mathcal{F}}t}$ in Eq. \ref{eq:5-11} can be acted on the operators $I_{j'}$. But as the $\hat{\mathcal{F}}$ is along $\hat{I}_{z'}$ (as that is how the rotated basis was defined above), it is only the operators $\hat{I}_{x'}$ and $\hat{I}_{y'}$ that will mix with each other in a simple fashion. For a given $k_{\tr{I}}$ the rotation on the operators are given by,
			\begin{equation}
				\begin{alignedat}{2}
					e^{i\omega_{\tr{cw}}\hat{\mathcal{F}}t} \Big(a_{j,x'}(k) \cdot \hat{I}_{x'}\Big) e^{-i\omega_{\tr{cw}}\hat{\mathcal{F}}t} &= a_{j,x'}(k) \Big(\hat{I}_{x'} \cos(\omega_{\tr{cw}}t) - \hat{I}_{y'} \sin(\omega_{\tr{cw}}t)\Big),\\
					e^{i\omega_{\tr{cw}}\hat{\mathcal{F}}t} \Big(a_{j,y'}(k) \cdot \hat{I}_{y'}\Big) e^{-i\omega_{\tr{cw}}\hat{\mathcal{F}}t} &= a_{j,y'}(k) \Big(\hat{I}_{y'} \cos(\omega_{\tr{cw}}t) + \hat{I}_{x'} \sin(\omega_{\tr{cw}}t)\Big),\\
					e^{i\omega_{\tr{cw}}\hat{\mathcal{F}}t} \Big(a_{j,z'}(k) \cdot \hat{I}_{z'}\Big) e^{-i\omega_{\tr{cw}}\hat{\mathcal{F}}t} &= a_{j,z'}(k) \hat{I}_{z'}.
				\end{alignedat}
			\label{eq:5-12}
			\end{equation}
			Replacing the cosines and sines in Eq. \ref{eq:5-12} with their exponential versions given by,
			\begin{equation}
				\cos(\omega_{\tr{cw}}t) = \frac{e^{i\omega_{\tr{cw}}t} + e^{-i\omega_{\tr{cw}}t}}{2},\\
				\sin(\omega_{\tr{cw}}t) = \frac{e^{i\omega_{\tr{cw}}t} - e^{-i\omega_{\tr{cw}}t}}{2i},
			\end{equation}
			and substituting Eq. \ref{eq:5-11} in Eq. \ref{eq:5-10}, the transformation of single spin operators can be neatly expressed as,
			\begin{equation}
				\hat{I}_{j} \xrightarrow{\hat{\mathcal{H}}_{\tr{rf+iso}}(t)} \sum_{j' = x,y,z}^{} c_{j,j'}(t) \hat{I}_{j'},
			\end{equation}
			where
			\begin{equation}
				\begin{alignedat}{2}
					c_{j,x'}(t) &= \sum_{k = -\infty}^{\infty} \sum_{\substack{l=-1\\l\neq0}}^{1} \frac{1}{2}(a_{j,x'}(k) - ila_{j,y'}(k))  e^{il\omega_{\tr{cw}}} e^{ik\omega_mt},\\
					c_{j,y'}(t) &= \sum_{k = -\infty}^{\infty} \sum_{\substack{l=-1\\l\neq0}}^{1} \frac{1}{2}(a_{j,y'}(k) + ila_{j,x'}(k))  e^{il\omega_{\tr{cw}}} e^{ik\omega_mt},\\
					c_{j,z'}(t) &= \sum_{k = -\infty}^{\infty} a_{j,z'}(k)  e^{il\omega_{\tr{cw}}} e^{ik\omega_mt}.
				\end{alignedat}
			\label{eq:5-16}
			\end{equation}
			
			The above formalism will be illustrated by describing C-symmetry pulse sequences, in particular the improved isotropic chemical shift compensation of the POST element of the sequence will be detailed.\\
			
		\section{C-symmetry Homonuclear Dipolar Recoupling Experiments}
		\label{sect:5-c7}
\begin{figure}[!h]
				\centering
				\includegraphics[width=\linewidth]{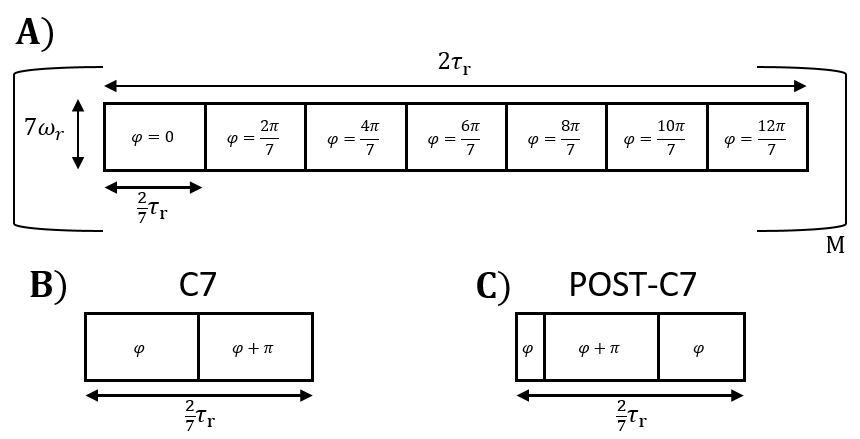}
				\caption{A) A schematic representation of the C7 pulse scheme, where a basic pulse element with a defining phase $\phi$ is repeated seven times over a duration of two rotor periods, each time with an increment in the phase value by $2\pi/7$. Each equally long pulse element has a constant rf field amplitude, equal to seven times the MAS spinning rate thereby performing a . The entire pulse sequence is repeated $M$ times in an experiment to achieve polarisation transfer between two homonuclear coupled nuclei. The basic pulse element in the pulse scheme is either B) C7, where the phase of the pulse is $\phi$ for the first half and $\phi+\pi$ for the second half, with each half performing a $2\pi$ rotation, or C) POST-C7, where the element is divided into three parts performing rotations of $\pi/2$, $2\pi$ and $3\pi/2$ with phase $\phi$, $\phi+\pi$ and $\phi$ respectively.}
				\label{fig:C7_Pulse_Scheme}
			\end{figure}

			In the C7 pulse scheme\cite{lee1995efficient}, a basic pulse sequence element with phase $\phi$ is repeated seven times over a span of two rotor periods, each time with the phase incremented by $\frac{2\pi}{7}$. This is schematically illustrated in Fig. \ref{fig:C7_Pulse_Scheme}A. The basic element of the original C7 pulse sequence, as shown in Fig. \ref{fig:C7_Pulse_Scheme}B, consists of two pulses of flip angle $2\pi$ each, and phases of $\phi$ and $\phi+\pi$. Citing experimental evidence for deficiencies in the dipolar recoupling when the rf carrier frequency is not close to the mean of the isotropic Larmor frequencies of the two nuclei, M. Hohwy et al. performed \textit{high-order} error term analysis to design of the POST element of C7 and to show its improved robustness with respect to chemical shift offsets\cite{hohwy1998broadband}. POST-C7 basic element is merely a permutation of pulses within the C7 basic element and consists of a $\pi/2$ pulse with phase $\phi$, a $2\pi$ pulse with phase $\phi+\pi$ and a $3\pi/2$ pulse with phase $\phi$, in order. This is illustrated in Fig. \ref{fig:C7_Pulse_Scheme}C. Employing the formalism described in the previous section, the effect of isotropic chemical shift on transfer efficiency of the C7 pulse scheme, using both C7 and POST-C7 basic elements will be shown below, using \textit{only} the first-order effective Hamiltonian.
			
			\subsubsection*{\underline{System Hamiltonian:}}
			Consider two homonuclear coupled spin$-\frac{1}{2}$ nuclei, $I_1$ and $I_2$. The rotating frame Hamiltonian, is given for the present problem by,
			\begin{equation}
				\hat{\mathcal{H}}(t) = \hat{\mathcal{H}}_{\tr{int}}(t) + \hat{\mathcal{H}}_{\tr{rf}}(t),
			\end{equation}
			with internal Hamiltonian $\hat{\mathcal{H}}_{\tr{int}}(t)$ given by,
			\begin{equation}
				\begin{alignedat}{2}
					\hat{\mathcal{H}}_{\tr{int}}(t) &= \sum_{n=-2}^{2} \omega^{(n)}_{\tr{I$_1$}} e^{in\omega_rt} \cdot \hat{I}_{1z} + \sum_{n=-2}^{2} \omega^{(n)}_{\tr{I$_2$}} e^{in\omega_rt} \hat{I}_{2z} \\
					&\hspace*{15pt}+ \sum_{n=-2}^{2} \omega^{(n)}_{\tr{I$_1$I$_2$}} e^{in\omega_rt}
					\cdot (2\hat{I}_{1z}\hat{I}_{2z} - \hat{I}_{1x}\hat{I}_{2x} - \hat{I}_{1y}\hat{I}_{2y}),
				\end{alignedat}
			\end{equation}
			and the rf Hamiltonian $\hat{\mathcal{H}}_{\tr{rf}}(t)$
			\begin{equation}
				\begin{alignedat}{2}
					\hat{\mathcal{H}}_{\tr{rf}}(t) &= \omega_{\tr{rf}}(t) \cdot e^{-i\phi(t) (\hat{I}_{1z}+\hat{I}_{2z})} (\hat{I}_{1x} + \hat{I}_{2x}) e^{i\phi(t) (\hat{I}_{1z} + \hat{I}_{2z})}
				\end{alignedat}	
			\end{equation}
			where $\omega_r$ is the MAS spinning rate and the spatial part of the interactions are given as Fourier components $\sum_{n=-2}^{2}\omega_{\lambda}^{(n)}e^{in\omega_rt}$, with $\lambda = \tr{I}_1\tr{I}_2$ for the dipolar coupling and, $\lambda = \tr{I}_1$ and $\lambda = \tr{I}_2$ for the chemical shift of the nuclei $I_1$ and $I_2$ respectively. For simplicity, only the isotropic component of the chemical shift Hamiltonian is considered in this work. However, the calculation of effective Hamiltonian including anisotropic chemical shift Hamiltonian is straightforward, just that it also involves second-order terms. For the both C7 and POST-C7 basic pulse element, the rf field strength is a constant, $\omega_{\tr{rf}}(t) = 7\omega_r$, while the phase $\phi(t)$ is a step function with increments of $2\pi/7$, every $2\tau_r/7$ period.\\
			
			\subsubsection*{\underline{Frequency Domain:}}
			The total Hamiltonian can be represented in an interaction frame defined by the rf field and the isotropic chemical shift of each spin given by,
			\begin{equation}
				\hat{\mathcal{H}}_{\tr{rf+iso}}(t) = \sum_{q=1}^{2}\omega_{\tr{rf}}(t) \cdot e^{-i\phi(t)\hat{I}_{qz}}\hat{I}_{qx}e^{i\phi(t)\hat{I}_{qz}} + \omega^{(0)}_{\tr{I}_q}\cdot\hat{I}_{qz},
			\label{eq:5-20}
			\end{equation}
			where the index $q$ represents the two nuclei. The combined rf field and isotropic chemical shift Hamiltonian is periodic with the same period as that of the rf field Hamiltonian, which is two rotor periods, i.e., $\tau_m = 2\tau_r$. The propagator, as given in Eq. \ref{eq:5_2}, for the combined rf field and isotropic chemical shift Hamiltonian at $\tau_m$ is
			\begin{equation}
				\hat{U}^{{(q)}}_{\tr{rf+iso}}(\tau_m) = e^{-i\omega^{{(q)}}_{\tr{cw}}\tau_m\hat{\mathcal{F}}_q}
			\label{eq:5-21}
			\end{equation}
			where $\hat{\mathcal{F}}_q$ is the axis about which the single-spin operators of spin $\hat{I}_q$ are rotated by flip angle $\omega^{{(q)}}_{\tr{cw}}\tau_m$, every $\tau_m$ period.\\
			
			The spin-part of the interaction frame Hamiltonian, for each spin, is now given by
			\begin{equation}
				\hat{I}_{qj} \xrightarrow[q=\{1,2\}]{\hat{\mathcal{H}}_{\tr{rf+iso}}(t)} \sum_{j'}^{} c^q_{j,j'}(t) \hat{I}_{qj'},
			\end{equation}
			where the index $j$ represents the conventional basis operators $\{x,y,z\}$, while the index $j'$ represents the rotated basis operators $\{x',y',z'\}$ such that $\hat{F}_q$ is along $z'$, when $\omega^{{(q)}}_{\tr{cw}}\neq0$ and represents the conventional basis operators $\{x,y,z\}$, when $\omega^{{(q)}}_{\tr{cw}}=0$. It is noted here that the rotated basis operators, if present, are defined independently for the two spins. The components $c^q_{j,j'}(t)$ are represented as Fourier series given by Eq. \ref{eq:5-5}, when $\omega^{{(q)}}_{\tr{cw}}=0$ and by Eq. \ref{eq:5-16}, when $\omega^{{(q)}}_{\tr{cw}}\neq0$.\\
			
			The total interaction frame Hamiltonian can now be represented as,
			\begin{equation}
				\hat{\tilde{\mathcal{H}}}(t) = \sum_{n=-2}^{2}\sum_{k_1=\infty}^{\infty}\sum_{l_1}^{}\sum_{k_2=\infty}^{\infty}\sum_{l_2}^{}\hat{\tilde{\mathcal{H}}}^{(n,k_1,l_1,k_2,l_2)} e^{in\omega_rt} e^{ik_1\omega^{(1)}_mt} e^{il_1\omega^{(1)}_{\tr{cw}}t} e^{ik_2\omega^{(2)}_mt} e^{il_1\omega^{(2)}_{\tr{cw}}t},
			\label{eq:5-23}
			\end{equation}
			where the index $l_q=0$, when $\omega^{{(q)}}_{\tr{cw}}=0$ and $l_q = \{-1,1\}$, when $\omega^{{(q)}}_{\tr{cw}}\neq0$. It is admittedly tedious, but straightforward to see from Eq. \ref{eq:5-5} and \ref{eq:5-16}, as the case may be, that the Fourier components $\hat{\tilde{\mathcal{H}}}^{(n,k_1,l_1,k_2,l_2)}$ work out to,
			\begin{equation}
				\hat{\tilde{\mathcal{H}}}^{(n,k_1,l_1,k_2,l_2)} = \omega^{(n)}_{\tr{I}_1\tr{I}_2} \hat{\tilde{I}}_1 \hat{\tilde{I}}_2,
			\end{equation}
			with
			\begin{equation}
				\hat{\tilde{I}}_q =
				\begin{cases} 
						\sum\limits_{j,j'}^{} a^{(q)}_{j,j'}(k_q) \hat{I}_{qj'}; \hspace*{10pt} \tr{if } l_q=0
						\begin{cases}
							j' = \{x',y',z'\} & \omega^{{(q)}}_{\tr{cw}} = 0,\\
							j' = \{z'\} & \omega^{{(q)}}_{\tr{cw}} \neq 0,
						\end{cases}&\\
						\frac{1}{2} \sum\limits_{j}^{} \big(a^{(q)}_{j,x'}(k_q) \mp ia^{(q)}_{j,y'}(k_q)\big)\hat{I}_{qx'} + \big(a^{(q)}_{j,y'}(k_q) \pm ia^{(q)}_{j,x'}(k_q)\big)\hat{I}_{qy'}; &  \tr{if } l_q=\pm1,
				\end{cases}
			\end{equation}
			where the summation index $j$ runs over the conventional basis operators $\{x,y,z\}$ and $l=\pm1$ exists only when $\omega^{(q)}_{\tr{cw}} \neq 0$.\\
			
			\subsubsection*{\underline{Effective Hamiltonian:}}
\begin{figure}[!h]
				\centering
				\includegraphics[width=\linewidth]{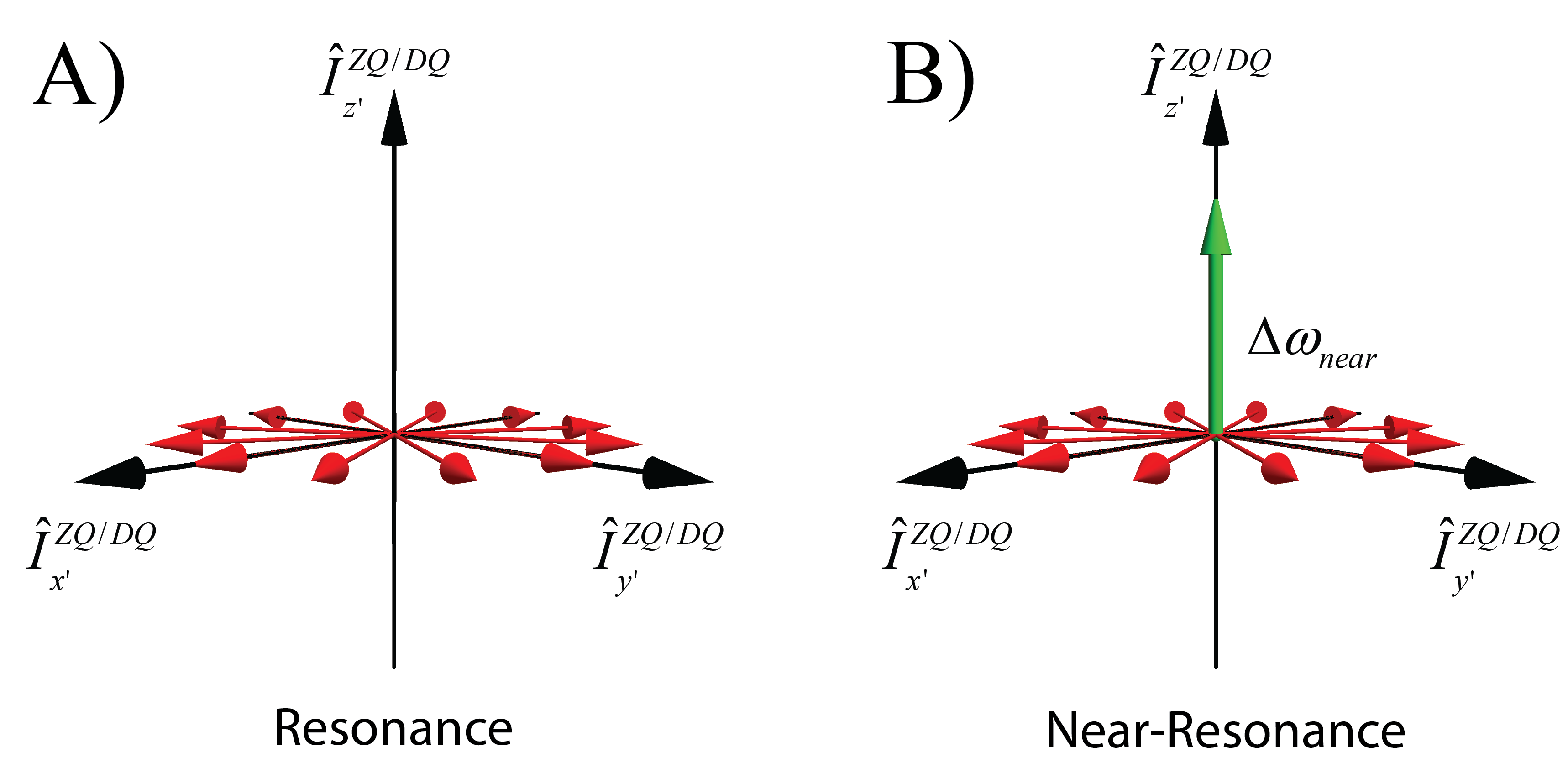}
				\caption{The zero-/double-quantum subspace for the effective Hamiltonian in the rotated basis with red arrows representing the of the effective dipolar coupling Hamiltonian for different crystal angles. In (A) the resonance condition, defined in Eq. \ref{eq:5-26}, is fulfilled whereas in (B) near-resonance condition, defined in Eq. \ref{eq:5-near-res} is fulfilled. The green arrow represents $\Delta\omega_{\tr{near}}$, indicating how close the resonance conditions are fulfilled. This term will be along the z-axis in the zero-/double-quantum subspace, defined in Eq. \ref{eq:5-28}.}
				\label{fig:ZQ-DQ_subspace}
			\end{figure}

			The time-dependent effective first-order Hamiltonian is now given by,
			\begin{equation}
				\hat{\overline{\tilde{\mathcal{H}}}}^{(1)} = \sum_{n,k_1,l_1,k_2,l_2}^{} \hat{\tilde{\mathcal{H}}}^{(n,k_1,l_1,k_2,l_2)},
			\label{eq:5-26}
			\end{equation}
			such that $n\omega_r + k_1\omega^{(1)}_m + l_1\omega^{(1)}_{\tr{cw}} + k_2\omega^{(2)}_m + l_2\omega^{(2)}_{\tr{cw}} = 0$, resonance condition is satisfied. In the C7 pulse scheme, with either C7 or POST-C7 basic element, $\omega^{(1)}_m = \omega^{(2)}_m = \omega_r/2$, whereas $\omega^{(q)}_{\tr{cw}}$ depends on isotropic chemical shift in addition to the rf field. The effective Hamiltonian can be represented in zero-quantum (ZQ) / double-quantum (DQ) subspace with fictitious spin$-\frac{1}{2}$ operators defined as
			\begin{equation}
				\begin{alignedat}{4}
					\hat{I}^{ZQ}_{z'} &= \frac{1}{2}\big(\hat{I}_{1z'} - \hat{I}_{2z'}\big), &\hspace*{40pt} \hat{I}^{DQ}_{z'} &= \frac{1}{2}\big(\hat{I}_{1z'} + \hat{I}_{2z'}\big),\\
					\hat{I}^{ZQ}_{y'} &= \hat{I}_{1y'}\hat{I}_{2x'} - \hat{I}_{1x'}\hat{I}_{2y'}, &\hspace*{40pt} \hat{I}^{DQ}_{y'} &= \hat{I}_{1x'}\hat{I}_{2y'} + \hat{I}_{1y'}\hat{I}_{2x'},\\
					\hat{I}^{ZQ}_{x'} &= \hat{I}_{1x'}\hat{I}_{2x'} + \hat{I}_{1y'}\hat{I}_{2y'}, &\hspace*{40pt} \hat{I}^{DQ}_{x'} &= \hat{I}_{1x'}\hat{I}_{2x'} - \hat{I}_{1y'}\hat{I}_{2y'}.
				\end{alignedat}
			\label{eq:5-27}
			\end{equation}
			In this basis, the effective first-order Hamiltonian given in Eq. \ref{eq:5-26} will be on the ZQ/DQ $x-y$ plane. This is illustrated in Fig. \ref{fig:ZQ-DQ_subspace}A, as red arrows representing the direction for different crystals. The length of red arrows should ideally be of different lengths in agreement with differing strengths of the Hamiltonian for different crystals, but for clarity, they are represented with equal lengths. Such a Hamiltonian enables the transfer over a mixing time $\tau_{mix}$,
			\begin{equation}
				\hat{I}_z = \hat{I}^{ZQ}_{z'} + \hat{I}^{DQ}_{z'} \xrightarrow[\sqrt{c^2 + d^2}t_{mix} = \pi]{c\hat{I}^{ZQ/DQ}_{x'} + d\hat{I}^{ZQ/DQ}_{y'}}
				\begin{cases}
					-\hat{I}^{ZQ}_{z'} + \hat{I}^{DQ}_{z'} = \hat{I}_{2z'}; & \tr{zero-quantum},\\
					\hat{I}^{ZQ}_{z'} - \hat{I}^{DQ}_{z'} = -\hat{I}_{2z'}; & \tr{double-quantum}.
				\end{cases}
			\label{eq:5-28}
			\end{equation}
			
			\subsubsection*{\underline{Near-resonance conditions:}}
			The summation in Eq. \ref{eq:5-26} is carried only over such quintuples $(n,k_1,l_1,k_2,l_2)$ such that the resonance condition, $n\omega_r + k_1\omega^{(1)}_m + l_1\omega^{(1)}_{\tr{cw}} + k_2\omega^{(2)}_m + l_2\omega^{(2)}_{\tr{cw}} = 0$, is satisfied. However when the condition is close to be zero, but not exactly zero, i.e.,
			\begin{equation}
				n\omega_r + k_1\omega^{(1)}_m + l_1\omega^{(1)}_{\tr{cw}} + k_2\omega^{(2)}_m + l_2\omega^{(2)}_{\tr{cw}} = \Delta\omega_{\tr{near}}
			\label{eq:5-near-res}
			\end{equation}
			such that $|\Delta\omega_{\tr{near}}|$ is small\footnote{How small is small enough is discussed a little later in the text.}, it is not intuitive to claim that there is absolutely no contribution whatsoever. In such a circumstance, $\Delta\omega_{\tr{near}}$ can be subtracted from either\footnote{Subtracted from the one such that $l_q = 1$. This is merely for mathematical convenience, and is not a limitation, as presence of $l_q = \pm1$ in the resonance conditions implies that $l_q = \mp1$ is also a part of the resonance conditions.} of the $\omega^{(q)}_{\tr{cw}}$. To do so, say, spin $I_1$ is chosen and the combined rf field and isotropic chemical shift Hamiltonian of spin $I_1$, as given in Eq. \ref{eq:5-20}, can be updated to,
			\begin{equation}
				\hat{\mathcal{H}}_{\tr{rf+iso}}(t) = \underbrace{\omega_{\tr{rf}}(t) \cdot e^{-i\phi(t)\hat{I}_{1z}} \hat{I}_{1x} e^{i\phi(t)\hat{I}_{1z}}+ \omega^{(0)}_{I_1} \cdot \hat{I}_{1z} - \Delta\omega_{\tr{near}}\hat{\mathcal{F}}_{1}} + \Delta\omega_{\tr{near}}\hat{\mathcal{F}}_{1}.
			\label{eq:5-29}
			\end{equation}
			Updating the definition of interaction frame transformation to the bracketed terms, Eq. \ref{eq:5-21} for spin $I$, is now given by,
			\begin{equation}
				\hat{U}^{(1)}_{\tr{rf+iso}}(\tau_m) = e^{-i(\omega^{(1)}_{\tr{cw}} - \Delta\omega_{\tr{near}})\tau_m\hat{\mathcal{F}}_1},
			\end{equation}
			provided $\Delta\omega_{\tr{near}}$ is small enough in magnitude to not change the rotation axis, here, much from the originally calculated axis, $\hat{\mathcal{F}}_1$, in Eq. \ref{eq:5-21}. This requirement translates to $|\Delta\omega_{\tr{near}}|\ll\overline{\omega_{\tr{rf}}(t)}$. The interaction frame Hamiltonian, given in Eq. \ref{eq:5-23}, now additionally contains the last term of Eq. \ref{eq:5-29}, which is not part of the interaction frame transformation definition. It figures as it is in the interaction frame Hamiltonian, since it is along $\hat{\mathcal{F}}_1$. The effective first-order Hamiltonian is correspondingly updated to,
			\begin{equation}
				\hat{\overline{\tilde{\mathcal{H}}}}^{(1)} = \sum_{n,k_1,l_1,k_2,l_2}^{} \hat{\tilde{\mathcal{H}}}^{(n,k_1,l_1,k_2,l_2)} + \Delta\omega_{\tr{near}}\hat{\mathcal{F}}_1,
			\label{eq:5-32}
			\end{equation}
			where $n\omega_r + k_1\omega^{(1)}_m + l_1(\omega^{(1)}_{\tr{cw}}-\Delta\omega_{\tr{near}}) + k_2\omega^{(2)}_m + l_2\omega^{(2)}_{\tr{cw}} = 0$. The additional single-spin operator in the effective Hamiltonian, $\Delta\omega_{\tr{near}}\hat{\mathcal{F}}_1$ can be viewed as $I_{z'}$ of the ZQ/DQ subspace, defined in Eq. \ref{eq:5-27}, as illustrated with a green arrow in Fig. \ref{fig:ZQ-DQ_subspace}B. The total effective Hamiltonian is then no longer on the ZQ/DQ $x-y$ plane, but shifted out of the plane by $\Delta\omega_{\tr{near}}$ amount, and this adversely affects the transfer described in Eq. \ref{eq:5-28}.\\
			
			\subsubsection*{\underline{Simulations}}
\begin{figure}[!h]
				\centering
				\includegraphics[width=\linewidth]{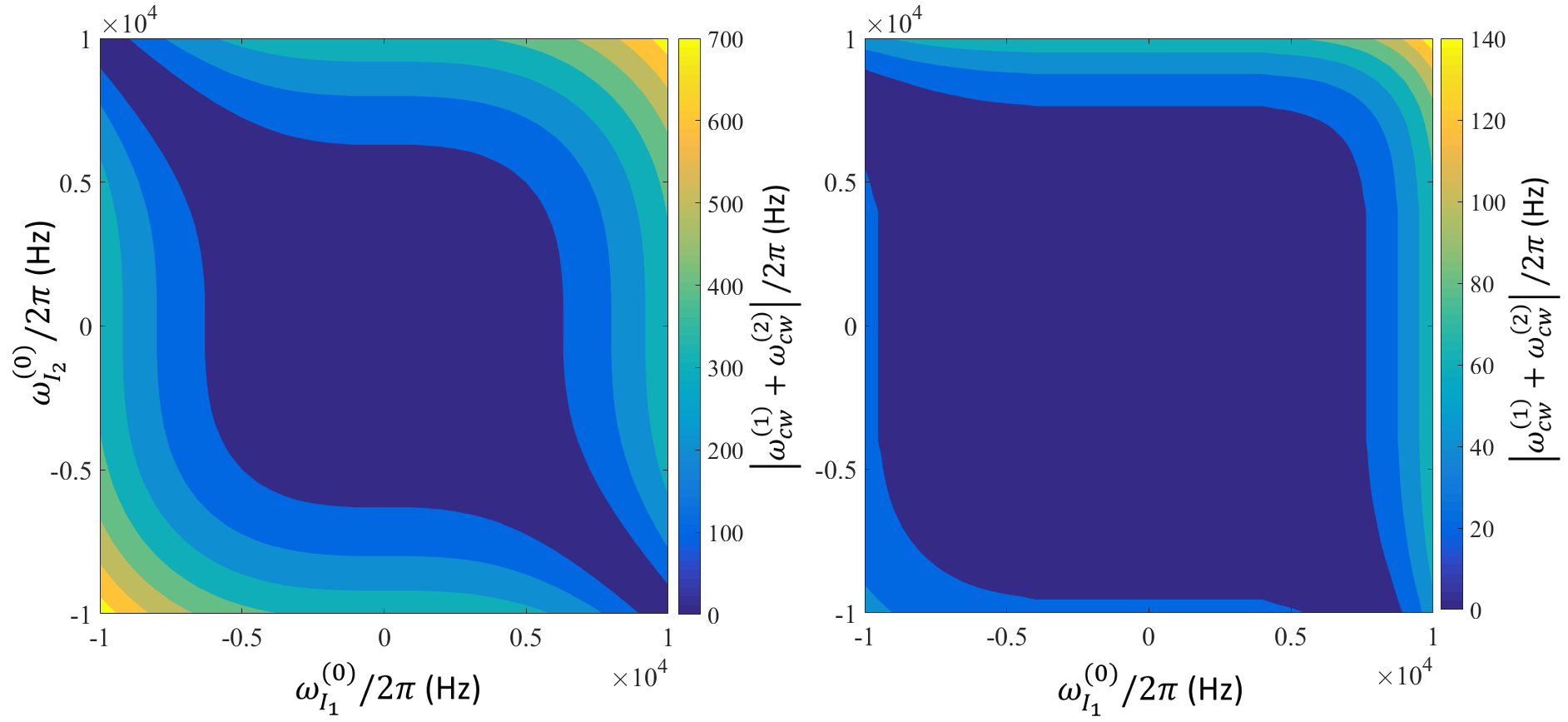}
				\caption{Absolute sum of effective field frequencies of the two spins, $|\omega^{(1)}_{\tr{cw}} + \omega^{(2)}_{\tr{cw}}|$ represented with varying isotropic chemical shift offset on either spin, for C7 (left) and POST-C7 basic pulse elements. This is a measure of how much off the total effective Hamiltonian is, from the x-y plane of the DQ subspace defined in Eq. \ref{eq:5-27}.}
				\label{fig:sum_of_freq}
			\end{figure}

			The effective Hamiltonian, using the description above, was found for the C7 pulse scheme using both C7 and POST-C7 basic elements with $\omega_r/2\pi = 5$ kHz. This corresponds to a constant rf field strength of $\omega_{\tr{rf}}=7\omega_r = 35$ kHz. As $|\Delta\omega_{\tr{near}}| \ll \overline{\omega_{\tr{rf}}(t)} \implies |\Delta\omega_{\tr{near}}| \ll 35$ kHz, the threshold for allowed $|\Delta\omega_{\tr{near}}|$ was set to $<$1 kHz, which happens to be of the same order as the dipolar strength $b_{I_1I_2}/2\pi = 1$ kHz. For both the basic pulse elements, $\omega^{(1)}_m = \omega^{(2)}_m = \omega_r/2 = 2.5$ kHz and therefore the resonance condition can be rewritten as
			\begin{equation}
				n\omega_r + k_1\omega^{(1)}_r/2 + l_1\omega^{(1)}_{\tr{cw}} + k_2\omega^{(2)}_r/2 + l_2\omega^{(2)}_{\tr{cw}} = \Delta\omega_{\tr{near}}.
			\label{eq:5-33}
			\end{equation}
			This implies that if the generated $\omega^{(q)}_{\tr{cw}}$ are smaller relative to $\omega_r/2$, then the $\omega^{(q)}_{\tr{cw}}$ have to cancel among themselves to satisfy the resonance condition, i.e., $\omega^{(1)}_{\tr{cw}} \pm \omega^{(2)}_{\tr{cw}} = \Delta\omega_{\tr{near}}$. This is the case for above mentioned parameters along with isotropic chemical shift in the range of $\pm 10$ kHz. The first term in Eq. \ref{eq:5-32}, which is the recoupled dipolar coupling Hamiltonian, is found to be predominantly DQ, and therefore only $\omega^{(1)}_{\tr{cw}} + \omega^{(2)}_{\tr{cw}} = \Delta\omega_{\tr{near}}$ is of relevance. Fig. \ref{fig:sum_of_freq} shows $|\omega^{(1)}_{\tr{cw}} + \omega^{(2)}_{\tr{cw}}|$, with varying isotropic chemical offset of the two nuclei, for C7 (left) and POST-C7 (right) basic elements. The plot for the basic C7 pulse element suggests that $|\Delta\omega_{\tr{near}}|$, and in turn the deleterious second term in Eq. \ref{eq:5-32}, is significant when the sum of the isotropic chemical shift of the two nuclei is farther from zero, thereby explaining the experimental observation stated by Howhy et al.,
			\begin{center}
					\textquotedblleft \textit{$\dots$ there is experimental evidence for deficiencies in the dipolar recoupling when the rf carrier frequency is not close to the mean of the isotropic Larmor frequencies of the two involved nuclei.} \textquotedblright
			\end{center}
			The suppression of $|\Delta\omega_{\tr{near}}|$ for the entire considered range of isotropic chemical shift values, for POST-C7 basic element, is evident from Fig. \ref{fig:sum_of_freq}B. The maximum value of $|\Delta\omega_{\tr{near}}|$ for POST-C7, reached at isotropic chemical shift values $\omega^{(0)}_{I_1} = \omega^{(0)}_{I_2} = +10$ kHz, is only a fifth of the maximum value for C7 reached at $\omega^{(0)}_{I_1} = \omega^{(0)}_{I_2} = \pm 10$ kHz. The asymmetric nature of the plot obtained for POST-C7, is due to the choice made in the C7 pulse scheme where the $\phi$ is incremented. If instead, the time series of $\phi(t)$ was decremented from $12\pi/7$ to $0$ in steps of $2\pi/2$, a mirrored plot where the maximum $|\Delta\omega_{\tr{near}}|$ is observed at $\omega^{(0)}_{I_1} = \omega^{(0)}_{I_2} = -10$ kHz, would have been obtained. Also, the generated $|\Delta\omega_{\tr{near}}| = |\omega^{(1)}_{\tr{cw}} + \omega^{(2)}_{\tr{cw}}|$ are also smaller than $\omega_r/2$, thus validating the assumption made below Eq. \ref{eq:5-33}.\\
			
\begin{figure}[!h]
				\centering
				\includegraphics[width=\linewidth]{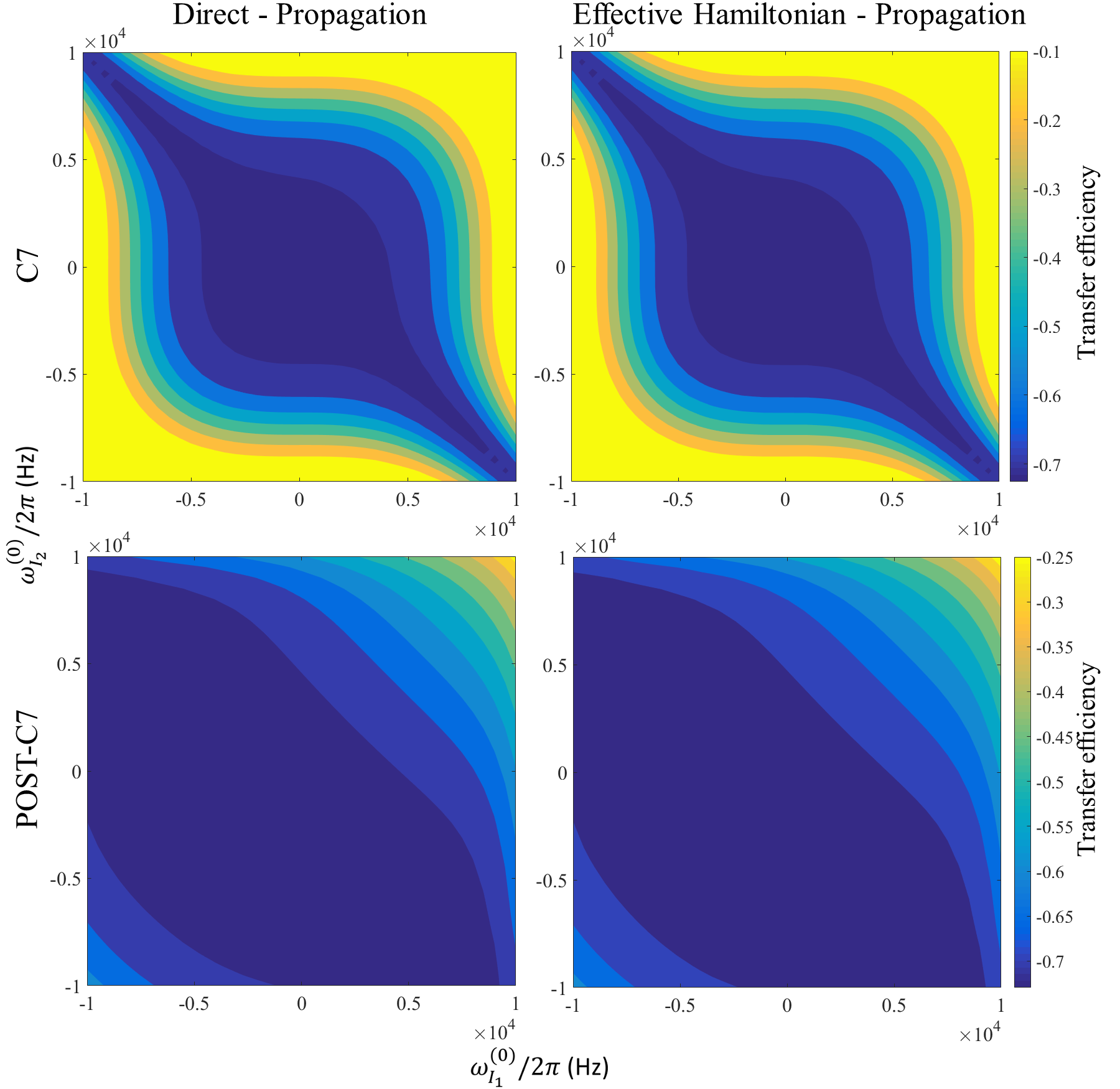}
				\caption{Numerical simulations for the $\hat{I}_1\rightarrow\hat{I}_2$ transfer efficiencies with varying isotropic chemical shift offsets of spins $I_1$ and $I_2$ for C7 basic pulse element (top row) and POST-C7 basic pulse element (bottom row). Simulations were performed using direct-propagation with SIMPSON software package (left column) and using the effective Hamiltonian (right column) given in Eq. \ref{eq:5-32}. All simulations were done for 5.0 kHz MAS with total mixing time set to 3.2 ms (M=8). The dipolar coupling constant was set to $b_{I_1I_2}/2\pi = 1$ kHz and powder averaging was performed using 11 $\gamma_{CR}$ and 320 $\alpha_{CR}, \beta_{CR}$ crystallite angles.}
				\label{fig:c7_prop}
			\end{figure}

			Fig. \ref{fig:sum_of_freq}, that shows how much off-plane is the total effective Hamiltonian from the DQ $x-y$ plane, is indicative of transfer efficiency. However to verify the prediction, direct propagation numerical simulations, where the initial density operator $\hat{I}_{1z}$ is evolved under the rf field and internal interaction Hamiltonians to a mixing time $\tau_{mix}$ and the resulting density operator is projected on $\hat{I}_{2z}$. Transfer efficiencies, so calculated, with varying isotropic chemical shift offsets of both nuclei, are shown in Fig. \ref{fig:c7_prop}(left) for C7 (top) and POST-C7 (bottom) basic pulse elements. $\tau_{mix}$ was constant for the entire grid, and chosen to be the time of maximum transfer for $\omega^{(0)}_{I_1} = \omega^{(0)}_{I_2} = 0$ kHz. The results are in good agreement with the prediction from Fig. \ref{fig:sum_of_freq}, which consists only the flip angles and the corresponding rotation axes effected on individual spins by one block of a repeating rf field combined with isotropic chemical shift of the spin. Calculation of the flip angles and the corresponding rotation axes for individual spins are simpler and faster, as they only involve single-spin operators.
			
			The transfer efficiencies were also computed using the total effective Hamiltonian, given in Eq. \ref{eq:5-32}, to verify the formalism described above and are shown in Fig. \ref{fig:c7_prop} (right) for C7 (top) and POST-C7 (bottom) basic pulse elements. Recalling discussion in Sec. \ref{sect:MultipleFrequencies}, the effective Hamiltonian computed here is valid only at times defined by the greatest common divisor of all five fundamental frequencies. However, it was also shown that propagator can be found at multiples of time defined only by the spinning and modulation frequencies, using only time-independent Hamiltonian, as given in Eq. \ref{eq:modProp},
			\begin{equation}
				\hat{U}(n\tau_c) = e^{-i\hat{\overline{\tilde{\mathcal{H}}}}^{(1)}\cdot n\tau_c} \cdot e^{i\sum\limits_{q=1}^{2}\omega^{(q)}_{\tr{cw}}\hat{\mathcal{F}}_q\cdot n\tau_c},
			\end{equation}
			where $\tau_c = \frac{2\pi}{\gcd(\omega_r,\omega^{(1)}_m,\omega^{(2)}_m)} = 2\tau_r$ and the effective first-order Hamiltonian $\hat{\overline{\tilde{\mathcal{H}}}}^{(1)}$ is defined in Eq. \ref{eq:5-32}. The transfer efficiencies so calculated, shown in Fig. \ref{fig:c7_prop} (right) are in great agreement with corresponding direct-propagation simulations (left).\\

			In summary, the better offset compensation of POST-C7 basic element over the original C7 basic element is shown using just effective first-order Hamiltonian. Howhy et al., resorted to finding higher-order terms to show the same, whose need was nullified here by treating the problem in an interaction frame defined by both rf field and isotropic chemical shift. The flip angle imparted by repeating a pulse element together with isotropic chemical shift on a spin has been shown to be of great value, in predicting the total effective Hamiltonian.

%% file: Chapters/chap6_Concl.tex
\chapter{Conclusion}
	In this work, the equivalence between AHT and Floquet theory with regard to finding effective time-independent Hamiltonian for a time-dependent Hamiltonian has been revisited. The equivalence is shown to rest on being able to represent the time-dependent Hamiltonian in the Fourier space with finite number of fundamental frequencies and a description to represent any time-dependent Hamiltonian with no more than two fundamental frequencies per involved spin is developed and detailed. The description has been put to use to describe, understand and design variants of established dipolar recoupling pulse sequences, with focus on the effects of chemical shift on magnetisation transfer.\\
	
	The RFDR pulse sequence, a homonuclear dipolar recoupling experiment, has been described to explain the effect of isotropic chemical shift on the transfer efficiency. The temporal placements of the $\pi$ pulses figure in the effective Hamiltonian and the form of it enabled the design of an adiabatic variant of the pulse sequence, where the placements of the $\pi$ pulses were shifted gradually throughout the mixing time, which was shown to significantly improve the transfer efficiency theoretically and the results are verified through experiments.\\
	
	The $^{\tr{RESPIRATION}}$CP pulse sequence, a heteronuclear dipolar recoupling experiment, was described using AHT without any assumptions on the constituent pulses, that limited the description in the original work. It was established that for the recoupling conditions to exist, it is enough to match the effective rotations imparted on the individual spins by the combined rf field and isotropic chemical shift Hamiltonians. This lead to a natural modification of the experiment to create a variant that sustains recoupling conditions over a larger range of isotropic chemical shift values. The new variant was verified experimentally to be significantly more broadband than the original $^{\tr{RESPIRATION}}$CP and DCP experiments. This enabled simultaneous recording of NCO and NC$_{alpha}$ experiments with modest rf power, otherwise possible only with ultra-high-power DCP or dual-band selective transfers.\\
	
	The above two experiments were described using a formalism developed for handling pulse sequences that are only amplitude modulated. Based on the findings in the BB-$^{\tr{RESPIRATION}}$CP work, a general formalism to describe experiments, where the pulse sequences are both amplitude and phase modulated, was developed and detailed. The general formalism was used to describe effect of isotropic chemical shift on C7 pulse schemes, and the better offset compensated POST-C7 element over the original C7 element was explained using just first-order calculations. In this case too, the simpler calculations of effective flip angles imparted by the combined rf field and isotropic chemical shift on the individual spins proved to be a useful preview of what to expect from the total effective Hamiltonian calculation.
	
	Though the description has been applied only to a select few pulse sequences to analyse and design improved variants, the theory has a larger potential for the design of advanced solid-state NMR experiments. In addition to the insights it provides, the effective Hamiltonian so calculated can also be applied to achieve faster numerical simulations, which will be apparent for larger spin systems, as the spins are treated individually and are combined to find two spin operators only in the end to construct the effective Hamiltonian matrix. Numerical optimisations of pulse sequences have mostly been focussed on optimising state-to-state transfers and effective Hamiltonian over the entire mixing time or only over such times where there is no net rotation imparted by the rf field on the corresponding spins. The theory detailed in this dissertation urges for a rethink of the pulse design optimisation approach, as the effective Hamiltonian can be found even for non-cyclic times and therefore allows more freedom and guidance for the optimisation procedures.

%% file: Chap99_Appendix/Append_adRFDR_sourcecode.tex
\appendix
\chapter*{Appendix A. SIMPSON script to find optimised adiabatic RFDR pulse sequence}
	Below is a script that can be executed via SIMPSON to find the best sweeps via a grid search. One can change the spin system directly in the spinsys section. In par section, an end user can change the spectrometer field (\texttt{proton$\_$frequency}), the MAS frequency (\texttt{spin$\_$rate}) and the duration of the $\pi$ pulse (\texttt{p180}). In the main section, it is possible to control the number of XY-8 blocks via an internal list (\texttt{lN}), the sweeps via an internal list (\texttt{lsweep}) and the tangential cut-off values via an internal list (\texttt{ltco}). Executing the program, a vdlist for implementing in topspin is generated for a given $N$. Likewise, the build-up curve is provided.
	\begin{lstlisting}[basicstyle=\scriptsize,
	breaklines=true,
	upquote=true,
	aboveskip={1.5\baselineskip},
	columns=fixed,
	showstringspaces=false,
	extendedchars=false,
	frame=single,
	showtabs=false,
	showspaces=false,
	showstringspaces=false,
	identifierstyle=\ttfamily,
	backgroundcolor={\color[rgb]{0.9,0.9,0.9}},
	keywordstyle={\color[rgb]{0,0,1}},
	commentstyle={\color[rgb]{0.026,0.112,0.095}},
	stringstyle={\color[rgb]{0.627,0.126,0.941}},
	numberstyle={\color[rgb]{0.205, 0.142, 0.73}},
	language=tcl]
spinsys {
## SET UP SYSTEM ACCORDING TO TABLE 2 IN BAK et al, JMR 154, 28 (2002)
	channels 13C
	nuclei 13C 13C
	shift 1 60p -76p 0.9 0 0 94
	shift 2 -60p -20p 0.43 90 90 0
	dipole 1 2 -2142 0 90 120.8
}
par {
	method cheby2
	proton_frequency 400e6
	start_operator I1z
	detect_operator I2z
	spin_rate 10000.0
	crystal_file rep66
	gamma_angles 9
	np -1
	# np gets changed according to N
	sw spin_rate/8.0
	## TIME
	variable tr 1.0e6/spin_rate
	variable p180 5.0
	## RF
	variable rf_p180 0.5e6/p180
	## SWEEP
	variable tsweep 0
	# CHANGE tsweep via lsweep in main
	variable tco 1
	# CHANGE tco via ltco in main
	## EXTRA
	string form %.2f
	# form gives number of digits in delay.
	string save_delays no
	# save or no
}
proc pulseq {} {
	global par vd_times
	
	## THIS PULSE SEQUENCE EMPLOYS XY-8 PHASE CYCLING OF THE PI PULSES
	reset
	store 0
	acq
	foreach lt $vd_times {
		set t1 [lindex $lt 0]; set t2 [lindex $lt 1]
		# only delays before pi pulses are used to make sure that prop is
		synchronized
		4
		reset
		delay $t1
		store 1
		reset [expr {$t1+$par(p180)}]
		delay [expr {$par(tr)-$t1-$par(p180)}]
		store 2
		reset
		delay $t2
		store 3
		reset [expr {$t2+$par(p180)}]
		delay [expr {$par(tr)-$t2-$par(p180)}]
		store 4
		reset
		prop 0
		prop 1
		pulse $par(p180) $par(rf_p180) x
		prop 2
		prop 3
		pulse $par(p180) $par(rf_p180) y
		prop 4
		
		prop 1
		pulse $par(p180) $par(rf_p180) x
		prop 2
		prop 3
		pulse $par(p180) $par(rf_p180) y
		prop 4
		prop 1
		pulse $par(p180) $par(rf_p180) y
		prop 2
		prop 3
		pulse $par(p180) $par(rf_p180) x
		prop 4
		
		prop 1
		pulse $par(p180) $par(rf_p180) y
		prop 2
		prop 3
		pulse $par(p180) $par(rf_p180) x
		prop 4
		
		store 0
		acq
		5
	}

}
proc main {} {
	global par vd_times
	
	## SET lN
	set lN [list 1 2 3 4 5 6 7 8 9 10]; puts "lN: $lN"
	# N is the number of XY-8 blocks
	## SET lsweep
	#set lsweep [list 0.0]
	set lsweep {}
	set sdelta 0.1
	
	for {set s 0} {$s <= 200} {incr s 1} {
		lappend lsweep [format $par(form) [expr {0.0+$s*$sdelta}]]
	};
	
	puts "lsweep: $lsweep"
	## SET ltco IN DEGREES
	#set ltco [list 1]
	set ltco {}
	
	for {set t 1} {$t <= 89} {incr t 1} {
		lappend ltco $t
	};
	
	puts "ltco: $ltco"
	set name COCa
	set name
	$name\_mas$par(spin_rate)\_$par(crystal_file)\g$par(gamma_angles)\_pf[format
	"%.0f" [expr {$par(proton_frequency)/1.0e6}]]MHz
	set log_time [open $name.log.time w]
	puts $log_time "\# N(all) Nmax max tsweep tco duration"
	
	foreach N $lN {
		set par(np) [expr {$N+1}];# np includes a zero-point
		set duration [expr {$N*8*$par(tr)}]
		set par(save_delays) "no"
		set transf_max -1e6
		set Npoint_max -1
		set sweep_max -1
		set tco_max -1
		
		## GRID SEARCH IN tsweep and tco FOR GIVEN N
		foreach par(tsweep) $lsweep {
			foreach par(tco) $ltco {
				set vd_times [p_list_time $par(tr) $par(tsweep) $par(tco)
				$par(p180) $N $par(form) $par(save_delays)]
				# vd_times GENERATES LISTS OF DELAYS
				set f [fsimpson]
				set lmax [p_find_max $f]
				set npoint [lindex $lmax 0]
				# npoint includes initial zero-time point
				6
				set transf [lindex $lmax 1]
				if {$transf > $transf_max} {
					set Npoint_max $npoint
					set transf_max $transf
					set sweep_max $par(tsweep)
					set tco_max $par(tco)
				}
				puts "N: $N (duration: $duration), tsweep: $par(tsweep),
				tco: $par(tco); transfer: $transf vs max: $transf_max "
				funload $f
			}
		}
	
		## RECALCULATE BEST SHAPE
		set par(tsweep) $sweep_max
		set par(tco) $tco_max
		set par(save_delays) "save"
		set vd_times [p_list_time $par(tr) $par(tsweep) $par(tco) $par(p180)
		$N $par(form) $par(save_delays)]
		set f [fsimpson]
		set lmax [p_find_max $f]
		set Npoint_max [lindex $lmax 0]
		set transf_max [lindex $lmax 1]
		puts $log_time "$N [expr {$Npoint_max-1}] $transf_max $sweep_max
		$tco_max $duration"
		fsave $f $name\_N$N\_max[format $par(form) $transf_max].fid
		fsave $f $name\_N$N\_max[format $par(form) $transf_max].xy -xreim
		funload $f
	}
	close $log_time
}
	
	proc p_list_time {tr tsweep tco p180 N form save} {
		global par
		if {$N < 1} {puts "$N should be >=1"; exit}
		
		set lsweep {}; # not same sweep as in main
		set lsweepBruker {}
		if {$N == 1} {
			# NORMAL RFDR
			set tFor [expr {($tr-$p180)/2.0}]; # TIME BEFORE FINITE PI-PULSE
			set tRev $tFor ; # TIME BEFORE FINITE PI-PULSE IN SECOND PERIOD
			lappend lsweep [list $tFor $tRev]
		}
		if {$N > 1} {
			set pi [expr {acos(-1.0)}]
			set tcoRad [expr {$pi/180.0*$tco}]
			for {set i 1} {$i <= $N} {incr i} {
				set xRad [expr {$tcoRad*(-1.0+2.0*($i-1)/($N-1))}]
				7
				set Dt [expr {$tsweep/2.0*tan($xRad)/tan($tcoRad)}]
				set tFor [format $form [expr {$tr/2.0+$Dt-$p180/2.0}]]; # DELAY
				BEFORE PI IN FIRST TR
				set tFor2nd [format $form [expr {$tr-$tFor-$p180}]]; # DELAY
				AFTER PI IN FIRST TR
				set tRev [format $form [expr {$tr/2.0-$Dt-$p180/2.0}]]; # DELAY
				BEFORE PI IN SECOND TR
				set tRev2nd [format $form [expr {$tr-$tRev-$p180}]]; # DELAY
				AFTER PI IN SECOND TR
				lappend lsweep [list $tFor $tRev]
				lappend lsweepBruker [list $tFor $tFor2nd $tRev $tRev2nd $tFor
				$tFor2nd $tRev $tRev2nd $tFor $tFor2nd $tRev $tRev2nd $tFor $tFor2nd $tRev
				$tRev2nd]
			}
		}
		if {[string equal $save "save"]} {
			set bruker_input COCa_tsw$tsweep\_tco$tco\_p180_$p180\_N$N.input
			set brukerList [open $bruker_input w]
			foreach lt $lsweepBruker {
				set t1 [lindex $lt 0]; set t2 [lindex $lt 1]
				set t3 [lindex $lt 2]; set t4 [lindex $lt 3]
				puts $brukerList "$t1\u\n$t2\u\n$t3\u\n$t4\u"
				puts $brukerList "$t1\u\n$t2\u\n$t3\u\n$t4\u"
				puts $brukerList "$t1\u\n$t2\u\n$t3\u\n$t4\u"
				puts $brukerList "$t1\u\n$t2\u\n$t3\u\n$t4\u"
			}
			close $brukerList
		}
		return $lsweep
	}

proc p_find_max {f} {
	# examines all np for max
	
	set np [fget $f -np]
	set sw [fget $f -sw]
	set np_max 1; set value_max [findex $f 1 -re]
	for {set i 2} {$i <= $np} {incr i} {
		set value_temp [findex $f $i -re]
		if {$value_temp > $value_max} {
			set value_max $value_temp
			set np_max $i
		}
	}
return [list $np_max $value_max]
}
	\end{lstlisting}